# Improving the State of the Art for Training Human-AI Teams

**Technical Report #2**

# Results of Researcher Knowledge Elicitation Survey

**March 2023**

*Jim McCarthy, Lillian Asiala, LeeAnn Maryeski, Dawn Sillars*

*Sonalysts, Inc.*

This Page Intentionally Blank

# TABLE OF CONTENTS

# LIST OF FIGURES

# LIST OF TABLES

# ACRONYMS

This Page Intentionally Blank

# ACKNOWLEDGEMENTS

We would like to acknowledge the members of the research community who donated their time and expertise to help us develop and complete this survey:

        Sarah Bibyk
        David Grimm
        Lixiao Huang
        Florian Jentsch
        Joan Johnston
        Andrea Krausman

We would also like to express our appreciation to those participants who asked to remain anonymous.

This Page Intentionally Blank

# ABSTRACT

A consensus report produced for the Air Force Research Laboratory (AFRL) by the National Academies of Sciences, Engineering, and Mathematics documented a prevalent and increasing desire to support human-Artificial Intelligence (AI) teaming across military service branches. Sonalysts has begun an internal initiative to explore the training of Human-AI teams. The first step in this effort is to develop a Synthetic Task Environment (STE) that is capable of facilitating research on Human-AI teams. Our goal is to create a STE that offers a task environment that could support the breadth of research that stakeholders plan to perform within this domain. As a result, we wanted to sample the priorities of the relevant research community broadly, and the effort documented in this report is our initial attempt to do so. We created a survey that featured two types of questions. The first asked respondents to report their agreement with STE features that we anticipated might be important. The second represented open-ended questions that asked respondents to specify their priorities within several dimensions of the anticipated STE. The research team invited nineteen researchers from academic and Government labs to participate, and 11 were able to complete the survey. The team analyzed their responses to identify themes that emerged and topics that would benefit from further analysis. The most significant finding of the survey was that a number of researchers felt that various open-source STEs that would meet our needs already exist. Researchers also emphasized the need for automated transcription and coding tools to ease the burden of assessing inter-team communications; the importance of robust data capture and export capabilities; and the desirability of extensive flexibility across many aspects of the tool.

This Page Intentionally Blank

# 1 BACKGROUND

In 2021, the United States Air Force Research Laboratory (AFRL) Human Performance Wing asked the National Academies of Sciences, Engineering, and Medicine (NASEM) to produce a consensus report to examine the role of Artificial Intelligence (AI), particularly as part of Human-AI teams. The goal of this effort was to allow AFRL to better support the design of future systems in which humans are teamed with intelligent agents to achieve mission objectives.

The NASEM report identified nine focus areas within the broader Human-AI-teaming domain:

1. Training Human-AI Teams
2. AI Transparency and Explainability
3. Human-AI Team Interaction
4. Trusting AI Teammates
5. Human-AI Teaming Processes and Effectiveness
6. Human-AI Teaming Methods and Models
7. Situation Awareness in Human-AI Teams
8. Identification and Mitigation of Bias in Human-AI Teams
9. Human-System Integration Processes and Measures of Human-AI Team Collaboration and Performance

Sonalysts has begun an internal initiative to explore the first of these Human-AI-teaming domains – Training Human-AI Teams. Table 1 reproduces a portion of the NASEM report in which the authors distributed the six research needs over three periods.

**Table 1: NASEM Training Research Needs over Time**

|  | Near-Term (1-5 Years) | Mid-Term (6-10 Years) | Far-Term (11-15 Years) |
|---|---|---|---|
| Training Human-AI Teams | 9.1: Developing Human-Centered Human-AI Team-Training Content | | |
|  | 9.2: Testing and Validating Traditional Team Training Methods to Inform New Methods | | |
|  | | 9.3: Training to Calibrate Human Expectations of Autonomous Teammates | |
|  | 9.4: Designing Platforms for Human-AI Team Training | | |
|  | | 9.5: Adaptive Training Materials Based on Differing Team Compositions and Sizes | |
|  | | 9.6: Training That Works Toward Trust in Human-AI Teaming | |

As documented elsewhere, the first step in our research plan is the development of a Synthetic Task Environment (STE) that will provide a validated research environment in which our Human-AI teams can perform. To arrive at a set of requirements that will maximize the applicability of our research, Sonalysts is engaging with leading researchers and Subject-Matter Experts (SMEs) within the field. This report summarizes the findings of the first of these outreach efforts with leading researchers in Human-AI teaming from universities and Government labs.

## 2  OVERVIEW

To increase operating tempo while simultaneously considering a much greater array of data and potential courses of action, it is likely that human decision-makers will be teamed with AI teammates much more extensively than they have been in the past. Further, we envision that this will be true "teaming." Our research will assume that autonomous agents will eventually be able to work with humans as peers/teammates and not merely as tools. To achieve this, we imagine that Human-AI Teams will involve **autonomous systems** that use AI or similar technologies to make decisions and/or take actions within a specified domain. These systems will have the ability to respond to novel performance challenges, and will be capable of coordination and cooperation. With the rise of not only independent, but **interdependent**, AI teammates, teaming dynamics are likely to shift, and new teaming principles will emerge. Thus, an effective teaming environment is needed to conduct research on the new principles. The purpose of the effort documented here was to work with members of the research stakeholder community to define the characteristics of such an environment.

## 3  SURVEY DEVELOPMENT AND ANALYSIS

The purpose of working with leading researchers in Human-AI teaming was to gain insight into the research questions, independent variables, and performance measures that interested researchers. This insight will guide the development of the teaming testbed and ensure that we develop it in ways that facilitate current and emerging research agendas. Our goal is to create a STE that offers a relevant task environment on a useful system architecture; has appropriate scenario authoring and modification capabilities; and facilitates data collection and analysis. In addition, since this is an autonomous teaming testbed, we wanted to use this opportunity to investigate the characteristics of autonomous teammates that would be useful for the studies that researchers are planning.

Using our research plan as a foundation, the research team identified various topics for which researchers might be able to provide useful inputs. Working from these knowledge elicitation objectives, the team then developed and reviewed a range of open-ended and Likert-style questions that would allow us to gather insights regarding the beliefs and priorities of the Human-AI teaming research community. Following our internal quality assurance process, the lead researchers "authored" the survey within the Survey Monkey delivery tool. Prior to releasing the survey, we conducted a pilot test during which a Human-AI teaming researcher completed the survey and provided feedback on the survey content. The pilot test led to a different approach to organizing questions for each survey topic; we adopted the strategy of asking respondents to provide a broad description of their experience with each survey topic, and then probed further by asking more specific open-answer questions and close-ended Likert type questions. For example, when asking about data collection in the STE, the survey asked respondents to list variables they were interested in collecting for studies on Human-AI teaming. After listing their variables, respondents were asked to rate their agreement using a scale from 1 (strongly disagree) to 5 (strongly agree) regarding the importance of specific variables and other data collection-related features of the STE. Each sub-section of the survey finished by asking participants for additional ideas, comments, or questions regarding the topic.

Following the pilot test, the research team released the survey to 19 Human-AI teaming researchers from universities and Government research labs. Each researcher received a unique link that allowed him/her to complete the survey across several sessions, if desired. The unique links also allowed us to monitor the progress of respondents as they completed the survey. The lead researcher provided reminder emails to encourage a high response rate. In the end, 11 respondents completed parts of the survey, and eight respondents completed the entire survey.

When the response period ended, the research team downloaded the results from Survey Monkey for offline analysis. The analysis of the survey results occurred in two threads. First, the two lead researchers conducted independent thematic analyses of the responses for each open-ended question. Within these analyses, each researcher reviewed respondents' answers to each question and captured them as a short phrase/theme. The goal was to identify ideas that recurred across answers, even if respondents used different wording. After completing the independent analyses, the researchers worked together to establish consensus between their lists to create one master list. We then mapped individual responses to the consensus list.

Second, in parallel with this qualitative analysis of the open-ended questions, the research team conducted a quantitative analysis of the various Likert-like items. These analyses focused on descriptive statistics summarizing the distribution of responses across the respondents.

The lead researchers then documented the survey findings, and our interpretation of those findings, in this technical report. The research team put the report through an internal quality assurance process, archived it, and distributed the report to stakeholders (including survey respondents).

The following sections summarize the inputs we received as they pertain to various elements of the STE. The raw survey responses are provided in Appendix A.

## 4 STE FEATURES

An analysis of the literature that we reviewed identified a number of features that might be useful to include within the STE. These features include aspects of authoring, architecture, data collection, etc. In this portion of the survey, we assessed the importance of these features globally. Later, we provide a more detailed review of each feature class.

### 4.1 Analysis of Discrete Features

In addition to our reading of the literature, Sonalysts' experience in modeling, simulation, and training allowed us to predict a range of features that we thought researchers would want included in the STE. Within the survey, we asked respondents to indicate the degree to which they agreed or disagreed with our hypotheses about which specific features would be important within the STE. Pie charts and descriptive statistics for each of the domain-related features are shown in Appendix B.

In this section, we focus attention on the items that had the greatest difference of opinion among the respondents. These items may represent topics that would benefit from further discussion. To establish a reasonable basis of comparison across the various sections, the research team decided to use a standard deviation of greater than or equal to 1.0 as our threshold. The team flagged items above that threshold for further consideration.

In this case, only one item met this standard. Respondents demonstrated a wide range of responses to "The STE should include support for a broad range of human and AI teammates." The standard deviation for this item was 1.43. The most frequent response was "strongly agree," followed by equal numbers of "agree," "neither agree nor disagree," and "disagree."

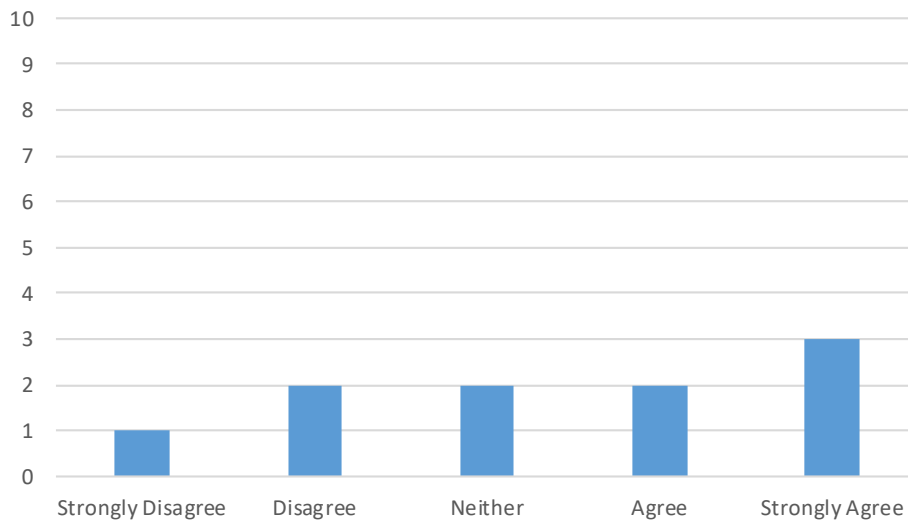

Figure 1. Agreement ratings for "The STE should include support for a broad range of human and AI teammates."

### 4.2   Analysis of Open-ended Questions

To augment our list of valuable features, we asked respondents to identify features that they thought would be important to include within a research testbed. We used two questions to solicit this input:

1. Please brainstorm up to 10 STE features that would be important to you as a researcher (*e.g.,* open source, performance assessment capabilities, flexible teaming configurations, *etc.*).
2. Do you have any additional ideas, comments, or questions regarding STE features? If so, please add them here.

Table 2 summarizes the themes that appeared in the respondents' answers to these questions.

Table 2: Analysis of Responses to Open-ended Questions Focused on Critical STE Features

| Question | Analysis |
|---|---|
| Please brainstorm up to 10 STE features that would be important to you as a researcher (e.g., open source, performance assessment | Most of the features that respondents suggested for the STE fell into broad themes that included system architecture, task domain, teaming, data collection and analysis, autonomy, and ease of use. |

| Question | Analysis |
|---|---|
| capabilities, flexible teaming configurations, etc.). | Among *system architecture* features, five respondents recommended an open-source architecture. Respondents also recommended flexibility (n=2), a modular infrastructure with local and distributed capabilities (n=1), and debrief functions (n=1). The importance of flexibility was echoed by respondents who desired an easily modifiable, modular task environment (n=3).<br><br>We also received the recommendation to use an existing testbed, possibly one with automated performance assessment capabilities.<br><br>When discussing the *task domain*, the notion of fidelity was present in several responses, such as incorporating "high fidelity task demands" (n=1) and military relevance (n=1). More generally, respondents felt the STE needed to be a good place for authentic training and practice (n=1). Respondents identified several features of the task environment that they felt would be important to include (n=1 in all cases):<br><br>• Presenting a theory-informed task<br>• Presenting opportunities for advanced technologies to assist decision-making<br>• Presenting task with sufficient complexity to result in sub-optimal performance of autonomous teammates<br>• Incorporating team stressors<br>• Providing the ability to manipulate information load<br><br>The third broad area that attracted respondents' attention was the nature of the *teaming* that the STE should support. For example, respondents noted that the task environment must require teaming interdependence (n=2), flexible teaming configurations (n=2), and multiple roles and stations for humans and AI teammates (n=1). Recommendations for team size ranged from smallest team recommendations at 4 (n=1) and 6 (n=6) to largest at 12 (n=2).<br><br>Suggested *data collection* features were broad with little overlap. The suggestion with the most overlap was that the STE should capture multiple data formats, including testbed logs, video, audio, and verbal communication (n=4). Two suggestions involved incorporating instrumentation and back-end analytics to support detailed data recording. |

| Question | Analysis |
|---|---|
| | Respondents recommended that ***autonomous teammates*** in the STE should have multiple autonomous capabilities (n=1) and "actual and scripted machine intelligence" (n=1). |
| | Some features suggested for the STE were also related to the STE's ***accessibility*** (n=3). One suggestion was to make displays intuitive. Another was to make the task easy to learn. The affordability of using the STE was a third aspect of accessibility suggested as a feature of the system. |
| | Finally, some responses reflected skepticism about the question of identifying STE features at this point in development. One respondent pointed out that features are arbitrary when one is not aware of the use-case. Another said automated capture is only as good as the development process that leads to it, and bad data collection design is worse than having nothing. |
| Do you have any additional ideas, comments, or questions regarding STE features? If so please add them here. | The goal of this question was to uncover suggestions that respondents had not made in prior questions. Some respondents suggested that teams within the STE should comprise at least one human and AI teammate (n=2). Another advised STE needed to be sensitive to the explosive growth of collected data. |

### 4.3  Summary of Findings

Not surprisingly, three of the four highest-scoring Likert items within this portion of our researcher survey focused on various aspects of data collection within the STE:

1. The STE must support capture and assessment of environmental states ($\bar{x}$ = 4.5).
2. The STE must support capture and assessment of operator actions ($\bar{x}$ = 4.6).
3. The STE must support capture and assessment of system states ($\bar{x}$ = 4.6).

The fourth member of that grouping focused on having a robust authoring capability to provide specific challenges that set the context for the data collection ($\bar{x}$ = 4.5). Section 11 includes a more detailed examination of scenario authoring.

The popularity of the authoring feature is in keeping with the emphasis on flexibility within the open-ended responses. At times, respondents expressed the desire for flexibility in terms of the use of an open, modular system architecture. At others, it focused on being able to create challenges and to structure teams in ways that met the theory-based objectives of a particular research project. Among the system architecture suggestions about modular, open-source testbeds, one recurring idea was for us to look into using an existing environment that met these conditions.

Another theme in the responses to these questions was the desire to make the STE a high-fidelity testing environment.

### 4.4  Topics for Further Exploration

The research team should use a follow-up event to explore the variation in response to the feature "The STE should include support for a broad range of human and AI teammates." This item may not have been entirely clear to respondents; a follow-up may help researchers understand how respondents conceptualized support, and what risks the respondents who reported disagreement with that feature perceived.

Several respondents noted that we should focus on using an existing open-source environment. Considering existing solutions opens up three options: (1) adopt an existing STE as-is; (2) adopt an existing open-source STE and extend it to add features that the community desires; or (3) build a STE from scratch that has the features that the community desires. A follow-up event will help researchers determine which of those approaches is most practical and which existing open-source STEs could fit our requirements. Follow-up investigations could also allow the research team to identify the capability gaps between an ideal solution and an existing one.

## 5 RESEARCH PRIORITIES

To support our research plan, we must have a STE that provides an appropriate testbed for our research. Moreover, in reviewing the foundational research in this area, it became apparent that each research group used a different task environment as the foundation for their research. This observation led us to believe that a more comprehensive STE might benefit the researcher community as a whole by enabling better replicability. Therefore, we wanted to explore the nature of the research that the community wants to conduct.

This portion of the survey characterized the type of research that our respondents anticipated conducting in this domain. We began by asking researchers to describe a hypothetical study they would like to conduct in the STE. Instructions encouraged them to consider the participants, tasks, variables they wanted to measure, and so forth. The goal of this exercise was to get respondents to consider what kind of research they wanted to accomplish in the STE, and, therefore, think about the features it needed to be a useful tool for conducting that research.

### 5.1 Analysis of Open-ended Questions

The survey included a pair of questions that explored the type of research that researchers might wish to conduct within the STE. The first explored the basic context or domain for the research, the second focused on the data that researchers would like to collect within the study:

1. In a few sentences, please describe a study that you might conduct using a hybrid teaming STE. Who are the participants? What is the task domain they engage with? What signifies success or failure of teams within that task domain? Task domains could be anything from conventional military operations to a fictionalized setting such as protecting Earth against alien invasion.
2. Think about a research study that you would like to conduct using the STE. What are the variables (or variable types) that you would like to collect in that study? Please list them below.

Table 3 summarizes the themes that appeared in the respondents' answers to these questions

**Table 3: Analysis of Open-ended Responses to Questions Focused on Research Priorities**

| Question | Analysis |
|---|---|
| In a few sentences, please describe a study that you might conduct using a | Responses to this question broke down into the following themes: task environment/setting, task, hypothesis and research questions, |

| Question | Analysis |
|---|---|
| hybrid teaming synthetic task environment. Who are the participants? What is the task domain they engage with? What signifies success or failure of teams within that task domain? Task domains could be anything from conventional military operations to a fictionalized setting such as protecting Earth against alien invasion. | participants, evaluation of performance, and general system and other concerns. We will discuss each.<br><br>Respondents who expressed a preference for a *setting* generally indicated that they would prefer an authentic military setting (n=5). One respondent suggested that a fictional setting might be appropriate for basic research.<br><br>Several respondents mentioned a desire for realism in the *task*. In keeping with this general theme, many respondents identified specific military tasks that would be of interest (n=7). These included Command, Control, Communications, Computers, Intelligence, Surveillance and Reconnaissance (C4ISR; n=3) and offensive and defensive operations in various domains (n=4). One respondent suggested tasks such as search-and-rescue or bomb disposal that are normally military in nature, but that have civilian counterparts. Others described general requirements of the task to exercise critical team-related skills. Here the focus was on including tasks that varied in complexity (n=1) while requiring a high degree of interdependency (including communication and coordination) and situation awareness (n=2).<br><br>The singular response to this question that dealt with a specific research *hypothesi*s was focused on how AI agent requirements in a human/autonomous team were affected by human teammate skill and task complexity. This response drives home the idea that research in the STE would offer an opportunity to not only measure and assess human teammate performance, but examine autonomous teaming dynamics and define useful requirements for autonomous agents as well.<br><br>As might be expected from the military-focus of the preceding responses, the respondents indicated that most of their research *participants* would be military personnel (n=6). Three respondents indicated that they include civilian participants; two specifically mentioned college students; and, in line with this, two respondents advocated for an unclassified environment.<br><br>Many respondents offered methods for how they would *evaluate performance* in research conducted in the STE. Two respondents mentioned process-based measures, and one mentioned outcome-based measures explicitly. However, respondents identified more specific processes and outcomes when reporting measures of interest. Examples included (n=1 unless otherwise noted) decision quality, decision speed, mission accomplishment (n=2), time to complete mission, enemy losses, and friendly losses (n=2). They also identified a few process measures of interest. These included a general mention of process measures (n=1), and more specific measures such as points earned (n=1), completion of sub-tasks (n=1), and measures of |

| Question | Analysis |
|---|---|
| | coordination (n=2). The survey addressed this issue in more depth in the following question.<br><br>***Additional recommendations*** included incorporating backend instrumentation in the STE (n=1), building the STE around a specific use-case (instead of a generic STE) (n=1), and creating an inherently unclassified environment with plugins that could be used to increase realism and classification (n=1). |
| Think about a research study that you would like to conduct using the STE. What are the variables (or variable types) that you would like to collect in that study? Please list them below. | In response to this question, participants provided answers that we grouped into seven themes: Teamwork Process Measures; Taskwork Process Measures; Team Outcome Measures; Individual Difference Measures; Team State Measures; Perception/Opinion Measures; and Agent Attributes.<br><br>Three respondents mentioned ***teamwork process*** measures generically, and an additional respondent mentioned them specifically. For example, one respondent listed the processes associated with Team Dimensional Training (including team self-correction), and several others identified specific measures of communication (n=5), such as (n=1 for all) written transcripts, audio transcripts, communication quality, communication content, communication timing, and communication and interaction patterns and effectiveness.<br><br>***Taskwork process*** measures included time on task (n=1), assessment of specific procedures (n=1), and measures of decision quality (n=1).<br><br>Only one response identified ***team/mission outcome*** as a measure of interest, but the answers to the preceding question indicate that this assessment would be of broader interest. Another researcher identified measures of learning (rate and quality) as being a focus of his/her research.<br><br>One respondent suggested that it would be good to capture ***individual differences***, and others mentioned specific individual differences that might be important such as demographic information, gaming proficiency, spatial ability, level of expertise, and level of knowledge (n=1 in all cases). Physiological measures such as heart rate (n=1) might also fall within this general theme. However, another respondent explicitly said he/she would prefer behavior data to biometric data.<br>Several respondents identified measures that we categorized within ***team state***. These states include workload (n=4), trust (n=3), cohesion (n=2), and situation awareness (n=2). A related set of measures might be considered ***perceptions or opinions***; these included individual perceptions of the team (*e.g.,* efficacy; n=2), impressions of the intent/benevolence of the agent (n=1), and assessments of the agents' perceived ethics (n=1). |

| Question | Analysis |
|---|---|
| | The final theme concerned an assessment of various *agent attributes* including trustworthiness, understanding, situation representation, and context sensitivity (n=1 in all cases). |

## 5.2 Summary of Findings

Generally, respondents felt that the task presented by the STE should focus on military operations that mimic genuine challenges of the military task environment, while maintaining an unclassified status. This preference ties in with the finding that the respondents were primarily interested in participants with a military background. As for what researchers wanted to study, two broad performance measurement categories were found: process measures and outcome measures. Under these general categories were three broad groupings of performance measurement: team state, team performance, and task performance. Separate from team-oriented dependent variables was the notion of measuring individual differences among participants (such as expertise, gaming experience, spatial ability, and other demographics) as quasi-independent variables.

## 5.3 Topics for Further Exploration

Respondents seemed to agree the STE should present an authentic military setting for the embedded task. It may be important to examine this more closely and explore whether there is a concern about task domains with high fidelity to military operations being classified and, therefore, off-limits for conducting research with civilians or publishing the results. An additional concern could be that the only population that could successfully work on tasks in the task environment are those with sufficient prior knowledge of the task domain (which may further limit populations that can successfully participate in studies using the STE).

## 6 DATA COLLECTION AND PERFORMANCE ASSESSMENT

Two related issues that will shape the development of the STE are the nature of the data that the STE should collect and the performance evaluation options that it should support. The research team feels that a core capability of the STE must be rich data collection and performance assessment capabilities. We believe that the STE must be able to assess and store mission and task outcomes as well as detailed task processes. In this section, we review the opinions of researchers regarding data collection and performance assessment features that the STE should include.

### 6.1 Analysis of Discrete Features

In the survey, we put forward a number of features we thought researchers might value regarding data collection in the STE. Within the survey, we asked respondents to indicate the degree to which they agreed or disagreed with each data collection feature to be incorporated within the STE. Pie charts and descriptive statistics for each of the domain-related features are shown in Appendix C.

As we did in the preceding section, we conducted an analysis to determine whether any items had standard deviations greater than or equal to 1.0. None of the Likert items associated with this section met that criterion.

## 6.2 Analysis of Open-ended Questions

The researchers who participated in this study were asked to respond to three prompts:

1. Are there human performance assessments that should be "baked in" to the STE? If so, please take a few sentences to describe them here.
2. Do you have any additional ideas, comments, or questions regarding data collection in the STE? If so, please add them here.

Table 4 summarizes the themes that appeared in the respondents' answers to these questions.

Table 4: Analysis of Responses to Data Collection and Performance Assessment Prompts

| Question | Analysis |
|---|---|
| Are there human performance assessments that should be "baked in" to the STE? If so, please take a few sentences to describe them here. | Respondents only made a handful of suggestions in response to this question, and one respondent pointed out that the performance assessments that would be most useful to be a part of the system depended on the specific research question. Among those was the recommendation to incorporate workload, attention, and engagement assessments within the STE (n=1). Another was the ability to measure the completion of certain tasks with timeliness (n=1) and incorporate measures of effective coordination into the STE (n=1). |
| Do you have any additional ideas, comments, or questions regarding data collection in the STE? If so, please add them here. | Responses to this question echoed the variables listed in response to the prior question about types of data researchers wanted to collect. However, one respondent added that they would find embedded survey capabilities useful in the STE. |

## 6.3 Summary of Findings

As indicated in Section 6.1, there was a great deal of consensus among the researchers on the Likert items associated with this section. It is also interesting to note that the five items with the greatest consensus (.44 ≤ sd ≤ .53) were also the five with the highest mean agreement scores:

1. The STE must gather data on team coordination process variables. ($\bar{x}$ = 4.44)
2. The STE must gather data on task process variables. ($\bar{x}$ = 4.56)
3. The STE must gather data on mission outcome variables. ($\bar{x}$ = 4.67)
4. The STE must gather data on task outcome variables. ($\bar{x}$ = 4.78)
5. The STE must support the capture of communication activities among teammates. ($\bar{x}$ = 4.78)

Building on the discussion of research priorities outlined in Section 5, when asked to describe the dependent variables for a study that respondents were interested in conducting in the STE, many of the responses could be grouped into team-related, individual-related, or autonomy-related variables. Team variables could be further broken down to state, performance, process, and communication sub-categories. Individual variables were described as outcome-related (such as measuring task-performance), individual differences, and biometric data. Autonomy-related variables focused on team and individual relationships with autonomous agents.

## 6.4 Topics for Further Exploration

Given the strong consensus among the respondents, it appears that we would gain little through a deeper exploration data collection and performance assessment features of the STE. It may be sufficient to note that the researchers sampled agreed with the research team's assessment that these capabilities represented the core of the STE.

## 7 DATA ANALYSIS AND VISUALIZATION

To complement the STE's extensive data collection and performance assessment capabilities, it may be useful to embed certain data analysis and visualization capabilities within the STE. This portion of the survey asked researchers to offer their perspective about the capabilities that should be built into the STE.

### 7.1 Analysis of Discrete Features

In the survey, we put forward a number of features we thought researchers might value regarding the data analysis and data visualization capabilities of the STE. We asked respondents to indicate the degree to which they agreed or disagreed with features related to the STE's ability to support data analytics and visualization. Pie charts and descriptive statistics for each of these features are shown in Appendix D.

As with other sections, we conducted an analysis to determine whether any items had standard deviations greater than or equal to 1.0. Quite a few items met this criterion. We discuss those items below.

The first set of items had to do with visualization of descriptive statistics within the STE. One item that had a range of agreement was that the STE should provide descriptive statistics for each variable (standard deviation = 1.31). These statistics included mean, median, mode, range, and standard deviation. As shown in Figure 2, respondents reported a range of opinions, from strong disagreement to strong agreement, with the most frequent response being indifference to the feature.

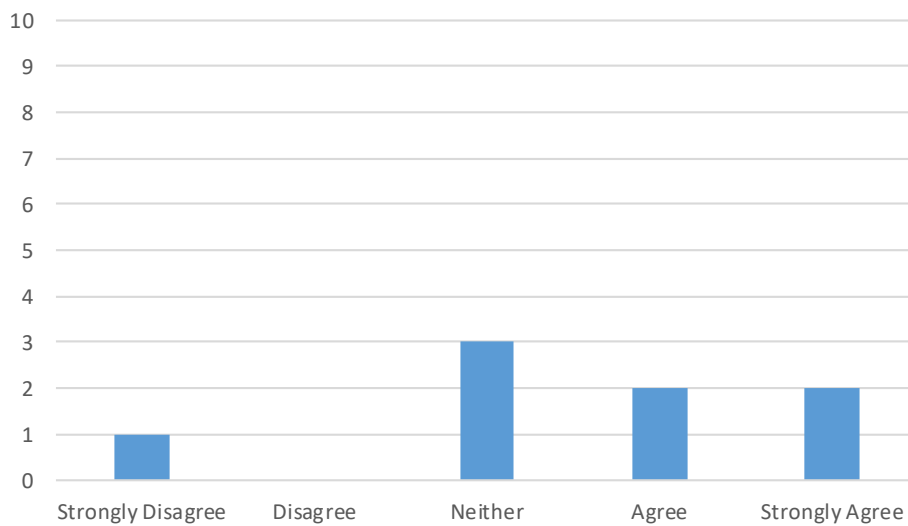

**Figure 2. Agreement ratings for "The STE should provide descriptive stats (means, mode, median, range, standard deviation) for each variable."**

A similar pattern of responses was observed for the item "The STE should provide bar charts of the means of variables." Responses to this item are displayed in Figure 3 (standard deviation = 1.19). Generally, respondents either agreed or were neutral on this front, with one respondent on the "strongly disagree" end of the scale.

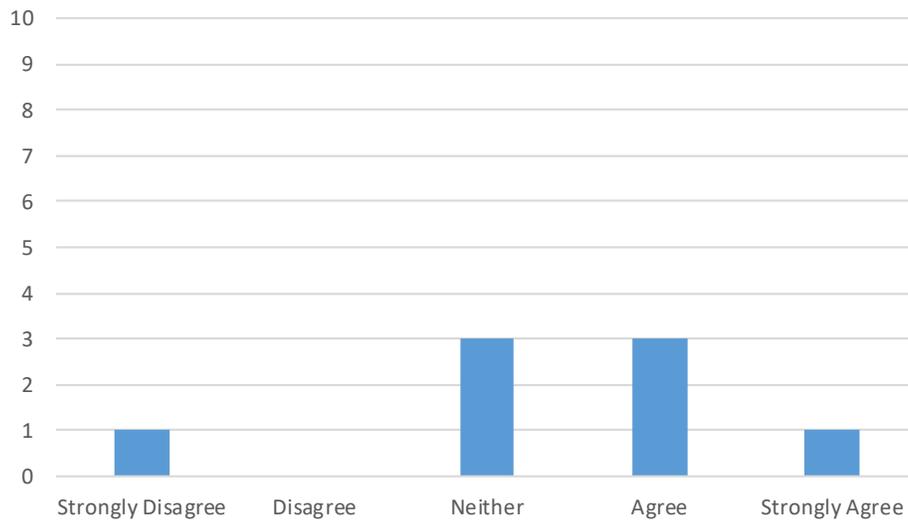

**Figure 3. Agreement ratings for "The STE should provide bar charts of the means of variables."**

The second set of items with a range of agreement responses had to do with features that would allow researchers to visualize relationships between variables. For example, a wide range of agreement was observed for offering scatterplots to view relationships between variables as a part of the STE (shown in Figure 4). General agreement and a neutral response to including this feature were equally frequent, and one respondent was observed on either end of the scale. This item had a standard deviation of 1.19.

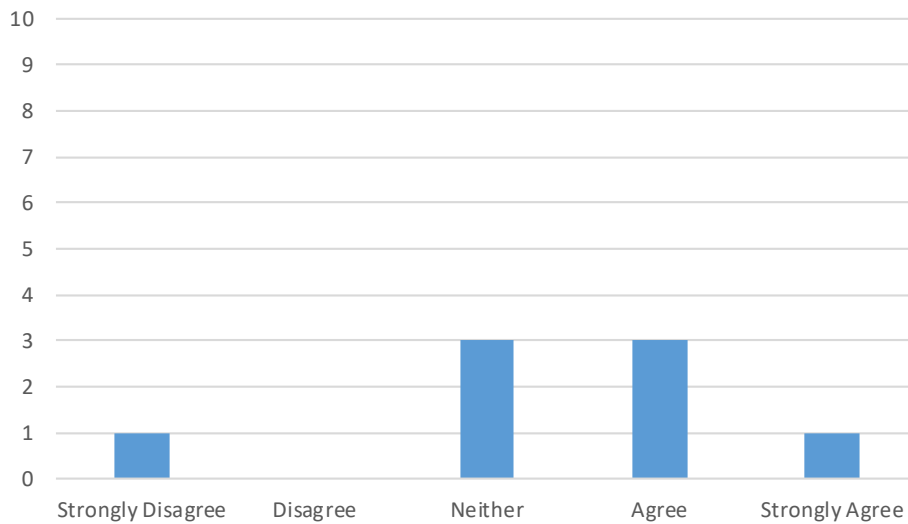

**Figure 4. Agreement ratings for "The STE should provide scatterplots that illustrate the association between pairs of variables."**

*Another data relationship visualization feature with the same pattern of response was the notion that the STE should provide histograms of variables. Responses to this item are shown in Figure 5 (standard deviation = 1.19). As with the other feature in this category, most respondents reported agreement or a neutral opinion on the inclusion of this feature, with one respondent on either extreme.*

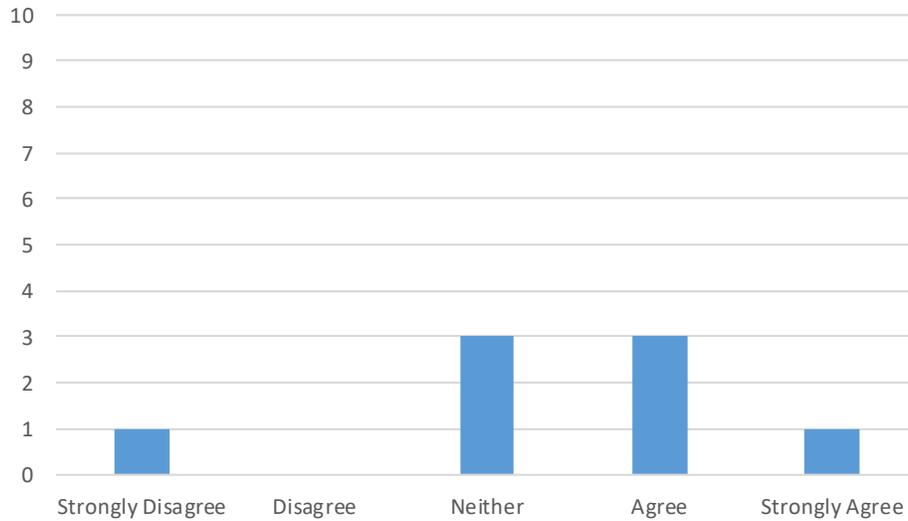

**Figure 5. Agreement ratings for "The STE should provide histograms (bar charts of frequency/counts) for variables."**

The third set of items pertaining to data analysis and visualization in the STE was related to conducting data analyses in the STE itself. In particular, items that asked whether the STE should enable t-tests (standard deviation = 1.16), ANVOA (standard deviation = 1.13), and regression analyses (standard deviation = 1.13) yielded a range of agreement responses. What is notable about these items is that the predominant response in all three was neutral. In other words, respondents neither agreed nor disagreed about whether the STE should involve data analysis features. Another similarity among all three items was that more respondents selected either "agree" or "strongly agree," while only one respondent selected "strongly disagree" for all three items. The bar charts for these items can be found in Figure 6, Figure 7, and Figure 8.

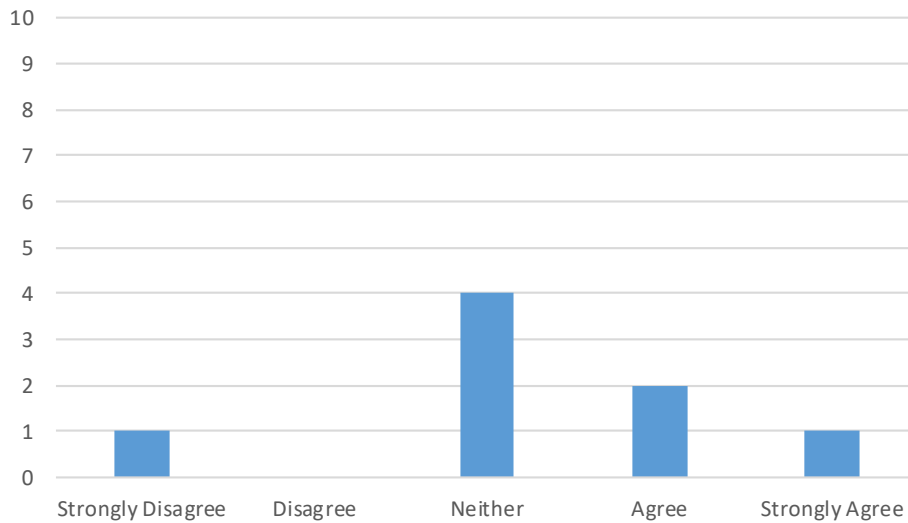

**Figure 6. Agreement ratings for "The STE should enable t-test analysis."**

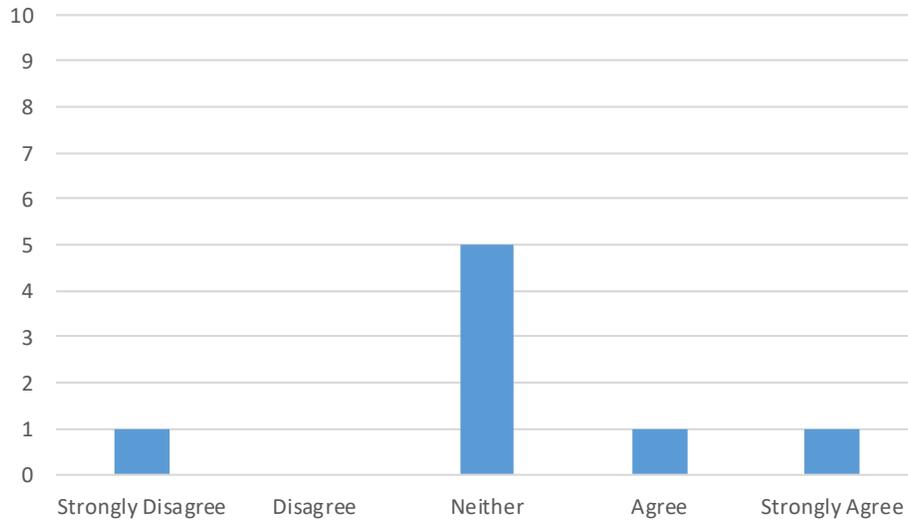
**Figure 7. Agreement ratings for "The STE should enable one-way ANOVA analysis."**

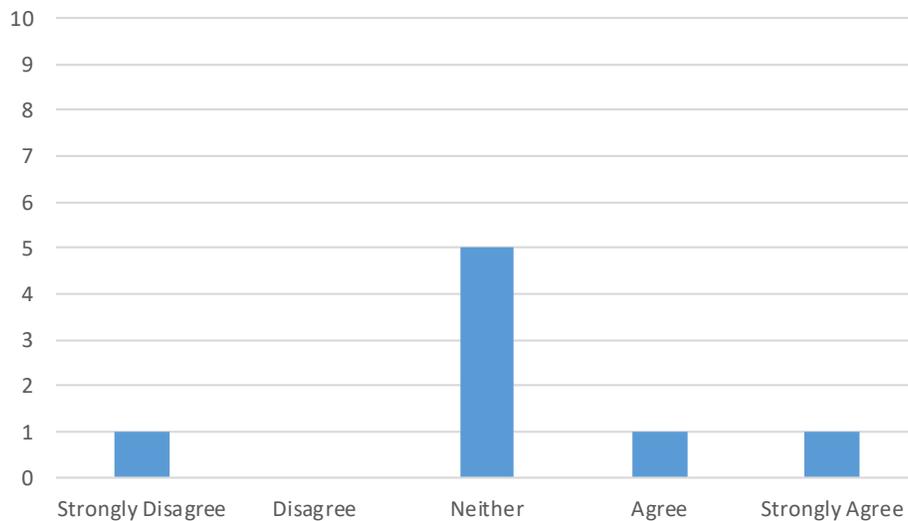
**Figure 8. Agreement ratings for "The STE should enable regression analysis."**

### 7.2 Analysis of Open-ended Questions

While the Likert items attempted to span the range of likely analysis and visualization features, the research team wanted to ensure that we did not neglect any features that would be important to the research community. To assess that, we followed the Likert items with a pair of open-ended prompts:

1. Using a couple of sentences, please describe what kind (if any) on-board data analytics and/or data visualization tools would be useful to include within the STE.
2. Do you have any additional ideas, comments, or questions regarding data visualization and analysis? If so, please add them here.

We have summarized our analysis of the answers we received in Table 5.

Table 5: Analysis of Responses to Data Analysis and Visualization Prompts

| Question | Analysis |
|---|---|
| Using a couple of sentences, please describe what kind (if any) on-board data analytics and/or data visualization tools would be useful to include within the STE. | A theme among a handful of responses to this question was the desirability of easy access to summary data. More specifically, respondents were interested in the ability to do "quick looks" at variables associated with team state (n=2), vehicle state (n=1), and relationships between variables (n=1). Respondents also mentioned the desire to use data analytic capabilities to support team debrief (n=2). There was no overlap among other recommendations. One respondent mentioned the ability to view intuitive displays of data trends over time. Someone else said they would value displays of decision-related data, criteria, and suggestions made. Another respondent emphasized flexibility as a useful feature. |
| Do you have any additional ideas, comments, or questions regarding data visualization and analysis? If so please add them here. | Responses to this question focused mainly on the integration of data analytics in the STE. Some participants were attuned to the potential risk of a trade-off between analytics and data export. As a result, there was some skepticism about incorporating data analytics. Two respondents emphasized the importance of data export over data visualization and analysis. One respondent explicitly said that data analysis should not be included as a part of the system at all. On the other hand, one respondent suggested including a number of advanced analytic methods within the STE, such as multivariate analysis, time series analysis, and stochastic analysis. |

## 7.3 Summary of Findings

As far as data analytics were concerned, respondents were primarily interested in being able to take "quick looks" at data, and support team debriefing. However, there was notable skepticism that offering data analytic services was a worthwhile endeavor, and what was more important was the ability to export data for offline analysis.

This focus on data export and offline analysis was reflected in the Likert items. The three items with the highest average agreement and the smallest standard deviation all focused on this capability:

1. The STE should allow researchers to export identified variables to Excel files.
2. The STE should allow researchers to export identified variables to comma separated value files.
3. The STE should allow researchers to export identified variables to database files.

On the other hand, respondents greeted all other analysis and visualization features with roughly equal levels of indifference ($3.125 \leq \overline{X} \leq 3.5$).

## 7.4 Topics for Further Exploration

Respondents appeared to be most divided on the utility of data analysis features available in the STE. While some had suggestions for data analytics, others were explicit that clean data export was of greater importance than ability to look at data immediately. This division was fairly minor, as all respondents seemed to agree that exporting data should be a top priority. The greatest difference was in the skepticism voiced regarding the incorporation of analysis features into the STE. A follow-up discussion would allow the researchers to explore this more and get a sense for the differing opinions associated with analysis-oriented capabilities.

## 8 TASK DOMAIN

A key element for the STE is the task domain. The task domain defines the mission of the teams and provides the context for the tasks that they will perform. This portion of the survey focused on identifying the features of the task domain that would enhance its credibility and the applicability of our research to real-world operations. This section of the survey began with a prompt that asked researchers to describe the features of an ideal task environment. Next, respondents recorded their agreement with a list of potential task domain features that ranged from the types of skills required by the task domain presented in the STE, to features like the ease of learning the task, to the classification of the task environment (classified or unclassified). The task domain portion of the survey ended with two open-ended questions for respondents to list additional features that were not present in that list of features and record any additional comments, ideas, or questions on the subject.

### 8.1 Analysis of Discrete Features

Preceding the survey, Sonalysts conducted a broad literature review. The review, coupled with our experience in a range of modeling and simulation efforts, allowed us to predict a range of task domain features that we thought might be useful to include. Within the survey, we asked respondents to indicate the degree to which they agreed or disagreed with our hypotheses about which specific features would be important within the STE. Pie charts and descriptive statistics for each of the domain-related features are shown in Appendix E.

As we did earlier, in this section we draw attention to items that surpass our threshold by having standard deviations greater than or equal to 1.0. Two of the STE features met that criterion.

One of these features was "the task environment must be unclassified to avoid CUI." The spread of responses ranged from strongly disagree to strongly agree, with one more agreement response than disagree response (standard deviation = 1.33).

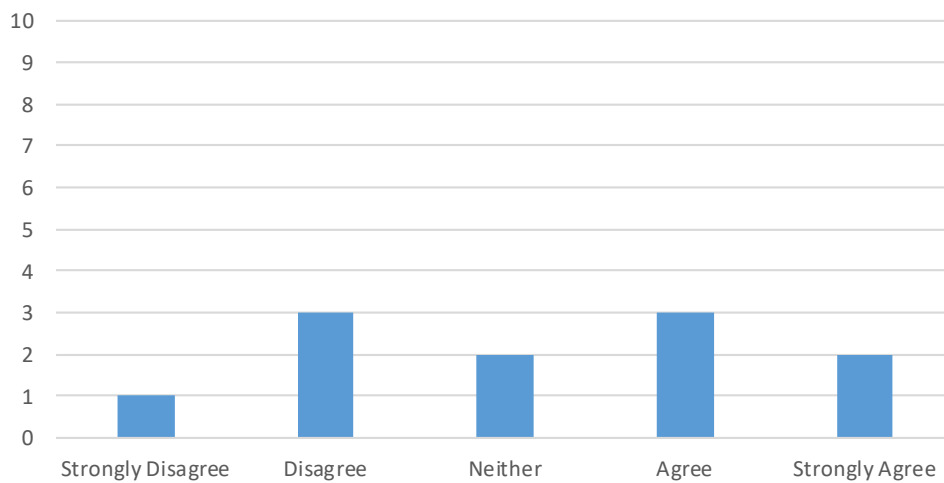

**Figure 9. Agreement ratings for "The task environment must be unclassified to avoid CUI."**

Another STE feature with a wide variation of response was "The task must be isomorphic with military operations." The standard deviation of this item was 1.27. This is somewhat surprising, given the number of open-ended responses regarding research priorities that recommended creating a task that presented a realistic military task environment. Responses spanned each node of the agreement scale, with the most frequent selections being "neither agree nor disagree" and "agree."

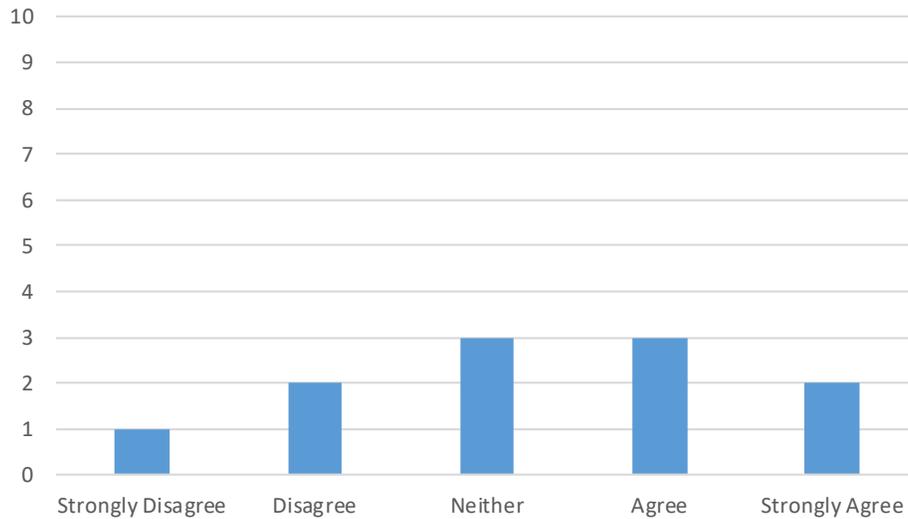

**Figure 10. Agreement ratings for "The task must be isomorphic with military operations."**

## 8.2 Analysis of Open-ended Questions

After reacting to a number of suggested task domain features, the survey presented open-ended prompts that would allow them to share additional perspectives. Table 6 summarizes the themes that appeared in the respondents' answers to these questions.

**Table 6: Analysis of Open-ended Responses to Task Domain Questions**

| Question | Analysis |
|---|---|
| Are there any features that you think are important to include in the task domain that are not listed above? Please list them here. | In response to this question, several respondents recommended features of STE itself, such as incorporating robust data collection and performance measurement (n=3). Other recommendations had to do with task, such as a need for the task to demand different levels of teamwork (n=1), the need for interdependency among team members (n=1), and the need for realistic adversarial warfighting scenarios (n=2). |
| Do you have any additional ideas, comments, or questions regarding the task domain? Please add them here. | Some of the listed features in response to this question were echoed by other respondents in prior answers. Moreover, the need for "realism" (*i.e.,* the need to incorporate realistic use-cases) was a theme that emerged in responses to this question (n=2). One respondent suggested that a generally useful STE may not be practical, and that it would be better to create a STE for a specific context. |

### 8.3 Summary of Findings

A significant theme in the Likert responses was a focus on ensuring that the STE required a high level of interdependency and coordination. The three items with the highest average Likert score all addressed this requirement:

1. The task environment must provide a venue for practicing and assessing coordination behaviors across the timeline (*i.e.*, from "preparation" activities to "adjustment" activities). ($\bar{x}$ = 4.18)
2. The task environment must include opportunities for cognitive coordination. ($\bar{x}$ = 3.91)
3. The task environment must include opportunities for physical/logistical interdependency (*e.g.*, you do this, then I do that). ($\bar{x}$ = 3.19)

These items also had relatively small standard deviations (.75 ≤ sd ≤ .94), indicating a strong level of agreement across respondents.

Within the open-ended items, respondents focused on how the STE would collect data. Several respondents mentioned robust data collection and performance measurement. Other responses emphasized task demands to create a performance environment that would yield the performance researchers want to measure. Some themes of task demands echoed some of the themes from the questions about research priorities in the earlier section, such as realism, a focus on adversarial environments, and (in keeping with the trend observed within the Likert items) opportunities to observe teamwork by creating a need for interdependency within the task.

### 8.4 Topics for Further Exploration

While a major theme emerging in this and the preceding section was overlap with military operations, no respondent specifically indicated interest in studying Joint All Domain Command and Control (JADC2). JADC2 is an emerging construct, but has the potential to be a significant use case for teaming with autonomous teammates. We may consider delving into this in a follow-up event with the research community, and ask specifically how much of their research focuses on the JADC2 setting.

Additionally, it will be important to understand the discrepancy between the lukewarm Likert agreement with the STE feature "The task must be isomorphic with military operations" and the trend in the short-answer responses to emphasize fidelity with military operations. We may consider asking how respondents interpreted the meaning of "isomorphic," and how that may have led to a discrepancy between agreement with that statement and the desire to ensure the task provided challenges for study participants that tapped the skills required to be successful in real military operations. One of the ways to understand this discrepancy might be to delve further into opinions regarding the classified nature of the STE, as feelings on this might have swayed researchers' responses to the question of making the task isomorphic to a military task.

## 9 COMMUNICATION

Many team coordination activities take the form of communications. As such, it will be important for the STE to provide support for a wide range of communication modalities, as well as the ability to capture and assess communication activities. This will certainly be a challenge, and development is likely to involve modular capabilities that evolve over time. To provide a foundation for this development, the research team wanted to explore the aspects of communications that researchers in this domain felt were important.

## 9.1 Analysis of Discrete Features

As with other categories of features discussed above, we presented respondents with features that we thought researchers might value regarding communication-related capabilities of the STE. We asked respondents to indicate the degree to which they agreed or disagreed with features related to communication in the STE. Pie charts and descriptive statistics for each of these features are shown in Appendix F.

As with other sections, we conducted an analysis to determine whether any items had standard deviations greater than or equal to 1.0. Three items met this criterion.

The items that had standard deviations greater than or equal to 1.0 were related to three different communication modalities: email, paper, and video. Researchers were mostly neutral on the support of email communications, although two respondents strongly agreed email communication should be supported, and two disagreed (standard deviation = 1.16; see Figure 11).

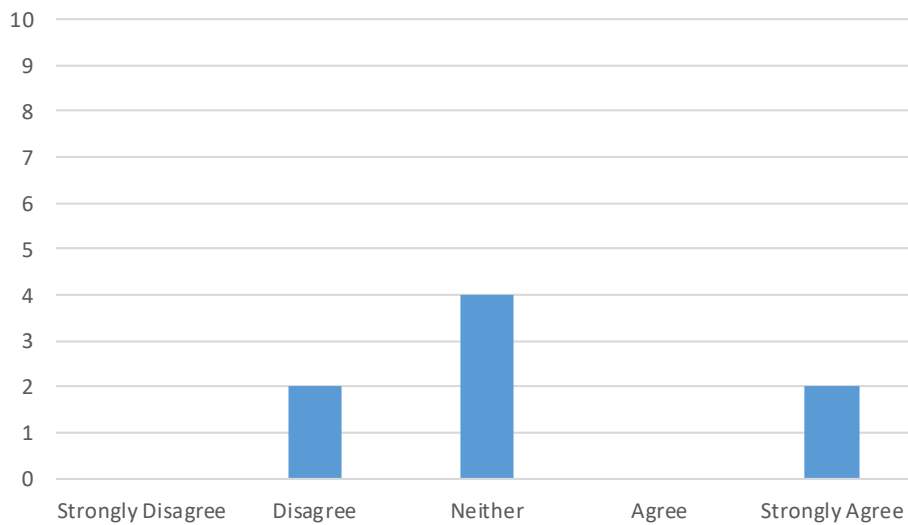

**Figure 11. Agreement ratings for "The STE must enable email communications among teammates."**

The primary response to support of paper-based communication in the STE was also neutral, but more respondents disagreed to some degree with this feature, than respondents who agreed with it (standard deviation = 1.13; see Figure 12).

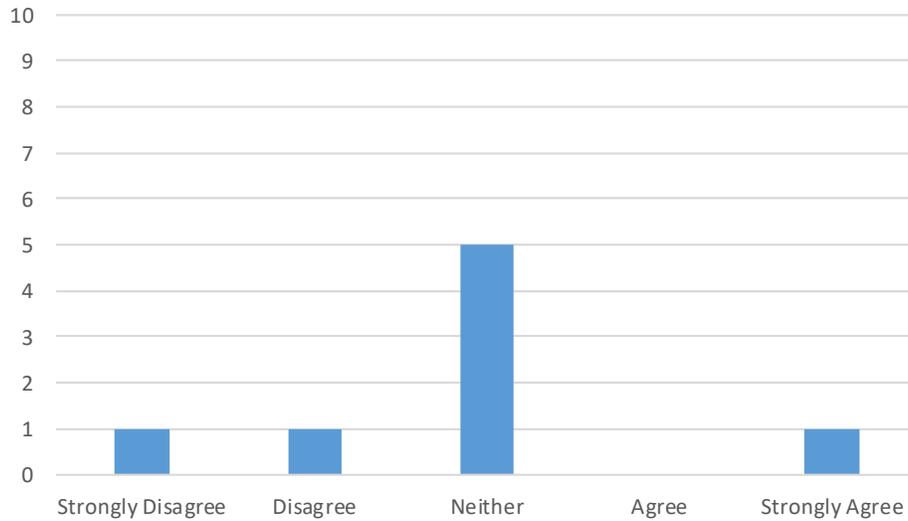

Figure 12. Agreement ratings for "The STE must enable paper-based communications among teammates."

*On the other hand, more respondents agreed to some degree with the support of video communication in the STE, with only one respondent disagreeing that it should be enabled by the system (standard deviation = 1.07; see Figure 13).*

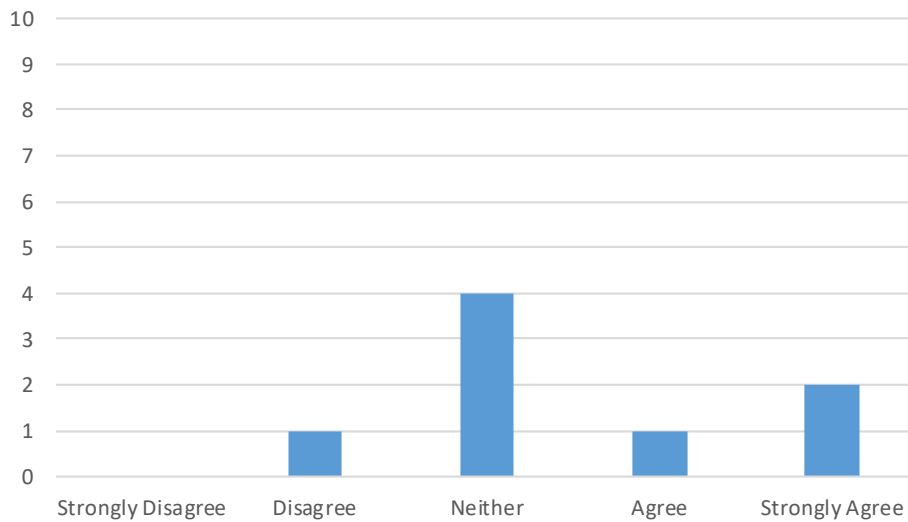

Figure 13. Agreement ratings for "The STE must enable video communications among teammates."

## 9.2 Analysis of Open-ended Questions

The research team wanted to develop a better appreciation for the challenges that researchers faced when conducting studies in existing testbeds. We also wanted to develop an appreciation for how these practitioners evaluated the communications that they observed within their studies. Therefore, we paired the discrete items with three open-ended items:

1. In a couple of sentences, please explain the challenges that you face facilitating, capturing, and analyzing communication using your current research tools.
2. How do you normally assess communication processes and outcomes?

3. Do you have any additional ideas, comments, or questions about measuring communication in the STE? If so, add them here.

Table 7 summarizes the themes that appeared in the respondents' answers to these questions.

Table 7: Analysis of Responses to Open-ended Communications Questions

| Question | Analysis |
|---|---|
| In a couple of sentences, please explain the challenges that you face facilitating, capturing, and analyzing communication using your current research tools. | Challenges to communication in research can be summarized as issues with crosstalk (n=1), capturing emotion (n=1), identifying the speaker (n=2) and target of communication (n=1), coding and transcription challenges (n=6), and automation problems (n=1). Coding and transcription problems were elaborated on by some respondents to include issues with "human" scoring (n=1), and the labor associated with collecting and analyzing communication (n=2; with one respondent expressing a desire to automate it). Three respondents mentioned timestamping communication as a necessary aspect of collecting data, one of whom mentioned that automating timestamp information about team communication with particular scenario events is critical for interpreting communication data.<br><br>The reports of communication capture challenges conflicted with reports that some respondents made about non-challenges. For example, another respondent said capturing communication was often successful in research, and did not represent a challenge. Another said accuracy of timestamping communication data, and identifying speakers, posed no challenge.<br><br>Regarding how researchers wanted teammates to communicate in the STE, a notable theme in responses was integrating text and chat communication as a part of the system (n=4). Suggestions also included the facilitation of voice communication (n=2), the flexibility to toggle back and forth between verbal and text-based chat (n=1), and the ability to support face-to-face communication (n=1) and non-verbal communication (n=1). A suggestion was also made to incorporate reports (via interface displays; n=1). A more general suggestion was to ensure high fidelity communication (n=1). |
| How do you normally assess communication processes and outcomes? | The responses for this question can be broadly grouped into themes of the assessment of communication processes and communication content, and methods of assessment.<br><br>Reported assessment of *communication processes* included analyzing who is speaking to who and when (n=2), the use of automated timestamps for behavioral assessment (n=2), and social network analyses (n=1). The method of assessing *communication content* was described as evaluating the proximity of captured communications to pre-defined |

| Question | Analysis |
|---|---|
| | communication criteria (n=1). One respondent described the use of Latent Semantic Analysis (LSA) and Natural Language Processing (NLP) to analyze communication content. <br><br> Respondents also provided a host of general communication *assessment methods* that were not specific to evaluating process and content. One respondent mentioned using a method in which two people code data and their results are compared. This respondent also described developing a coding book as method for coding data. The raw data to be coded was described as coming from transcripts of verbal communication (such as a zoom transcript), with the use of Google's speech-to-text application. One respondent also noted that a tablet worked more effectively for assessing communication than pencil/paper measures. |
| Do you have any additional ideas, comments, or questions about measuring communication in the STE? If so, add them here. | Suggestions included technologies associated with capturing and assessing communication, what to value in assessing communication, and the importance of assessing communication. One respondent suggested that the research team look into commercial and open source technologies for capturing and assessing communication. Another emphasized the need to assess communication quality (rather than sheer quantity). Another respondent said that communication measurement would be the most important aspect of the STE, implying an importance of quality communication capture and assessment capabilities. |

## 9.3 Summary of Findings

When considering the communication-focused features of the STE, there was near consensus that the STE should support both chat-based and spoken communications among teammates. The respondents were largely indifferent when assessing the majority of the remaining communication features ($3.25 \leq \bar{x} \leq 3.875$). One item, support for paper-based communication, fell on the negative end of the spectrum ($\bar{x} = 2.875$).

A big theme that came out of the open-ended questions was the desirability of support for coding and transcription associated with analyzing verbal communication. The process associated with transcribing and coding verbal information tends to be cumbersome and labor-intensive.

Another theme that emerged from responses to the open-ended questions was integration of text-based communication as a part of the system. Based on agreement ratings to the communication features in the survey, there was a wider range of opinions associated with enabling email communications, enabling paper-based communications, and enabling video-based communications. This indicates that while respondents universally desired a synchronous text-based communication in the form of chat logs, an asynchronous text-based communication method (email) or paper-based communication were not as valued.

### 9.4 Topics for Further Exploration

A good topic for follow-up regarding communication in the STE will be to understand the breadth of communication support in the STE. The research community was in agreement regarding support for spoken and electronic chat communications, but had a variety of perspectives on supporting other types of communication. A joint conversation among stakeholders in the research community could allow us to assess the priorities of the research community regarding communication, and rank the communication capabilities for a future capabilities document.

## 10 AUTONOMOUS AGENTS

The primary focus of this effort is to explore various hypotheses regarding the use of training to improve the performance of Human-AI teams by enhancing the ability of humans to function as part of those teams. To enable this research, especially in its early stages, it will be necessary to provide functioning autonomous systems within our task domain. Therefore, the research team wanted to explore the features and/or capabilities of autonomous teammates that are likely to be important to the research community.

In this section, we explore respondents' answers to survey questions that explored the desired characteristics of agents they would like to use within their research.

### 10.1 Analysis of Discrete Features

As we did for other dimensions of the STE, the research team presented respondents with features that we thought researchers might value regarding autonomous agent capabilities incorporated into the STE. We asked respondents to indicate the degree to which they agreed or disagreed with features of autonomous agents in the STE. Pie charts and descriptive statistics for each of these features are shown in Appendix G.

We conducted an analysis to determine whether any items had standard deviations greater than or equal to 1.0, and two items met this criterion.

One contentious item regarding autonomous agents in the STE is the idea that human teammates should perceive autonomous teammates as cognitively similar to them (see Figure 14 for the distribution of responses associated with this item). Respondents were primarily neutral on this front; however, responses spanned the full range of the scale, in equal numbers and with equal distribution (standard deviation = 1.20).

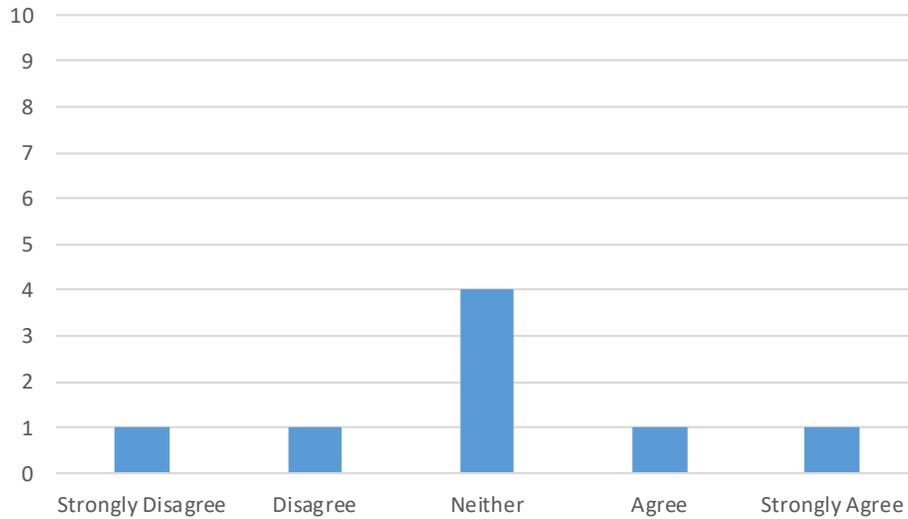

**Figure 14. Agreement ratings for "Autonomous teammates within the STE must be perceived as cognitively similar to their human teammates."**

The other somewhat contentious feature had to do with agreement about whether autonomous teammates established, maintained, and repaired common ground. The standard deviation of this item was 1.13. Respondents mostly agreed, and in fact "strongly agreed," with this feature; however, at least one respondent disagreed and one was neutral on the matter (see Figure 15).

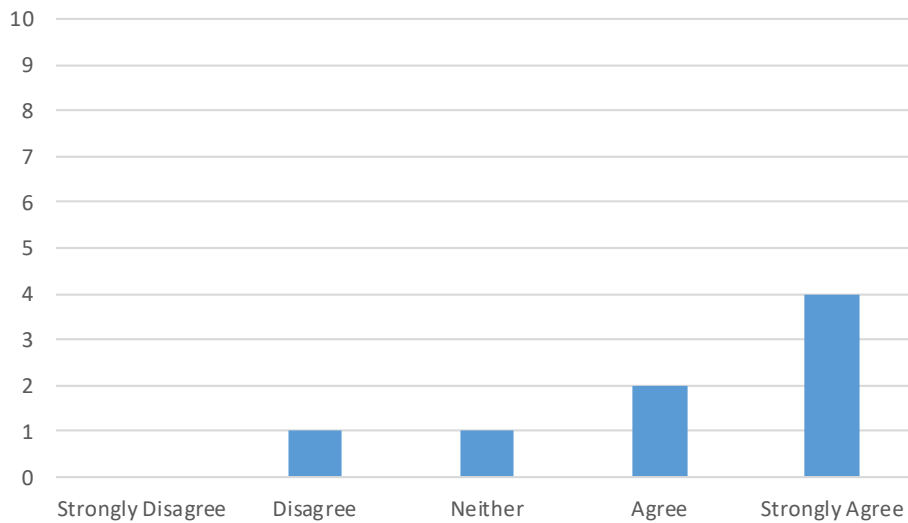

**Figure 15. Agreement responses to "Autonomous teammates within the STE must be able to establish, maintain, and repair common ground with other teammates."**

### 10.2 Analysis of Open-ended Questions

One topic of particular interest to the research team was whether the STE should include "built in" autonomous agents or support the ability to plug externally developed agents into the STE architecture. Alternative approaches would have significant implications for the underlying architecture. Similarly, we wanted to conduct a "deep dive" on the importance of "explainability" and related concepts within this research domain. Table 8 summarizes the themes that emerged from our analysis of the responses we received to these prompts.

Table 8: Analysis of Responses to Open-ended Questions Pertaining to Autonomous Agents

| Question | Analysis |
|---|---|
| In a couple of sentences, please describe the ideal autonomous teammates to support your research within the STE. Are they "built-in" or developed externally? What are their characteristics (e.g., transparency, reliability, etc.)? What dimensions of their performance are adjustable? | Descriptions of ideal autonomous teammates broke down into themes of teammate characteristics, teammate support capabilities, features that researchers should be able to adjust, and development of autonomous teammates.<br><br>***Characteristics*** of autonomous teammates were primarily put forward by one respondent, who suggested they be similar to human teammates, fill a recognized role on the team, complete tasks associated with that role, and coordinate with other team members (on teamwork and task-related behaviors). Another respondent described the ideal autonomous teammate as having human-like communication capabilities.<br><br>When considering important ***teammate support capabilities***, respondents focused on helping human teammates make the right decisions, take effective actions, and get the right information (n=1).<br><br>Respondents noted that it would be helpful if they could ***adjust a variety of features*** related to autonomous teammates, such as their transparency (n=1), reliability (n=2), trustworthiness (n=1), morality (n=1), and decision authority (n=1). Ideally, the roles filled by autonomous teammates (n=1), as well as their team-oriented intent (n=1), should be adjustable. Communication styles (n=1) and communication mediums (n=1) should also be adjustable. One respondent also suggested programming autonomous teammates to have adjustable trust repair strategies, adaptation capabilities, and interface features.<br><br>Respondents did not have strong opinions regarding whether autonomous teammates should be ***externally developed or built into*** the system. Two respondents said that either built in or externally developed teammates would be satisfactory. One respondent said externally developed teammates would be preferable, but that built-in teammates would be useful, too. Another said the STE would need to provide an Application Programming Interface (API) for the development of agents. |

| Question | Analysis |
|---|---|
| Should autonomous teammates be "explainable"? Describe what "explainable" means to you in this case. | Five respondents offered definitions of explainability. Elements of those definitions included:<br>• Understanding the decisions, intentions, and actions of autonomous systems (n=2)<br>• Understanding how the autonomous teammate did something, why they did it, and how well it was done (n=1)<br>• Understanding what the autonomous teammate can and cannot do (n=2)<br>• Understanding the purpose and role of the autonomous teammate in the team structure (n=1)<br><br>Respondents offered a spectrum of perspectives on the importance of explainability. Two respondents described it as a critical feature of autonomous teammates. Two more said it was important, but one of those respondents warned researchers not to "miss good in pursuit of the perfect." Two others said explainability is desirable, but not required as a feature. Two respondents also said that explainability might be a good feature to be able to adjust. One of the respondents who answered that explainability was desirable but not required noted that current AI technologies used today are not explainable.<br><br>There was some skepticism expressed by one of the survey respondents about developing autonomous agents for the STE. The development of autonomous agents did not seem important to the development of the testbed, and this respondent warned that trying to meet all expectations of autonomous teaming capabilities would be daunting and possibly unattainable. |
| Do you have any additional ideas, comments, or questions about autonomous teammates in the STE? If so, please add them here. | Additional ideas for autonomous teammates included the notion that teammates needed to learn and adapt (n=1), and that many features of agents should be adjustable (and research could determine optimal levels) (n=1). One respondent suggested that dealing with multiple externally developed agents might become difficult, and that it may be easier to deal with one autonomous agent at a time. Another respondent suggested that we select a use case with near-term data (implying a general purpose STE may not be possible). |

## 10.3 Summary of Findings

In considering specific discrete features of autonomous agents, clear patterns did not emerge. That is, the respondents generally agreed (albeit relatively weakly) that most of the listed features should be included in the STE ($4.00 \leq \overline{X} \leq 4.625$). Three items broke against this basic trend:

1. Autonomous teammates within the STE must be perceived as cognitively similar to their human teammates. ($\overline{x}$ = 3.00)
2. Autonomous teammates within the STE must include domain knowledge similar to that possessed by humans who would staff that role. ($\overline{x}$ = 3.625)
3. Autonomous teammates within the STE must include knowledge of the capabilities and limitations of other autonomous teammates. ($\overline{x}$ = 3.625)

Responses to the open-ended prompts focused on autonomous teammate characteristics, support capabilities, and adjustable features. Desirable characteristics of autonomous teammates included similarity to human teammates[1], the ability to fill a concrete role and perform the tasks associated with that role, and the ability to communicate and coordinate with other team members. Support capabilities that came up in responses included helping humans make the right decisions, taking effective actions, and collecting the right information. Respondents forwarded many adjustable features for autonomous teammates, including transparency, reliability, trustworthiness, morality, and decision authority (among others).

## 10.4 Topics for Further Exploration

A follow-up event would allow the research team to explore some of the discrepancies among the research stakeholder group on their expectations of the autonomous teammates. However, it may be prudent to limit the conversation to discrete feature ranking, rather than explore an in-depth conversation about similarity to humans, trust, or explainability. This is because the immediate goal for the STE will be to create a testbed that allows for productive teaming opportunities among hybrid teams, not the development of autonomous agents. It is therefore important that we focus on the features that are most likely to factor into the development of the system.

## 11 SCENARIO AUTHORING

The research team felt that the STE must include a robust scenario authoring capability to create specific "missions" within the broader task domain. The team envisioned a number of important authoring capabilities such as the ability to control the level of interdependency, difficulty, time pressure, and uncertainty.

In this section, we explore the researchers' assessment of various authoring features.

---

[1] It is interesting to note that this trend in the open-ended responses is at odds with the respondents answers to the Likert item assessing whether autonomous teammates within the STE must be perceived as cognitively similar to their human teammates. The open-ended responses might have led to an expectation that many researchers would agree with this statement. Surprisingly, it had the lowest average agreement score and the largest standard deviation.

## 11.1 Analysis of Discrete Features

Sonalysts' experience developing training software permitted insight on a number of features about scenario authoring that we thought researchers would want included in the STE. Within the survey, we asked respondents to indicate the degree to which they agreed or disagreed with our hypotheses about scenario authoring features important to include within the STE. Pie charts and descriptive statistics for each of the domain-related features are shown in Appendix H.

As before, we draw attention to items that surpass our threshold by having standard deviations greater than or equal to 1.0. Only one scenario-authoring feature surpassed our threshold: "The authoring environment should make it easy to specify which roles should be staffed by humans and which should be staffed by agents." This item has a standard deviation of 1.0. As shown in Figure 16, only one respondent disagreed with this item; all other respondents agreed or strongly agreed that the STE should incorporate this feature. It may be worthwhile to broach this topic during a follow-up to identify what thoughts respondents have on including it as a scenario authoring feature, and what reservations anyone may have on including it.

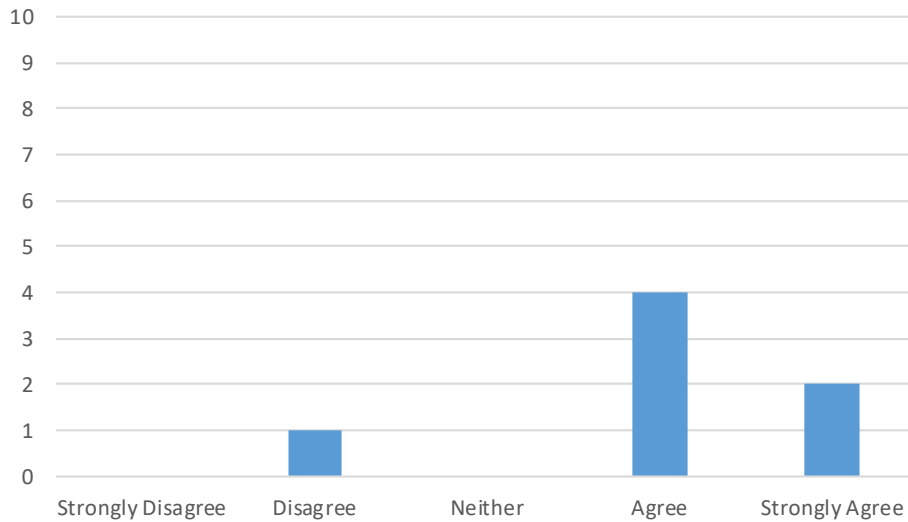

Figure 16. Agreement ratings for "The authoring environment should make it easy to specify which roles should be staffed by humans and which should be staffed by agents."

## 11.2 Analysis of Open-ended Questions

After presenting the Likert items, the survey presented the researchers with three prompts intended to explore other aspects of authoring within the STE:

1. In a couple of sentences, please describe the challenges that you currently face with scenario authoring in the research tools you use today. What scenario authoring features do you value? What scenario authoring features do you feel are lacking?
2. What's more important to you in an authoring environment- simplicity or power? Please explain your decision using one or two sentences.
3. Do you have any additional ideas, comments, or questions regarding scenario authoring? If so, please add them here.

Table 9 summarizes the themes that appeared in the respondents' answers to these questions.

Table 9: Analysis of Responses to Scenario Authoring Prompts

| Question | Analysis |
|---|---|
| In a couple of sentences, please describe the challenges that you currently face with scenario authoring in the research tools you use today. What scenario authoring features do you value? What scenario authoring features do you feel are lacking? | Experience with scenario authoring varied across the sample, with two respondents reporting no experience with authoring tools.<br><br>Of the valued scenario authoring features, the most common theme was simplicity (n=2). Aspects of simplicity involved the ability to build or modify existing scenarios rapidly, without the help of software engineers or SMEs. Other desirable features of scenario authoring included flexibility (n=1) and fidelity to real-world teaming challenges (n=1). Two respondents mentioned features related to setting challenges within a task. One of those respondents specifically described the utility of being able to link particular scenario events with the skills those events require and the mechanisms/standards used to assess performance of those skills.<br><br>As an example of scenario authoring tools that "worked," one respondent described a system similar to what could already be found in a commercial gaming engine (e.g., that most had Software Development Kits (SDKs) to facilitate authoring, and were flexible).<br><br>Most respondents did not say anything about challenges associated with scenario authoring, but one mentioned it was a challenge to come up with a scenario that was technically feasible and also satisfied the research question. |
| What's more important to you in an authoring environment- simplicity or power? Please explain your decision using one or two sentences. | Responses to this question were fairly evenly split among respondents. Two said they would prefer both simplicity and power, with one respondent claiming this was a false dichotomy. Three respondents selected power over simplicity, and two selected simplicity over power. |
| Do you have any additional ideas, comments, or questions regarding scenario authoring? If so, please add them here. | A theme for responses in this section that echoed a response in the question about valued scenario authoring features was the suggestion that we look into existing open source tools and engines for scenario authoring. One of those respondents also suggested that we adopt existing user experience guidelines into the development of the STE. |

## 11.3 Summary of Findings

A very clear insight can be gained by examining the features that were most valued by the researchers (as assessed via the Likert items). Two items had much higher agreement scores than the other five (the emphasis is added for clarity):

1. The authoring environment should make it easy to **modulate the "difficulty"** of a task (*i.e.*, the likelihood that a team will achieve a favorable outcome). ($\bar{x}$ = 4.43)

2. The authoring environment should make it easy to **modulate the "uncertainty"** inherent in a task (*i.e.*, the clarity or lack thereof regarding what is happening in the environment). ($\bar{x}$ = 4.43)

Our respondents were clear that the critical feature of any authoring environment was the ability to control the levels of difficulty and uncertainty easily within a given scenario.

The responses to the open-ended prompts provided the research team with additional insights. These responses indicated that authoring should set meaningful challenges for the participants and that the tool should help with design of scenarios to ensure direct linkages among the skills to be assessed, the events that demand those skills, and the strategies and standards that are used to assess those skills.

Another theme that emerged from responses was the desire for simplicity, specifically the simplicity to create scenarios without SME input and modify existing scenarios. This seems to tie in with the feature "the authoring environment should make it easy to specify which roles should be staffed by humans and which should be staffed by agents," which, while meeting our threshold, was primarily supported by respondents.

### 11.4 Topics for Further Exploration

The biggest discrepancy among open-ended responses and agreement ratings for features related to scenario authoring was between the notion that it should be easy to (1) modify scenarios and (2) specify which roles humans should staff and which agents should staff. In a follow-up event, it may be worthwhile to dedicate some time to understanding why this facet of ease of use may not be a desirable feature of the STE. One possibility might be that someone disagreeing with that feature might want all roles to be staffed by either human or autonomous teammates, and therefore this distinction is not relevant. Additional discussion about this will be required to understand these distinctions.

## 12 ARCHITECTURE GOALS

We anticipate that the STE we will develop will support many research studies over a span of many years. As a result, it will be important to develop an architecture that enables on-going improvement of the system. The research team anticipates that an open source modular architecture is likely to offer the best solution, but we wanted to explore this issue with the research community. That topic was the focus of this portion of the survey.

### 12.1 Analysis of Discrete Features

In the survey, we put forward a number of features we thought researchers might value regarding the architecture of the STE. Within the survey, we asked respondents to indicate the degree to which they agreed or disagreed with features related to the STE's ability to support distributed research teams, and update the system using a cloud-based or local approach. Pie charts and descriptive statistics for each of these features are shown in Appendix I.

As we did for other portions of the survey, we conducted an analysis to determine whether any items had standard deviations greater than or equal to 1.0. Items that satisfied these criteria were prime targets for further investigation. None of the architecture features met that criterion.

## 12.2 Analysis of Open-ended Questions

The survey asked researchers who participated in this study to respond to two prompts about system architecture. The first probed researchers' views of whether a cloud-based or local network approach would be preferable for meeting research needs. The second was to collect additional ideas researchers may have had about system architecture that did not fit into the cloud vs. local network distinction.

Responses to these questions are documented in Table 10.

Table 10. Analysis of Responses to Architecture Prompts

| Question | Analysis |
| --- | --- |
| A cloud-based STE would allow researchers at one location to make changes to the STE (e.g., software changes or scenarios definition changes) and have them simultaneously reflected at all locations. An alternative approach would treat each location as independent and allow isolated changes. Please share your thoughts on these alternatives. What are the strengths and weaknesses of these approaches? | Responses to this question resulted in a list of benefits of each approach. Respondents mentioned "pros" for the cloud-based architecture being better configuration control (n=2) and more support for interaction across labs (n=1). Answers focused on the benefits of a local system included lower expense and better security (n=1) and more reliability (n=1). Two respondents reported an explicit preference for a locally-hosted STE because that solution provided more flexibility for system modification at a single location.<br><br>One respondent stated that they did not know enough to answer the question, but recommended both a cloud-based configuration and independent location approach. |
| Do you have any additional ideas, comments, or questions regarding the architecture goals of the STE? If so please add them here. | Respondents did not report any additional ideas in this section of the survey. |

## 12.3 Summary of Findings

As shown in Appendix I, when responding to the Likert items respondents generally agreed with prompts that *favored cloud-based architectures* (the ability of geographically distributed labs to replicate a given task environment, and the ability to conduct studies on distributed teams; ($\bar{x} \geq 4$, for both prompts). Respondents also felt that their research groups included individuals with the software engineering skills required to make modifications to an open-source STE ($\bar{x} = 4.125$).

Interestingly, when responding to open-ended prompts, respondents may have shown a *slight preference for a local system*. Benefits of the local system included better security and more reliability for the co-located research group. However, respondents also drew attention to potential benefits of a cloud-based system, and some were proponents of both.

## 12.4 Topics for Further Exploration

It may be useful to conduct a follow-up event to assess how common multi-site research is for this community. Understanding the logistics of how labs conduct research and where participants are located will help us focus the development of the STE to meet these needs.

Earlier in our discussion of STE features, respondents also expressed interest in open-source STEs. Another question to broach with the group in a follow-up event would be whether the skills that are required to modify open-source solutions are available in the research community. As a part of the survey, we asked researchers to confirm whether their research teams included individuals with the software engineering skills needed to manage modification of an open-source STE. Participants generally agreed that they did, but some reported a neutral response to the question.

This Page Intentionally Blank

*Appendix A:*

*Researcher Knowledge Elicitation Survey Responses*

In this appendix, we list the respondents' answers to each question. The following conventions are used:

- Bold and italicized text indicates a survey question.
- Text in quotations indicates a response to the question.
- Line breaks are used to separate respondents.
- Answers that were "non-responsive" (*e.g.,* "I don't know") are not included.

***In a few sentences, please describe a study that you might conduct using a hybrid teaming synthetic task environment. Who are the participants? What is the task domain they engage with? What signifies success or failure of teams within that task domain? Task domains could be anything from conventional military operations, to a fictionalized setting such as protecting Earth against alien invasion.***

"A task that necessitates information exchange and communication between teammates, and one that possibly allows for different communication structures (e.g. hierarchical vs flat)."

"Military C2. Maritime Operations Center. Participants Undergraduate students through junior officers. Teams attempting to complete complex mission planning & mission management. Could be any context of military operations including HA-DR, logistics planning. Expeditionary Base Operations, etc. Success: 1) Better decisions faster. 2) Non-expert performers behave more like expert performers."

"My responses specifically target a military context using ground vehicles (manned and unmanned) as well as some aerial assets such as UAVs attached to the vehicles. Within this context, Soldiers teamed with autonomous systems could engage in a variety of missions and subsequent tasks. For example, reconnaissance and surveillance, movement to contact, offensive and defensive operations. Participants should be familiar with these types of missions (which may require a specific MOS) and how they are performed using legacy systems (without autonomy). And after being introduced to the concepts of how autonomy could be employed, provide feedback as to how these systems would be most beneficial during operational missions. With respect to specific tasks - they would depend on the type of mission but generally there are 3 high level tasks: move, shoot, and communicate. Success or failure most likely depends on the research questions and how you are employing the autonomy - for example, is just mission accomplishment enough? Or by incorporating autonomy do you want to see an improvement in mission completion time, or a reduction in decision time by the commander, or increased engagements with OPFOR and less damage taken for friendlys? Lots of things to think about here."

"I'd like to see human and machine teams on a variety of offensive and defensive tactical operations. These operations could be air or space related. Participants could be military personnel or if you made a more gamified STE could be used with mil and Civ participants. My view is the ste could be unclass or could be modular to add in characteristics That are more military while keeping some of the foundational ste components to support experiments and potential digital twinning. Success could be both process and outcome based and those criteria could be defined as we'd go. Backend instrumentation would be a key component of the ste infrastructure."

"Participants are normal adults over 18 years old. The task domain could be search and rescue, or bomb disposal, or move to contact on the battlefield. Success is when they complete the task within the allotted time and maximized their points (e.g., rescued most victims, disposed most bombs, found the most targets and moved from A to B safely). In addition, the injury at the end, the completion of subtasks, and whether the team processes are effective are also a criterion for success."

"I might conduct a study using a hybrid teaming STE which would simulate similar team skills to a real-life complex, high-stakes environment. This would be a task which would require a high degree of communication, coordination, and situation awareness. The participants could be anywhere from undergrad students to real military members, and success would be mission completion along with effective coordination, etc. I would prefer the task domain to be conventional military operations as opposed to a fictionalized setting."

"I would want to study the construct in as close to real-world environment, warfighter tasks, and systems as possible (e.g., difficulty is not a driver, unless it must mimic the real world). None of these answers under item 2 can be answered, truly, until you know your real-world use case you're wanting to target. If you don't have a use case, then what is the purpose of simulating a task environment?"

"I would conduct a study to determine the effect of human team member skill expertise (Low vs moderate vs high expertise) and task complexity (low vs moderate vs high) on synthetic team member task expertise requirements. Most human team members will have different levels of expertise due to changes in task requirements and member turnover. Adapting to team member changes may be easy given an easy task, but as task complexity increases, the adaptation will become more difficult. Synthetic team members should effectively adapt to this increase in complexity. In this study the task domain should require strong dependencies among team members for satisfactory task performance at each level of task complexity, for example, a ship combat team, or other command and control task environments in any of the military services."

"We do research involving everything from MUMT applications involving fighter pilots, ABMs, and ground operators interacting with heterogeneous automated assets, to AI assisted dogfighting, to space applications (i.e., satellite operations), to automated systems on actual fielded platforms, to teaming testbeds for distributed team members interacting in a multidomain context, to basic research using team space fortress or other toy scenarios. Bottom line, we operate in many domains and the testbeds used range from low fidelity to medium fidelity to high fidelity based on the program and type of question being asked. Basically, I'm saying this question cannot be answered like this."

"I would be most interested in tasks that require online learning and adaptation based upon uncertainty and dynamic changes in tasks, requirements, constraints, and communications. That learning should happen for both humans and machines. Military surface characteristics are less important than appropriate fidelity to create varying levels of complexity."

Are there any features that you think are important to include in the task domain that are not listed above? Please list them here.

"Flexibility with what data your collect - ability to collect subjective data, and data from other sensory modalities - but only if that is of interest to the researcher/research questions."

"Not sure what cognitive coordination means. I think there's lots in my ste environment above to consider here."

"The tasks can reflect good and bad teamwork; the tasks do not force teamwork but allow for optional teamwork. The task environment provides opportunities/challenges for the team to benefit from advanced technologies."

"The task domain allows for measurement of both team process (e.g., communication, efficient processes) and team performance. These things are most likely correlated, but they can be separate in certain cases (e.g., a team with poor process that may just get lucky can achieve high performance)."

"Anything that is actually found in the nearest term real-world warfighting use case."

"I recommend the STE development should build on many years of past research on combat teams and it should effectively enable valid and reliable electronic assessments of team performance."

"Depends on the task… what are you trying to study? HAT platforms need opportunities for teaming (i.e., working together toward a shared goal), which is different than just interaction or coordination. This kind of assumes more than one role, some level of machine intelligence or capability to act on behalf of that role, a shared goal of sorts, teaming affordances - ability to communicate and coordinate, and interdependence in a context that matters (you might say one in which there is some cost or risk for the interaction - but also potential benefit)."

"Adversarial environments are a desirable feature."

Do you have any additional ideas, comments, or questions regarding the task domain? Please add them here.

"None."

"I think you could start with what the fundamentals of a robust flexible ste are and build on those. For me, specific task activity and outcomes instrumentation will be one key thing. Modularity is another. Application in mil and Civ contexts is another."

"It is pretty difficult to design such a task environment and collect a large dataset."

"In general, I would prefer an STE to be as realistic as possible to real-life operations."

"If you don't know where the real-world operational tools and data are likely to support this type of H-M teaming integration, then it's too early for selecting a use case. If your use case doesn't leverage near-term data, then you picked wrong."

"None."

"I think you're being too broad - I think their would be more fruitful if you pared it down a bit more. What HAT context or task are you thinking of? Then we could talk specifics."

Please brainstorm up to 10 STE features that would be important to you as a researcher (e.g., open source, performance assessment capabilities, flexible teaming configurations, etc.).

- "Open source"
- "Open source"
- "Flex"
- "Performance Assessment Capabilities"

- "This would be arbitrary if you don't know where real-world potential exists for the type of data you need to test this in real world ops."
- "Build on past research programs and findings to leverage reliable and valid measures."
- "Ability to create interdependence between humans and machines"
- "Detailed data recording and performance metrics"
- "Communication capture/documentation"
- "Flexible teaming configurations"
- "Modularity"
- "Optional teamwork"
- "High Fidelity Task Demands"
- "Try to leverage an existing STE that is in 6.3/6.4/6.5 development as a transition testbed and is open source"
- "Military relevance"
- "Language-based communication"
- "Flexible teaming configurations"
- "Ability to collect multiple data sources"
- "Instrumentation and back end analytics"
- "Enough teamwork datapoints"
- "Requires Communication and Coordination from Teammates"
- "Get transition sponsor support"
- "Some actual and scripted machine intelligence"
- "Sufficient complexity that prevents optimal AI systems"
- "Multiple autonomous capabilities"
- "Open data and model infrastructure"
- "Opportunities for advanced technologies to help with decision making"
- "Task Informed by Theory"
- "Pick a team task that requires from 6 up to 12 team members"
- "If you are doing basic research with academic collaborators the more open it needs to be but honestly that is less important the more relevant the question you are seeking to answer"
- "Intuitive displays"
- "Local and distributed capable"
- "Captures various data format, testbed logs, video, audio, verbal communication etc."
- "Flexible Teaming Configurations"
- "Identify how the team under study functions relate to functions outside the team and address study artificialities as a result of this"
- "Scenarios need to be able to be created/changed relatively quickly. If it takes months to change something you are dead on arrival"
- "Simulation easy to learn"
- "Multiple roles and stations for humans and the AI team mates"
- "Flexible teaming configurations"
- "Task can work with both Human and AI teammates"
- "Identify relevant synthetic team member tasks and task performance met"

- "Automated data capture is only as good as those who developed it. Great if it works, terrible if it doesn't - better off having nothing than bad data."
- "Capacity to create specific tasks and metrics"
- "Task modules that supports various task types"
- "Open Source"
- "Flexible teaming configurations would be beneficial as synthetic team members may influence flexibilities"
- "Debrief functions for researchers and participants"
- "Good for training, practice, and actual task performance evaluation"
- "Affordability (if not open source)"
- "Use known team task stressors to identify where team performance breaks down and how synthetic team members can help manage breakdowns"
- "Use known team task stressors to identify where team performance breaks down and how synthetic team members can help manage breakdowns"
- "Capability to Hold Remote Experiments due to possible restrictions (e.g., Covid)"
- "Get a dedicated group of credible subject matter task experts to help with defining tasks and scenarios"
- "Has the virtual version of the task environment and corresponding physical environment"
- "Use a testbed that may already have proven automated performance assessment capabilities"

What is the minimum number of human teammates the STE should accommodate?

- 4
- 7
- 1
- 3
- 3
- 6
- 2
- 2

What is the maximum number of teaming entities that the STE should accommodate?

- 6
- >9
- 2
- >9
- 6
- >9
- >9
- >9

Do you have any additional ideas, comments, or questions regarding STE features? If so, please add them here.

"To further clarify my answer to (7), it may be desirable to have the ability to have two human teammates and no AI, but assuming there is an AI teammate, then 1 human would be the minimum."

"Minimum number of human teammates really depends on the context, but making this as flexible as possible would be a great asset. Keep in mind that as the number of participants increase so does the quantity of data being collected."

"Smallest team would be a human and an AI teammate."

Think about a research study that you would like to conduct using the STE. What are the variables (or variable types) that you would like to collect in that study? Please list them below.

- "Communication content"
- "Team trust/cohesion"
- "Demographic information"
- "Performance Scores"
- "Team Task Performance (e.g., detect to engage sequence)"
- "Trust"
- "Learning rate"
- "Communication timing"
- "Team state"
- "Gaming proficiency"
- "Communication transcripts (e.g., chat format, words)"
- "Teamwork Performance (e.g., communication quality, team leadership/initiative, information exchange, backup)"
- "HAT performance (task performance)"
- "Learning quality"
- "Time on task"
- "Mission outcomes"
- "Spatial ability"
- "Audio transcripts of communication"
- "Self-reported and objectively measured levels of team cohesion, efficacy, team performance, team processes"
- "Workload"
- "Communication and interaction patterns and effectiveness"
- "Task performance"
- "Vehicle state"
- "Individual performance"
- "Measures of trust"
- "Self- and objectively measured levels of expertise and knowledge"
- "Trustworthiness"
- "Understanding"
- "Team communication"
- "Team performance"
- "Situation Awareness queries or tests"

- "Event based assessment of variables within and across scenarios"
- "Perceived ethics"
- "Situation representation"
- "Workload"
- "Teamwork effectiveness"
- "Subjective Workload (NASA-TLX)"
- "Changes in teamwork and task work performance over time"
- "Machine perceptions (i.e., intent)"
- "Context-sensitivity"
- "Team processes"
- "Video recordings of behavior"
- "Measures of team self-correction during and after scenarios"
- "Individual differences"
- "Team states"
- "Physiological measures (heart rate, EEG, etc.)"
- "Teaming perceptions/processes"
- "Team workload/bottleneck"
- "Subjective ratings of each their own team's performance"
- "Behaviors (task-specific)"
- "Team situation awareness"
- "Machine performance (independent of human)"

Are there human performance assessments that should be "baked in" to the STE? If so, please take a few sentences to describe them here.

"Unsure."

"I would say it depends on the research questions and context of interest."

"Workload, attention/ engagement."

"If a team can complete certain tasks, that indicates they are a good team; if they cannot complete the tasks or take too long to finish, that indicates they are a bad team."

"It would be great if measures of effective coordination could be baked into the STE. This is because coordination is harder to measure in a quantifiable manner than other constructs such as trust, workload, and communication."

Do you have any additional ideas, comments, or questions regarding data collection in the STE? If so please add them here.

"None at this time."

"Lots here already."

"Process variables like coordination and communication are important and would be my primary measure of interest."

"Consider conducting a comparison between groups on whether a structured AAR is used after each scenario."

"Embedded survey capabilities are a nice to have (when they work) but paper and pencil works just fine too and rarely breaks."

Using a couple of sentences, please describe what kind (if any) on-board data analytics and/or data visualization tools would be useful to include within the STE.

"As long as the data are easily exportable, on-board data analytics and visualization tools would be nice, but not required."

"Visualization of team state during the mission, data regarding vehicle state during the mission - the specific visualization would depend on the variable of interest as well as the user - but in general on-board data analytic should be easily understood with the ability to do 'quick looks.'"

"If you have real-team analytical data to show the weakness of the team, display it in a certain way to the commander so the commander knows how to fix that weakness. If the system has processed information to help the operators make a difficult decision, show the number and criteria, and suggestions."

"Any analytics which show relationships between variables at a quick glance would be useful."

"Onboard data analytics tools should collect and evaluate pre-existing validated human and team performance measures at critical incident events throughout STE scenarios. The data visualization tools should provide easily understandable visualizations of team member performance and interactions over time, for both for research analysis, and after action reviews to team members."

"Debrief capabilities where appropriate - or rehearsal."

"Ultimately, flexibility will be the key. So, tools for integrating analytics by end users that are designed for specific research purposes."

Do you have any additional ideas, comments, or questions regarding data visualization and analysis? If so, please add them here.

"Some variables and visualizations would be more important to look at during the mission, and some after the mission. For example, any of the analyses listed above would probably be more effective after the mission because they would take more time to interpret. These will also require more computing power depending on frequency that these data are output."

"Normally STEs provide the testbed environment and data, not processed data and visualization."

"It would be nice to have the STE have statistics embedded within the system, but even more important to allow users to export data for their own analyses."

"Because team interactions involve assessing multiple related variables, the STE should ALSO enable Multivariate Analysis of Variance and Multivariate Analysis of Covariance. Because team interactions change and develop over time, STE should also enable stochastic analyses."

"STE should not be used for analysis but it should enable data export for subsequent analysis."

A cloud-based STE would allow researchers at one location to make changes to the STE (e.g., software changes or scenarios definition changes) and have them simultaneously reflected at all locations. An alternative approach would treat each location as independent and allow isolated changes. Please share your thoughts on these alternatives. What are the strengths and weaknesses of these approaches?

"A strength of isolated is that if there is any concern over network connectivity, the isolated STE will be able to function even if connectivity is lost; I assume if the STE is only in the cloud, connectivity is always required? But otherwise, I assume having the cloud-based option allows for faster implementation of updates."

"I'm not as familiar with the architecture side of things, but it might be a good idea to allow both types of changes - some you may want to be more isolated so testing can be done before pushing to all locations. You may also want different locations to have different capabilities depending on what you are interested in studying."

"Maybe. Depends on the nature of the tasks and their sensitivity."

"The team at each location has its research questions and hypotheses on its timeline. Unless all the teams related to the STE are working on the same program, otherwise, they would not want others to change their testbed."

"I would prefer that each location is independent and can enact their own isolated changes. While the cloud based approach is more convenient in ideal cases, there can be instances where a team of distributed teams would want manipulations unique to their own local location. I would prefer flexibility over convenience in this case."

"A cloud-based STE is more effective at maintaining control over the distributed STE systems and would ensure a more stable architecture for teams interacting together in real-time across distributed network. The independent location approach may be ok (and possibly cost effective) if each location is focused on studying teams at THAT location. Both the cloud based and local STEs would have to be secured to ensure human data collected is protected according to Human Subjects research policies. The cloud based system may be more vulnerable to leaking human subjects data, but I am not an expert on this."

Do you have any additional ideas, comments, or questions regarding the architecture goals of the STE? If so, please add them here.

None.

In a couple of sentences, please describe the challenges that you currently face with scenario authoring in the research tools you use today. What scenario authoring features do you value? What scenario authoring features do you feel are lacking?

"This is not something I have a lot of experience with. I've primarily designed fairly simple, experimental psychology paradigms that provide control over the scenario through the simplicity of the task (i.e. limited variance in behaviors and outcomes). Anything that would allow me to more closely simulate real-world team interactions would honestly be an improvement."

"This is not my area - so I'm unable to comment."

"Most commercial game engines have decent SDKs for this. Flexible too."

"I value (1) setting up the challenges and difficulties in the task processes, (2) choosing input types on the human-machine interface, and (3) allowing participants to give feedback to the displays."

"I am only recently beginning to author scenarios for an STE. One challenge I can think of is coming up with a variety of scenarios that can be both technically feasible, while at the same time satisfying a research question that I may have."

"My past efforts (as I am retired now) with building STEs has always been hampered by unreliable/incompatible scenario authoring systems. There is always some work around that has to happen when adopting an existing system. A lot of money is spent incorporating more features into an authoring system. A lot of time is spent with subject matter experts in the design of scenarios. I value the ability to a) rapidly build event based scenarios with minimal inputs from SMES, b) automatically link team skills/competencies with scenario event design, c) easily modify existing scenarios, d) link scenario events to trigger performance assessments, and e) rapidly generate scenario event team performance analyses to visualizations."

What's more important to you in an authoring environment- simplicity or power? Please explain your decision using one or two sentences.

"There's a lot of focus currently on JADC2 - simulated that kind of environment (or what we may envision that kind of environment to be like) I expect will necessitate a lot of control over how the environment is configured."

"Both it's not an either or imho."

"I am not sure what you mean by simplicity or power. I guess simplicity is related to user experience, and power is related to functionality. I would prefer power. If the function does not work, then the format does not matter."

"Power is more important in authoring an environment. I would be willing to trade simplicity for the ability to create a variety of scenarios which can answer a research question."

"Many authoring environments focus on computing power to produce complicated/complex scenarios that can last for hours and not crash. I think that's really important. But they don't get used enough because SMEs are left to figure how to use the most arcane interfaces. The authoring tools interface NEEDS to be simple - that's more important than emphasizing power."

"Simplicity if good for basic platforms - for doing online studies for example when you are just trying to answer a lot of questions quickly with large n studies. The closer you get to high fidelity the less this is important and more psychological fidelity is important regardless of complexity."

"You need both. Simple options to allow typical end users to create scenarios and variation, but also functionality so that "power users" can extend the capabilities and make more complicated changes."

Do you have any additional ideas, comments, or questions regarding scenario authoring? If so please add them here.

"This is not my area - so I'm unable to comment."

"Check out where available. There are some excellent open source tools and engines perfect for this."

"Changing one of the members to be AI or human seems more complex than simply switching. So it is not required in a short-term project."

"Great strides have been made in the military to move from power to simplicity, pointing to the US Army's work on the generalized framework for Intelligent Tutoring (GIFT) system. There are plenty of guidelines for designing simple/user friendly interfaces. So I don't see this as difficult problem to solve."

In a couple of sentences, please explain the challenges that you face facilitating, capturing, and analyzing communication using your current research tools.

"Getting audio data from multiple speakers that is of sufficient quality to facilitate auto-transcription can be challenging - especially if the participants are co-located and have the ability to talk over one another. Participants that are co-located will also leverage visual cues to communicate, which can mean that verbal communications alone are impoverished for reconstructing the full dialogue (video is one solution, but this then requires time-intensive annotation of the video files). Time-locking of communications to task events can also be critical for fully interpreting communications; having this automated would make a huge difference. Being able to identify individual speakers is also crucial; some operational data I have access to does not automatically track individual speakers."

"It really depends on what you want to do with the communication data, but some of the challenges we have faced are: crosstalk (picking up comms from one team member at another crew station - due to close proximity), being able to tease out who is talking (e.g., what crew station) and who they are speaking to, but probably the greatest challenge is transcribing speech to text in real-time with an acceptable level of accuracy."

"None we've had good success here."

"The current technologies cannot process natural language quickly and accurately. Transcribing verbal communication is labor intensive."

"Voice transcription is a common problem. I think the STEs I have worked with are very good at collecting timestamp data, and speaker identification data, but higher level measures like transcription, or even emotion, are harder to collect."

"So many people have to be involved in physically collecting and analyzing communications that this daunting challenge must be mitigated with automated, artificially intelligent capabilities. Capturing and assessing team communications is based on a trained cadre of observers/SMEs listening to comms, recording/noting the comms on electronic tablets or paper, ensuring they are time stamped, and then analyzing findings to assess performance. The reliability of these data points rests on everyone doing an exemplary job. So, for research moving forward this has to be automated - namely speech recognition to assessment."

How would you like teammates to communicate within the STE?

"Flexibility would be very appealing (e.g. the ability to toggle between verbal and chat, and co-located and distributed comms with the additional flexibility to have either open channels or push-to-talk for verbal communication."

"Chat comms are easier to work with, but a lot of comms within military vehicles is verbal. Obviously collecting and analyzing these data becomes more complex as the team size grows."

"Same ways we'd expect them to real world."

"Text chat, radio, and reports embedded on the interface."

"I would prefer spoken communication, since it is the most natural."

"Via chat, electronically supported comms, ftf comms, nonverbal and verbal comms."

How do you normally assess communication processes and outcomes?

"Most of my work requires being able to analyze and interpret the content of specific communications."

"It depends on the research questions - there are different methods we have used depending on if we are interested in the flow of comms within the team (social network analysis), or if you're interested in content we have used LSA, NLP, etc."

"We define criterion and look for proximity to that in what's captured."

"Use Zoom transcript/or Google speech-to-text app to get the text, clean it, develop a code book, and have two people code the data."

"I analyze data on who was speaking to whom and when (speaker ID's, timestamps, etc.)."

"Using a tablet-based device loaded with the scenario, and that automatically timestamps when an assessment is made. Using a stylus or finger to tap a predetermined set of team communication behaviors. I have used plenty of paper based checklists keyed to specific events, but the tablet works better."

Do you have any additional ideas, comments, or questions about measuring communication in the STE? If so add them here.

"None at this time."

"Check commercial and open sources here."

"It is important to measure the quality of communication, not just the quantity of communication."

"Communication measurement would be the most important aspect of an STE."

"Not at this time."

In a couple of sentences, please describe the ideal autonomous teammates to support your research within the STE. Are they "built-in" or developed externally? What are their characteristics (e.g., transparency, reliability, etc.)? What dimensions of their performance are adjustable?

"Being able to have externally developed agents dropped into the STE would be most desirable, but "built-in" would also be useful. I'd be particularly interested in agents that can communicate through human (or human-like) language."

"The ideal autonomous teammates would act similar to human teammates and engage in missions in specific tasks, roles and coordinate with other team members (e.g. teamwork and taskwork behaviors).

Although these capabilities do not currently exist - there is definitely interest and movement in this area. As far as built-in or developed externally, I'm not sure, I think either would work."

"I don't know. I'd have to see some first I think honestly."

"Focus on supporting the difficult tasks of humans to help make the right decisions, take effective actions, and get the right information."

"It personally doesn't matter to me if the autonomous teammate is built-in or developed externally, I think both instances would work. Transparency would be important depending on the type of research. It would be ideal if researchers could adjust things like reliability and trustworthiness, as I think these could be helpful to answering a variety of research questions related to how humans work with autonomous teammates."

"Please see my responses to the survey below."

"Features that we would like tweak include: roles, decision authority, team-oriented intent (e.g., benevolence), communication styles and content, available comm mediums, reliability, trust repair strategies, morality, adaptation capabilities, interface features."

"Any included AI teammate capability is likely to be simplistic. Providing an API for development of agents and autonomy will probably be necessary."

Should autonomous teammates be "explainable"? Describe what "explainable" means to you in this case.

"Yes, if "explainable" means that the human teammates can develop an accurate, or nearly accurate mental model of the teammates capabilities/limitations."

"Explainable here I think means that the decisions, intentions, and actions of an autonomous system/AI is understandable to their human team members. This is also a critical component of human-machine teaming."

"Yes but you're asking for a LOT in this section. Let's not miss the hood in pursuit of the perfect."

"Explainability, to my understanding, means understanding what the teammate can or cannot do and how they did it and why they did it, and how well they did it."

"In this case, explainable would mean that the human teammate can understand why the autonomous teammate made the decision that it made, and can adjust their own behavior accordingly. I think these teammates don't necessarily have to be explainable, since much of the AI technologies we work with aren't that explainable. However, the flexibility to have it would be nice, especially since some researchers may be interested in manipulating explainability."

"Explainable means team members should be able to explain the inherent purpose and role of the autonomous team members in the current team structure. Knowledge of the roles of one's other teammates and how the team task is achieved with team inputs is critical to good team performance. Therefore, understanding and being able to explain the autonomous member's role is critical."

"Different affordances for explanation could be useful for studying HAT using STE."

"I don't see most of this section as being part of the STE. Meeting all the diverse expectations for AI capabilities in a complex task will be a daunting and unattainable pursuit. The STE should create opportunity, not solve all the problems."

Do you have any additional ideas, comments, or questions about autonomous teammates in the STE? If so please add them here.

"I had difficulty choosing the appropriate level for many of the ratings under 30. I was tempted to put "strongly agree" for nearly all of them, but that seems impossibly far out to develop an agent or agents with all of these options. Some of the options seemed related also, like the ability to "acknowledge inputs" I would consider to part of 'establishing common ground.'"

"You may have already captured this in one of the capabilities above, but another important one is for autonomous teammates to learn and adapt."

"No."

"Question 30 talked about autonomous teammates interacting with other autonomous teammates. If there are different types of autonomous teammates (AI), it will be okay to monitor the other agent's input, but it requires multiple software agents to be able to read each other's data. If you have plug-in software agents, we would need more technical support to make them compatible. So it is very challenging to have multiple live agents to establish a shared mental model and display information to human operators appropriately. It may be easier to deal with one autonomous agent at a time."

"Pick a use case where possible near-term data to support this is feasible, then answer all of these questions accordingly."

"For some of the items I answered I wasn't sure that characteristic was absolutely necessary for an autonomous team member. That tells me that I don't know the research well enough on autonomous team members to answer the question and/or there is still research needed to be done. I think each of the Likert questions above should be seen as research questions. That is each item should be considered a behavioral variable that can be electronically manipulated along a continuum or behavioral dial whereby an autonomous team member 'has none of this characteristic' to 'has all of this characteristic.'"

This Page Intentionally Blank

*Appendix B*

*Results for STE Feature Likert Items*

The STE should include support for a broad range of human and AI teammates.

| | |
|---|---|
| Valid | 10 |
| Missing | 1 |
| Mean | 3.40 |
| Median | 3.50 |
| Mode | Strongly Agree |
| Std. Deviation | 1.430 |
| Minimum | 1 |
| Maximum | 5 |

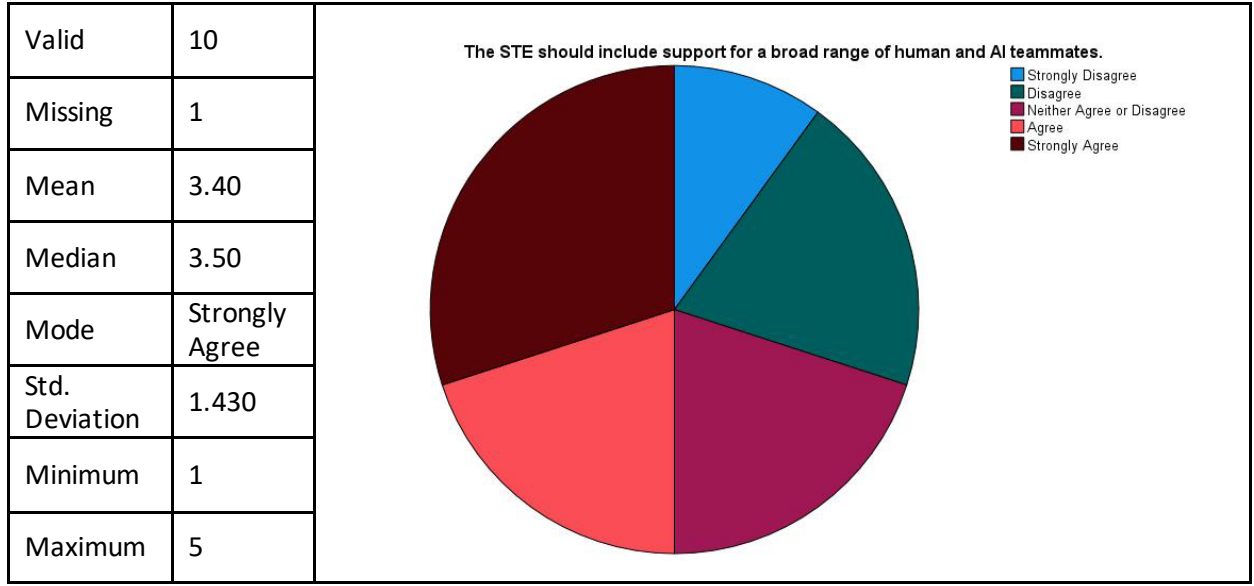

| | | Frequency | Percent | Valid Percent | Cumulative Percent |
|---|---|---|---|---|---|
| Valid | Strongly Disagree | 1 | 9.1 | 10.0 | 10.0 |
| | Disagree | 2 | 18.2 | 20.0 | 30.0 |
| | Neither Agree nor Disagree | 2 | 18.2 | 20.0 | 50.0 |
| | Agree | 2 | 18.2 | 20.0 | 70.0 |
| | Strongly Agree | 3 | 27.3 | 30.0 | 100.0 |
| | Total | 10 | 90.9 | 100.0 | |

The STE should support a variety of hybrid teaming configurations (for example, one human teammate with several AI teammates, several human teammates with one autonomous teammate, or several human and several autonomous teammates).

| Valid | 10 |
|---|---|
| Missing | 1 |
| Mean | 4.00 |
| Median | 4.00 |
| Mode | Agree |
| Std. Deviation | .667 |
| Minimum | 3 |
| Maximum | 5 |

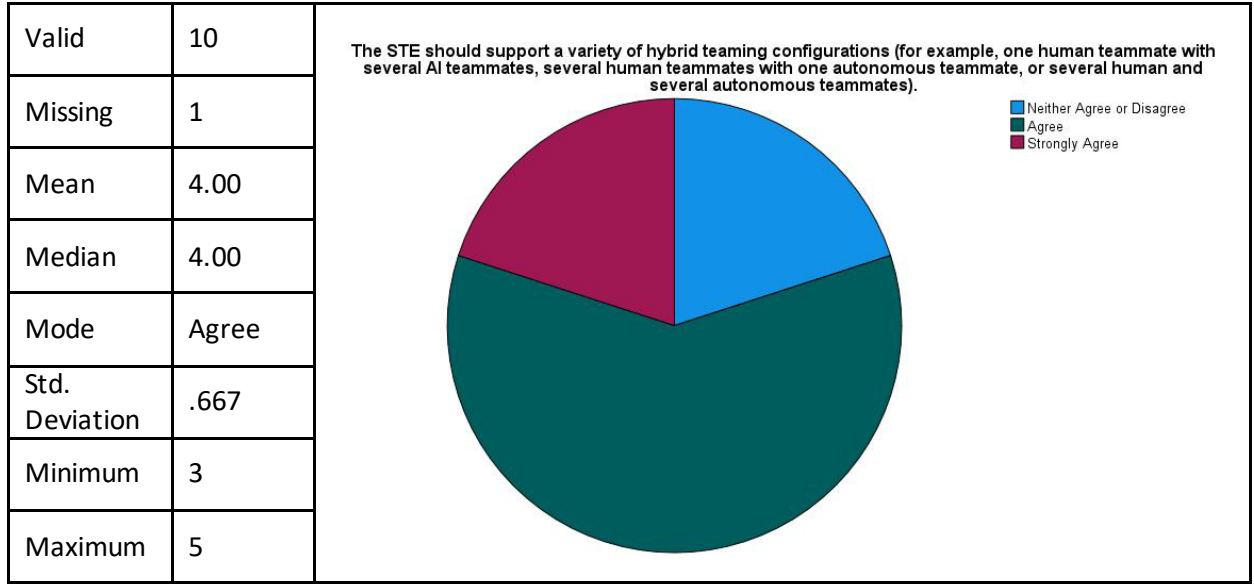

| | | Frequency | Percent | Valid Percent | Cumulative Percent |
|---|---|---|---|---|---|
| Valid | Strongly Disagree | | | | |
| | Disagree | | | | |
| | Neither Agree nor Disagree | 2 | 18.2 | 20.0 | 20.0 |
| | Agree | 6 | 54.5 | 60.0 | 80.0 |
| | Strongly Agree | 2 | 18.2 | 20.0 | 100.0 |
| | Total | 10 | 90.9 | 100.0 | |

The STE should include support for Wizard of Oz experimentation.

| | |
|---|---|
| Valid | 10 |
| Missing | 1 |
| Mean | 3.60 |
| Median | 4.00 |
| Mode | Agree |
| Std. Deviation | .843 |
| Minimum | 2 |
| Maximum | 5 |

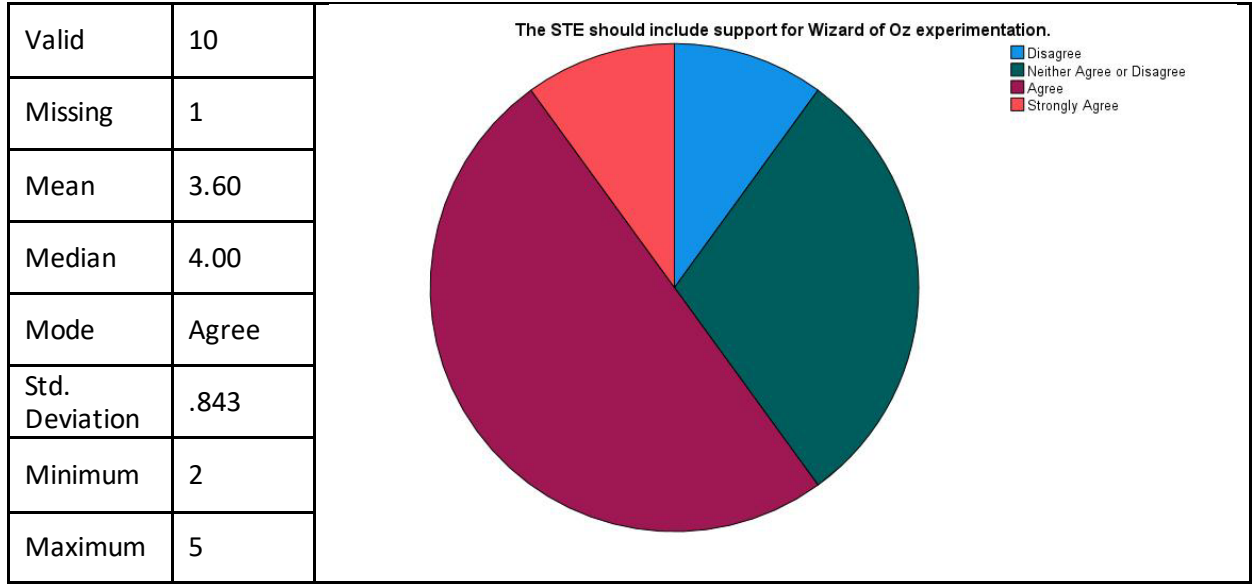

| | | Frequency | Percent | Valid Percent | Cumulative Percent |
|---|---|---|---|---|---|
| Valid | Strongly Disagree | | | | |
| | Disagree | 1 | 9.1 | 10.0 | 10.0 |
| | Neither Agree nor Disagree | 3 | 27.3 | 30.0 | 40.0 |
| | Agree | 5 | 45.5 | 50.0 | 90.0 |
| | Strongly Agree | 1 | 9.1 | 10.0 | 100.0 |
| | Total | 10 | 90.9 | 100.0 | |

The STE should be implemented as an "open source" tool.

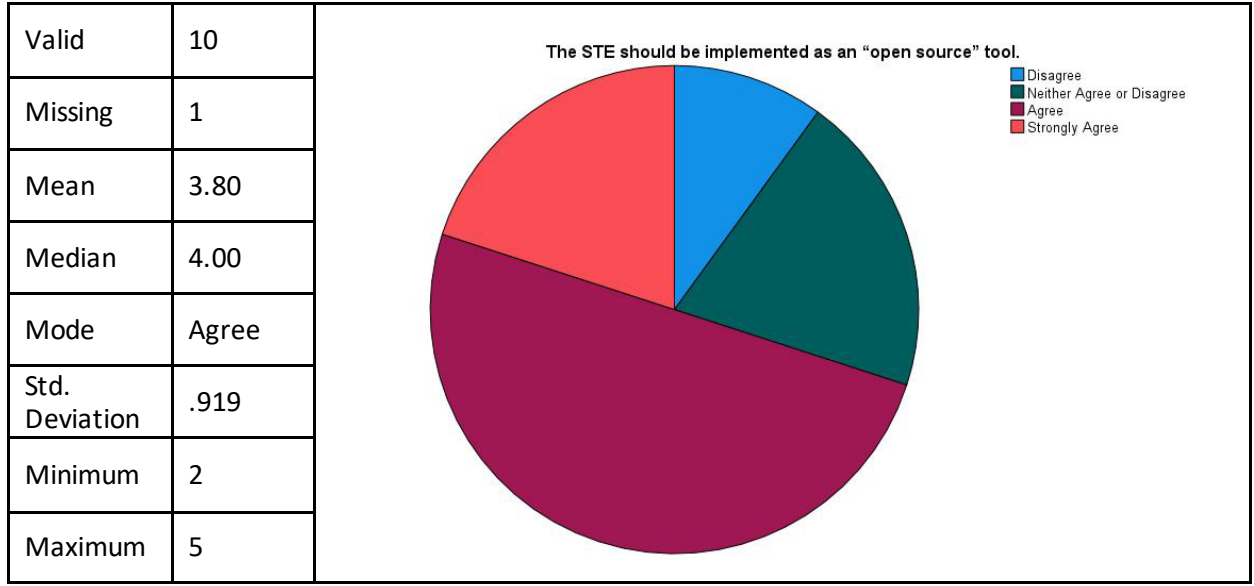

| Valid | 10 |
|---|---|
| Missing | 1 |
| Mean | 3.80 |
| Median | 4.00 |
| Mode | Agree |
| Std. Deviation | .919 |
| Minimum | 2 |
| Maximum | 5 |

| | | Frequency | Percent | Valid Percent | Cumulative Percent |
|---|---|---|---|---|---|
| Valid | Strongly Disagree | | | | |
| | Disagree | 1 | 9.1 | 10.0 | 10.0 |
| | Neither Agree nor Disagree | 2 | 18.2 | 20.0 | 30.0 |
| | Agree | 5 | 45.5 | 50.0 | 80.0 |
| | Strongly Agree | 2 | 18.2 | 20.0 | 100.0 |
| | Total | 10 | 90.9 | 100.0 | |

The STE must support capture and assessment of operator actions.

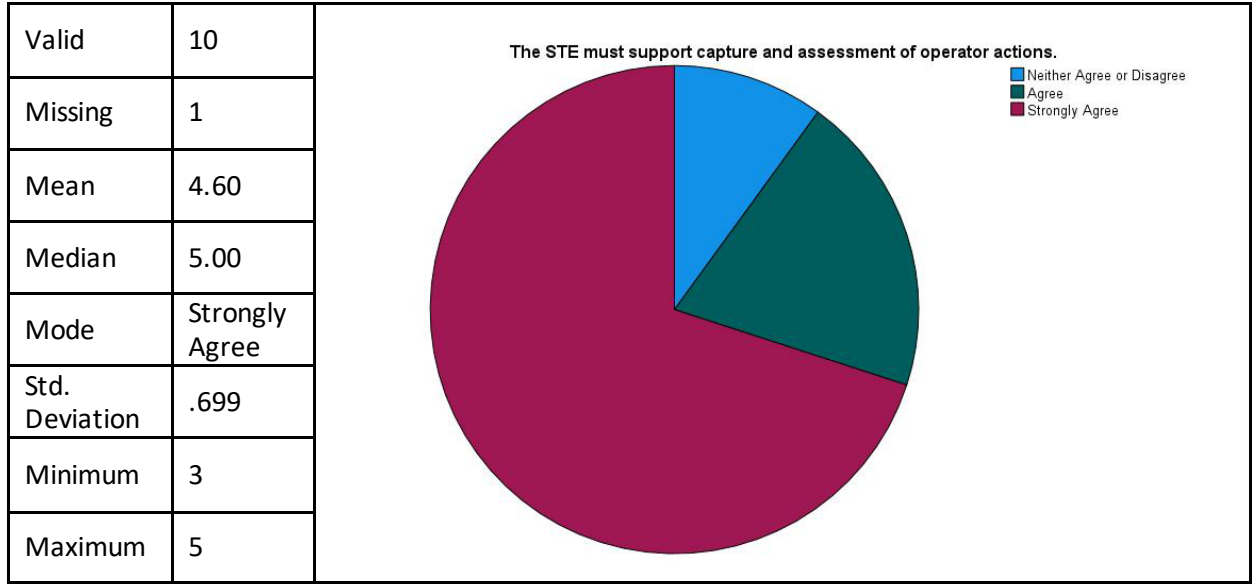

| | |
|---|---|
| Valid | 10 |
| Missing | 1 |
| Mean | 4.60 |
| Median | 5.00 |
| Mode | Strongly Agree |
| Std. Deviation | .699 |
| Minimum | 3 |
| Maximum | 5 |

| | | Frequency | Percent | Valid Percent | Cumulative Percent |
|---|---|---|---|---|---|
| Valid | Strongly Disagree | | | | |
| | Disagree | | | | |
| | Neither Agree nor Disagree | 1 | 9.1 | 10.0 | 10.0 |
| | Agree | 2 | 18.2 | 20.0 | 30.0 |
| | Strongly Agree | 7 | 63.6 | 70.0 | 100.0 |
| | Total | 10 | 90.9 | 100.0 | |

The STE must support capture and assessment of system states.

| | |
|---|---|
| Valid | 10 |
| Missing | 1 |
| Mean | 4.60 |
| Median | 5.00 |
| Mode | Strongly Agree |
| Std. Deviation | .699 |
| Minimum | 3 |
| Maximum | 5 |

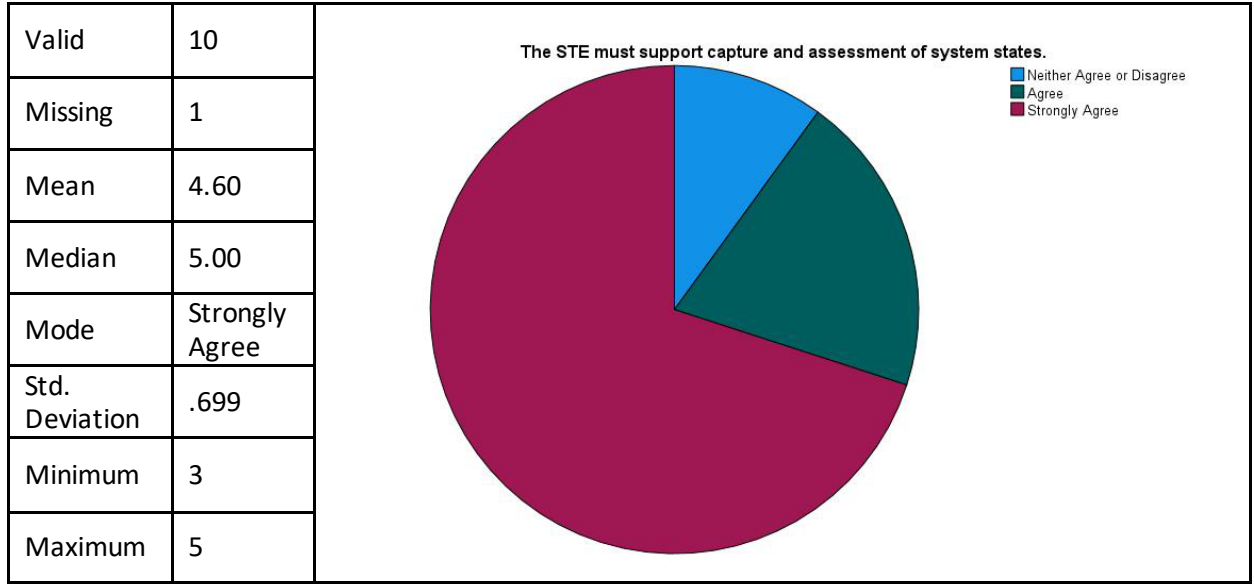

| | | Frequency | Percent | Valid Percent | Cumulative Percent |
|---|---|---|---|---|---|
| Valid | Strongly Disagree | | | | |
| | Disagree | | | | |
| | Neither Agree nor Disagree | 1 | 9.1 | 10.0 | 10.0 |
| | Agree | 2 | 18.2 | 20.0 | 30.0 |
| | Strongly Agree | 7 | 63.6 | 70.0 | 100.0 |
| | Total | 10 | 90.9 | 100.0 | |

The STE must support capture and assessment of environmental states.

| | |
|---|---|
| Valid | 10 |
| Missing | 1 |
| Mean | 4.50 |
| Median | 5.00 |
| Mode | Strongly Agree |
| Std. Deviation | .850 |
| Minimum | 3 |
| Maximum | 5 |

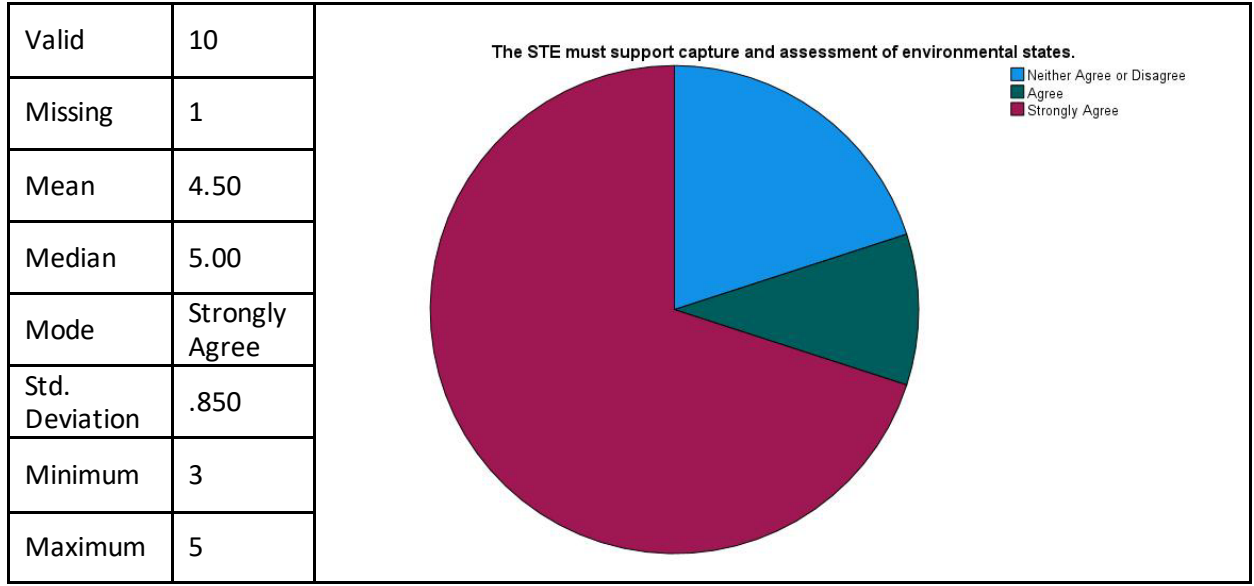

| | | Frequency | Percent | Valid Percent | Cumulative Percent |
|---|---|---|---|---|---|
| Valid | Strongly Disagree | | | | |
| | Disagree | | | | |
| | Neither Agree nor Disagree | 2 | 18.2 | 20.0 | 20.0 |
| | Agree | 1 | 9.1 | 10.0 | 30.0 |
| | Strongly Agree | 7 | 63.6 | 70.0 | 100.0 |
| | Total | 10 | 90.9 | 100.0 | |

The STE must include a robust scenario authoring capability to create specific "missions" within the broader task domain.

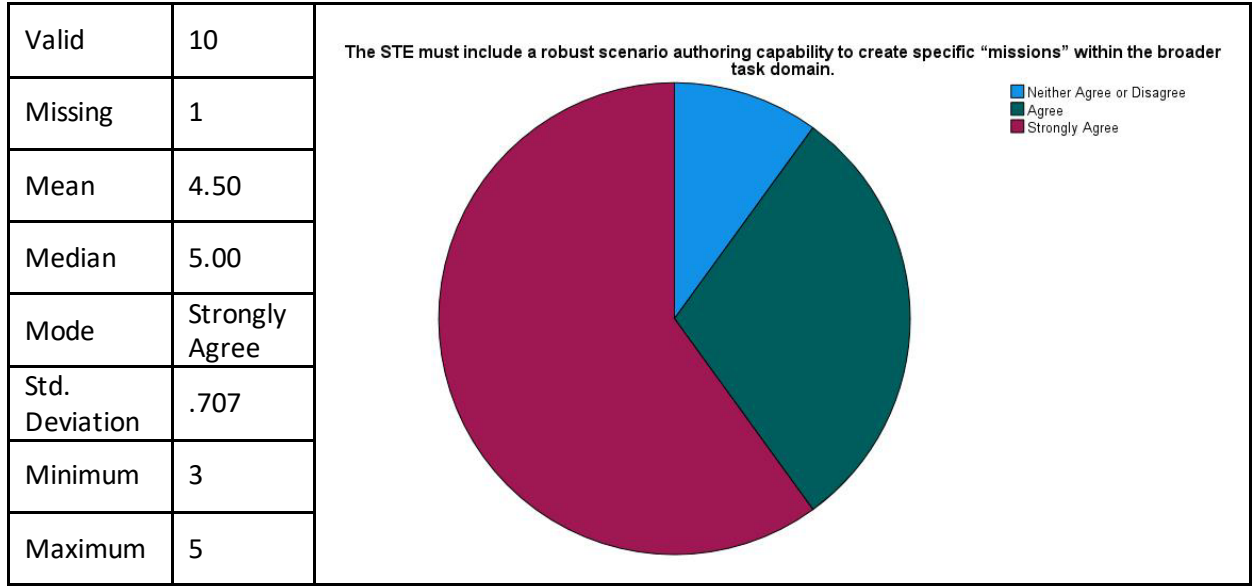

| Valid | 10 |
|---|---|
| Missing | 1 |
| Mean | 4.50 |
| Median | 5.00 |
| Mode | Strongly Agree |
| Std. Deviation | .707 |
| Minimum | 3 |
| Maximum | 5 |

|  |  | Frequency | Percent | Valid Percent | Cumulative Percent |
|---|---|---|---|---|---|
| Valid | Strongly Disagree |  |  |  |  |
|  | Disagree |  |  |  |  |
|  | Neither Agree nor Disagree | 1 | 9.1 | 10.0 | 10.0 |
|  | Agree | 3 | 27.3 | 30.0 | 40.0 |
|  | Strongly Agree | 6 | 54.5 | 60.0 | 100.0 |
|  | Total | 10 | 90.9 | 100.0 |  |

The STE must include the ability to control the level of interdependency across tasks.

| | |
|---|---|
| Valid | 10 |
| Missing | 1 |
| Mean | 4.10 |
| Median | 4.00 |
| Mode | Strongly Agree |
| Std. Deviation | .876 |
| Minimum | 3 |
| Maximum | 5 |

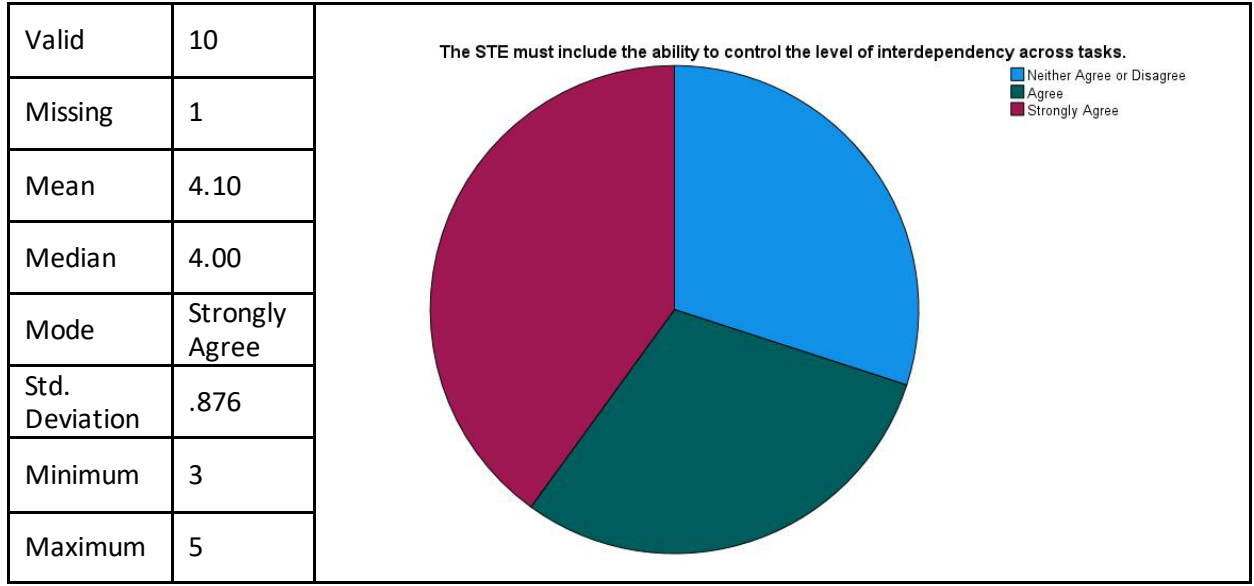

| | | Frequency | Percent | Valid Percent | Cumulative Percent |
|---|---|---|---|---|---|
| Valid | Strongly Disagree | | | | |
| | Disagree | | | | |
| | Neither Agree nor Disagree | 3 | 27.3 | 30.0 | 30.0 |
| | Agree | 3 | 27.3 | 30.0 | 60.0 |
| | Strongly Agree | 4 | 36.4 | 40.0 | 100.0 |
| | Total | 10 | 90.9 | 100.0 | |

The STE authoring capabilities must support the ability to modulate task difficulty.

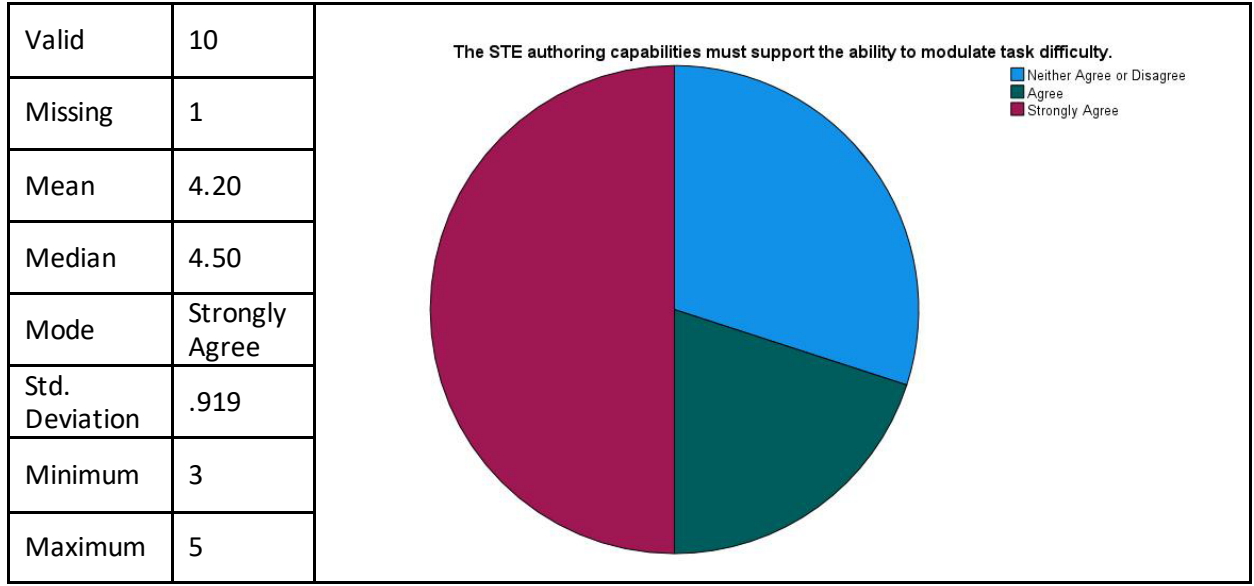

| | |
|---|---|
| Valid | 10 |
| Missing | 1 |
| Mean | 4.20 |
| Median | 4.50 |
| Mode | Strongly Agree |
| Std. Deviation | .919 |
| Minimum | 3 |
| Maximum | 5 |

| | | Frequency | Percent | Valid Percent | Cumulative Percent |
|---|---|---|---|---|---|
| Valid | Strongly Disagree | | | | |
| | Disagree | | | | |
| | Neither Agree nor Disagree | 3 | 27.3 | 30.0 | 30.0 |
| | Agree | 2 | 18.2 | 20.0 | 50.0 |
| | Strongly Agree | 5 | 45.5 | 50.0 | 100.0 |
| | Total | 10 | 90.9 | 100.0 | |

The STE authoring capabilities must support the ability to modulate task time pressure.

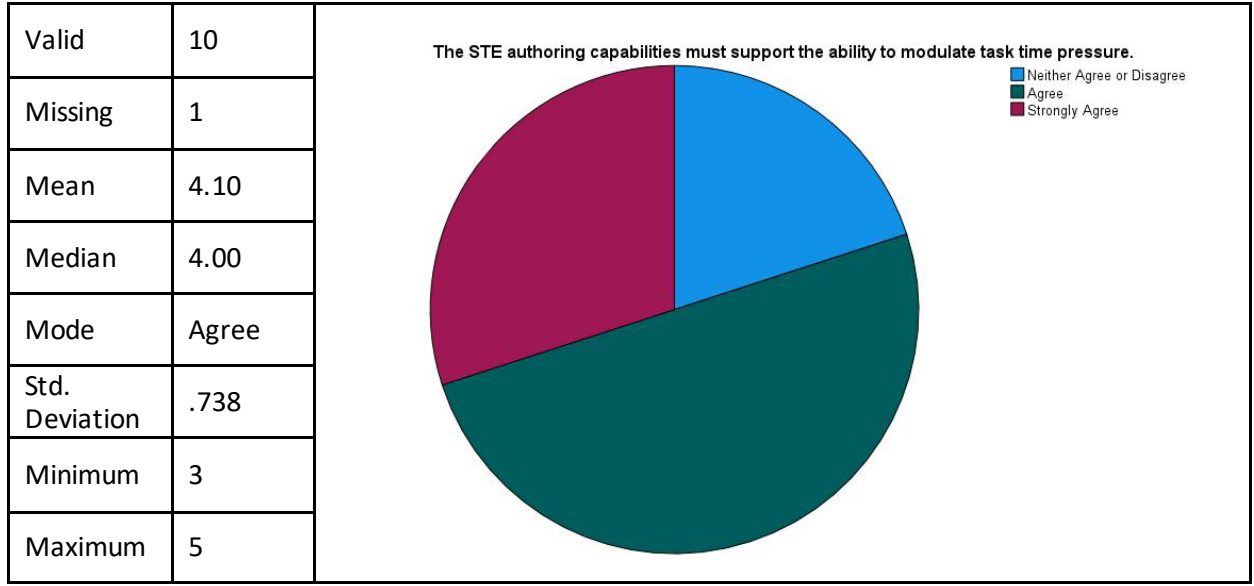

| | |
|---|---|
| Valid | 10 |
| Missing | 1 |
| Mean | 4.10 |
| Median | 4.00 |
| Mode | Agree |
| Std. Deviation | .738 |
| Minimum | 3 |
| Maximum | 5 |

| | | Frequency | Percent | Valid Percent | Cumulative Percent |
|---|---|---|---|---|---|
| Valid | Strongly Disagree | | | | |
| | Disagree | | | | |
| | Neither Agree nor Disagree | 2 | 18.2 | 20.0 | 20.0 |
| | Agree | 5 | 45.5 | 50.0 | 70.0 |
| | Strongly Agree | 3 | 27.3 | 30.0 | 100.0 |
| | Total | 10 | 90.9 | 100.0 | |

The STE authoring capabilities must support the ability to modulate task uncertainty.

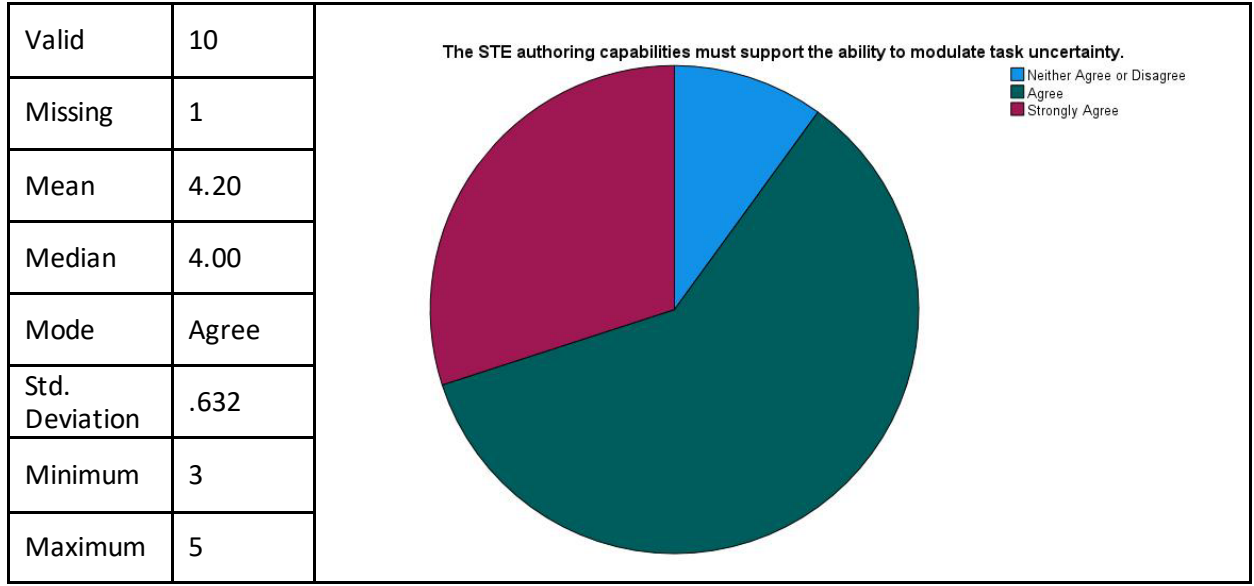

| | |
|---|---|
| Valid | 10 |
| Missing | 1 |
| Mean | 4.20 |
| Median | 4.00 |
| Mode | Agree |
| Std. Deviation | .632 |
| Minimum | 3 |
| Maximum | 5 |

| | | Frequency | Percent | Valid Percent | Cumulative Percent |
|---|---|---|---|---|---|
| Valid | Strongly Disagree | | | | |
| | Disagree | | | | |
| | Neither Agree nor Disagree | 1 | 9.1 | 10.0 | 10.0 |
| | Agree | 6 | 54.5 | 60.0 | 70.0 |
| | Strongly Agree | 3 | 27.3 | 30.0 | 100.0 |
| | Total | 10 | 90.9 | 100.0 | |

The STE must be able to capture audio/video recordings of performance sessions.

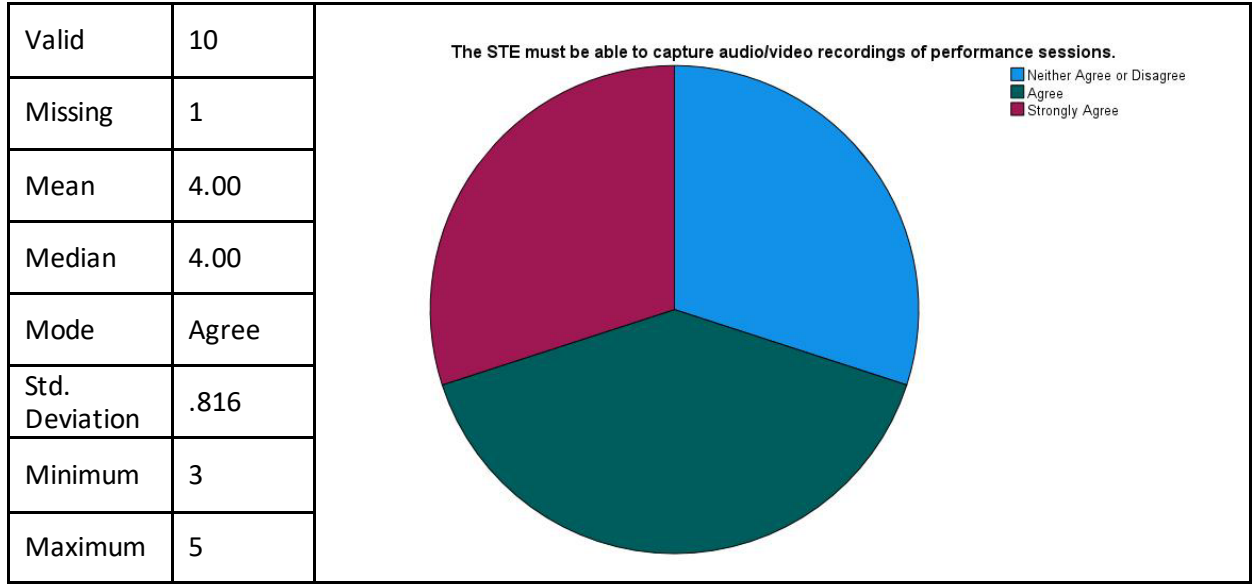

| Valid | 10 |
|---|---|
| Missing | 1 |
| Mean | 4.00 |
| Median | 4.00 |
| Mode | Agree |
| Std. Deviation | .816 |
| Minimum | 3 |
| Maximum | 5 |

|  |  | Frequency | Percent | Valid Percent | Cumulative Percent |
|---|---|---|---|---|---|
| Valid | Strongly Disagree |  |  |  |  |
|  | Disagree |  |  |  |  |
|  | Neither Agree nor Disagree | 3 | 27.3 | 30.0 | 30.0 |
|  | Agree | 4 | 36.4 | 40.0 | 70.0 |
|  | Strongly Agree | 3 | 27.3 | 30.0 | 100.0 |
|  | Total | 10 | 90.9 | 100.0 |  |

The STE must allow participants to "bookmark" key events within performance sessions.

| | |
|---|---|
| Valid | 10 |
| Missing | 1 |
| Mean | 3.90 |
| Median | 4.00 |
| Mode | Neither Agree or Disagree |
| Std. Deviation | .876 |
| Minimum | 3 |
| Maximum | 5 |

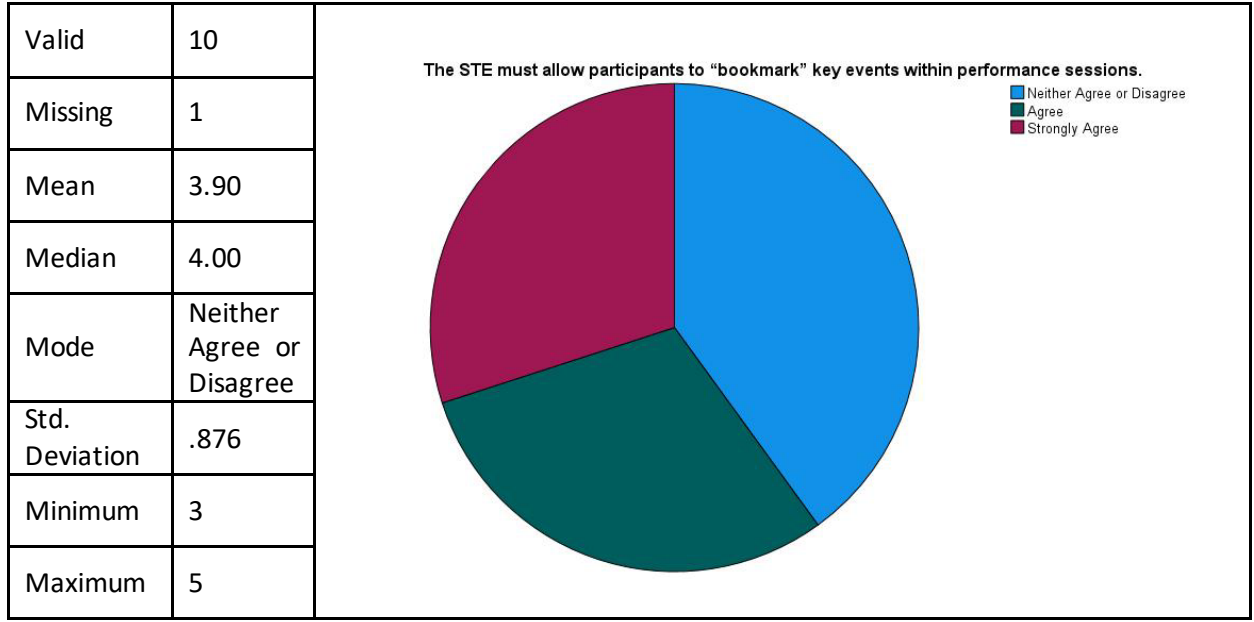

| | | Frequency | Percent | Valid Percent | Cumulative Percent |
|---|---|---|---|---|---|
| Valid | Strongly Disagree | | | | |
| | Disagree | | | | |
| | Neither Agree nor Disagree | 4 | 36.4 | 40.0 | 40.0 |
| | Agree | 3 | 27.3 | 30.0 | 70.0 |
| | Strongly Agree | 3 | 27.3 | 30.0 | 100.0 |
| | Total | 10 | 90.9 | 100.0 | |

This Page Intentionally Blank

*Appendix C*

**Results for Data Collection and Performance Assessment Likert Items**

The STE must gather data on mission outcome variables.

| | |
|---|---|
| Valid | 9 |
| Missing | 2 |
| Mean | 4.67 |
| Median | 5.00 |
| Mode | Strongly Agree |
| Std. Deviation | .500 |
| Minimum | 4 |
| Maximum | 5 |

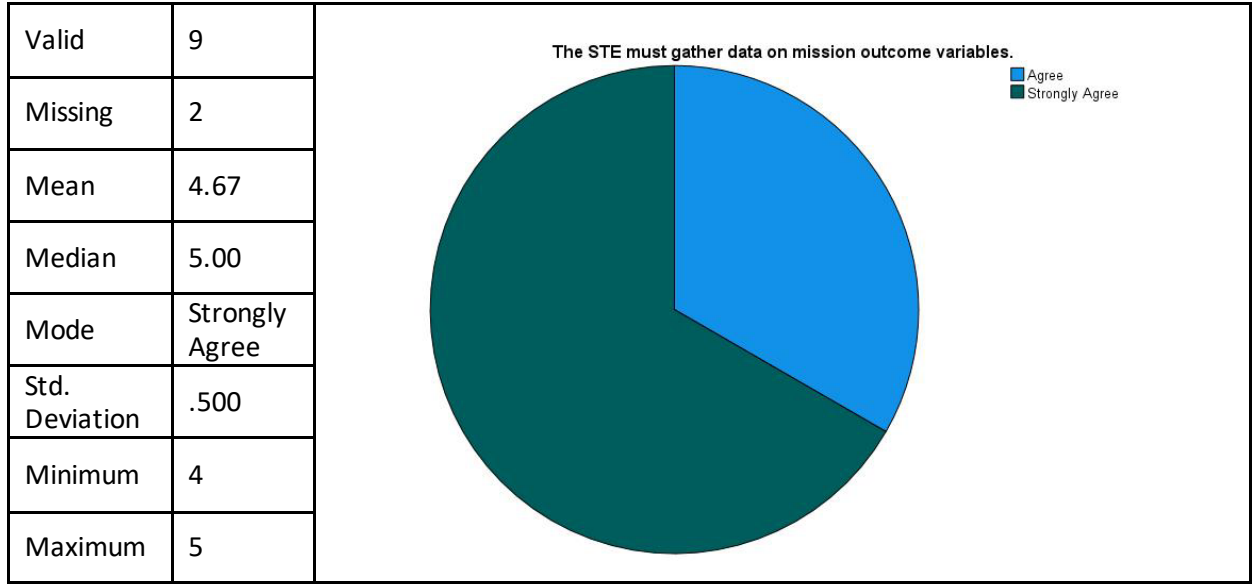

| | | Frequency | Percent | Valid Percent | Cumulative Percent |
|---|---|---|---|---|---|
| Valid | Strongly Disagree | | | | |
| | Disagree | | | | |
| | Neither Agree nor Disagree | | | | |
| | Agree | 3 | 27.3 | 33.3 | 33.3 |
| | Strongly Agree | 6 | 54.5 | 66.7 | 100.0 |
| | Total | 9 | 81.8 | 100.0 | |

The STE must gather data on task outcome variables.

| | |
|---|---|
| Valid | 9 |
| Missing | 2 |
| Mean | 4.78 |
| Median | 5.00 |
| Mode | Strongly Agree |
| Std. Deviation | .441 |
| Minimum | 4 |
| Maximum | 5 |

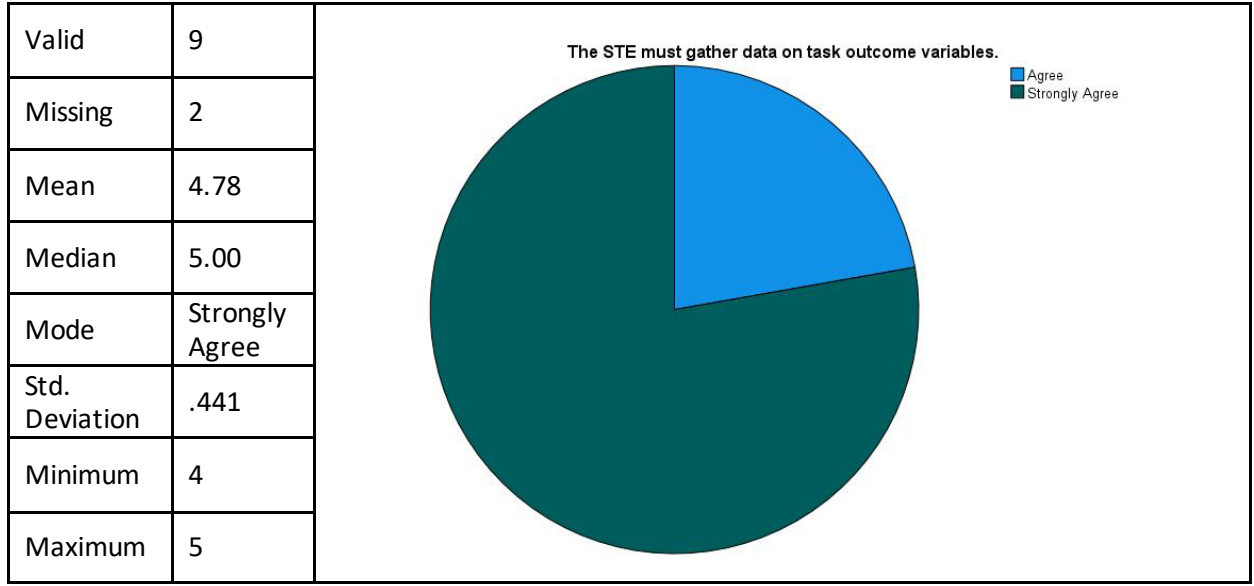

| | | Frequency | Percent | Valid Percent | Cumulative Percent |
|---|---|---|---|---|---|
| Valid | Strongly Disagree | | | | |
| | Disagree | | | | |
| | Neither Agree nor Disagree | | | | |
| | Agree | 2 | 18.2 | 22.2 | 22.2 |
| | Strongly Agree | 7 | 63.6 | 77.8 | 100.0 |
| | Total | 9 | 81.8 | 100.0 | |

The STE must gather data on task process variables.

| | |
|---|---|
| Valid | 9 |
| Missing | 2 |
| Mean | 4.56 |
| Median | 5.00 |
| Mode | Strongly Agree |
| Std. Deviation | .527 |
| Minimum | 4 |
| Maximum | 5 |

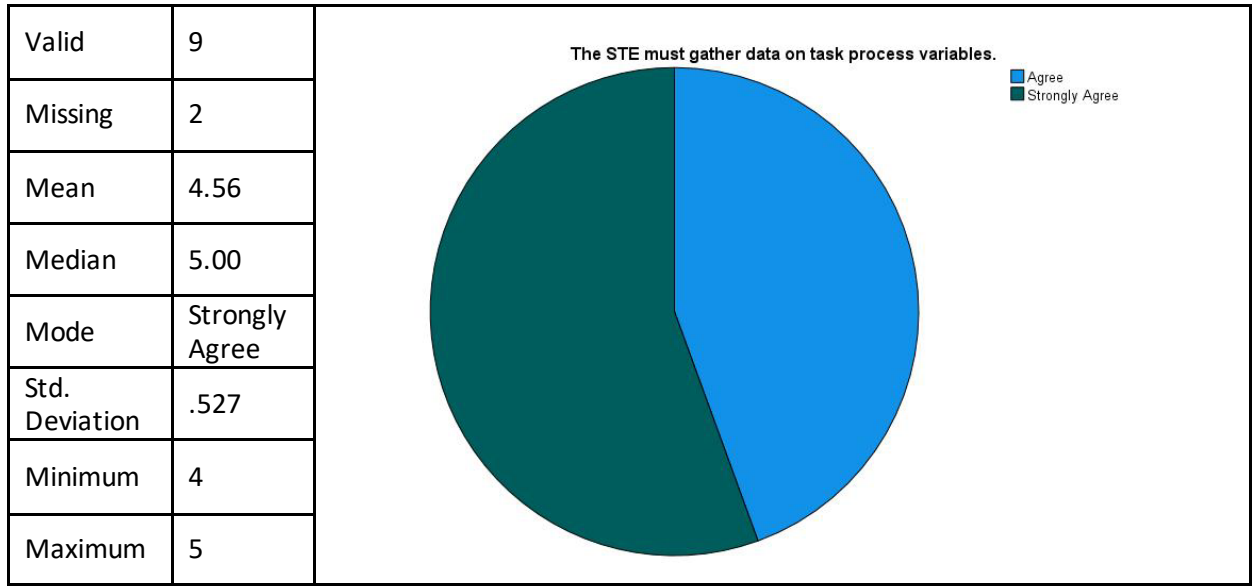

| | | Frequency | Percent | Valid Percent | Cumulative Percent |
|---|---|---|---|---|---|
| Valid | Strongly Disagree | | | | |
| | Disagree | | | | |
| | Neither Agree nor Disagree | | | | |
| | Agree | 4 | 36.4 | 44.4 | 44.4 |
| | Strongly Agree | 5 | 45.5 | 55.6 | 100.0 |
| | Total | 9 | 81.8 | 100.0 | |

The STE must gather data on team coordination process variables.

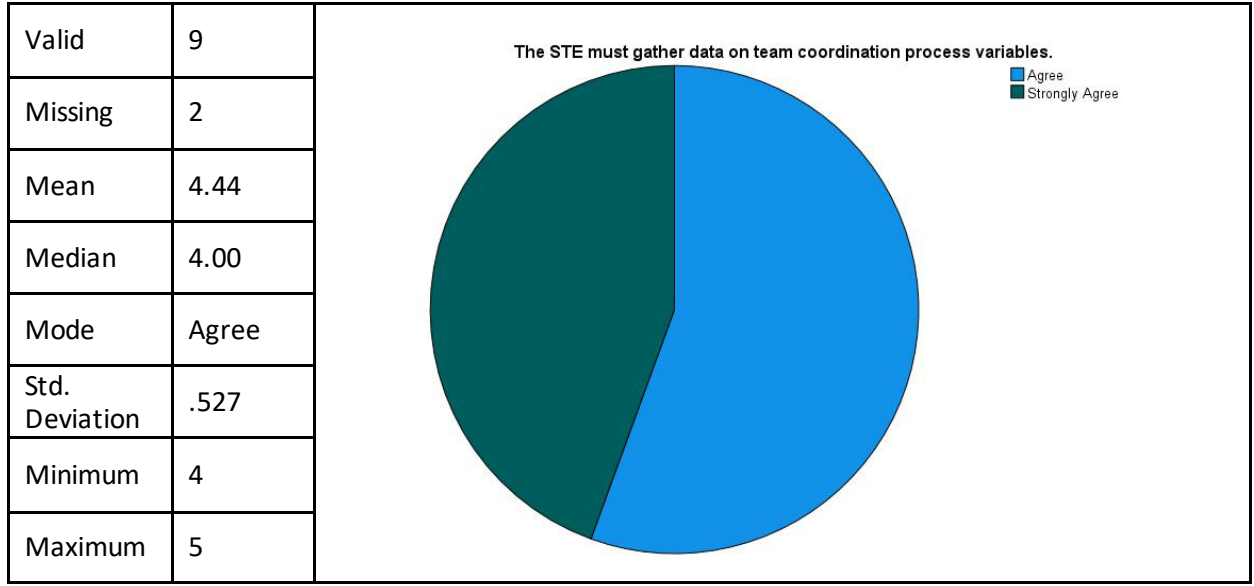

| | |
|---|---|
| Valid | 9 |
| Missing | 2 |
| Mean | 4.44 |
| Median | 4.00 |
| Mode | Agree |
| Std. Deviation | .527 |
| Minimum | 4 |
| Maximum | 5 |

| | | Frequency | Percent | Valid Percent | Cumulative Percent |
|---|---|---|---|---|---|
| Valid | Strongly Disagree | | | | |
| | Disagree | | | | |
| | Neither Agree nor Disagree | | | | |
| | Agree | 5 | 45.5 | 55.6 | 55.6 |
| | Strongly Agree | 4 | 36.4 | 44.4 | 100.0 |
| | Total | 9 | 81.8 | 100.0 | |

The STE must gather data on team coordination knowledge variables.

| | |
|---|---|
| Valid | 9 |
| Missing | 2 |
| Mean | 3.89 |
| Median | 4.00 |
| Mode | Agree |
| Std. Deviation | .782 |
| Minimum | 3 |
| Maximum | 5 |

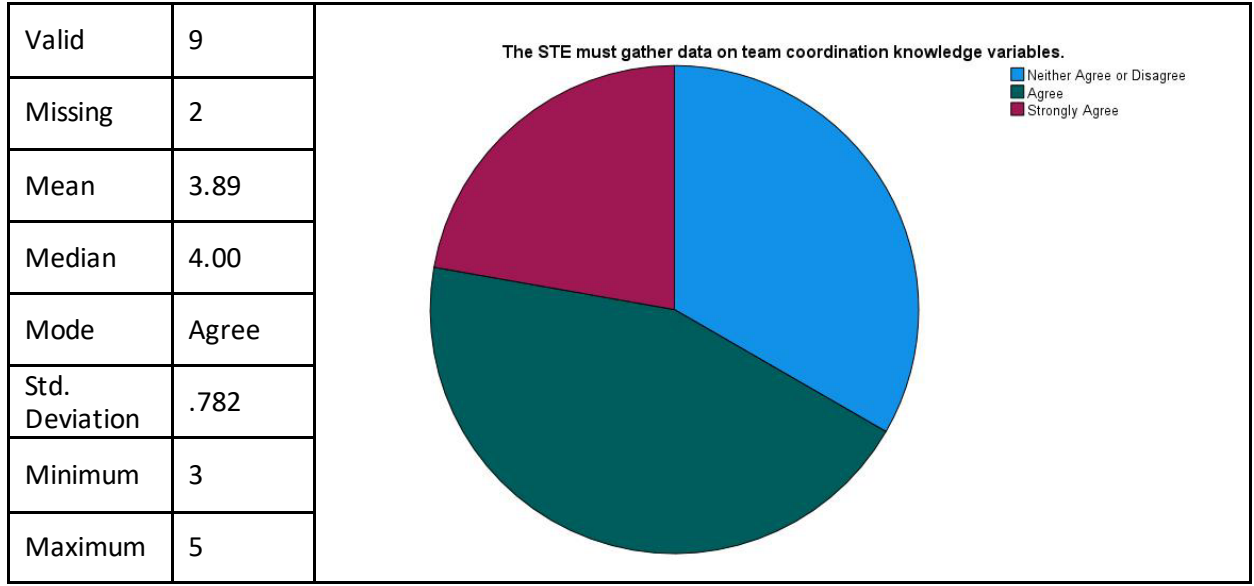

| | | Frequency | Percent | Valid Percent | Cumulative Percent |
|---|---|---|---|---|---|
| Valid | Strongly Disagree | | | | |
| | Disagree | | | | |
| | Neither Agree nor Disagree | 3 | 27.3 | 33.3 | 33.3 |
| | Agree | 4 | 36.4 | 44.4 | 77.8 |
| | Strongly Agree | 2 | 18.2 | 22.2 | 100.0 |
| | Total | 9 | 81.8 | 100.0 | |

The STE must include the ability for researchers to define performance assessment variables using simulation "primitives" (e.g., windows opening/closing, time passing, values entered).

| | |
|---|---|
| Valid | 9 |
| Missing | 2 |
| Mean | 4.33 |
| Median | 4.00 |
| Mode | Agree[2] |
| Std. Deviation | .707 |
| Minimum | 3 |
| Maximum | 5 |

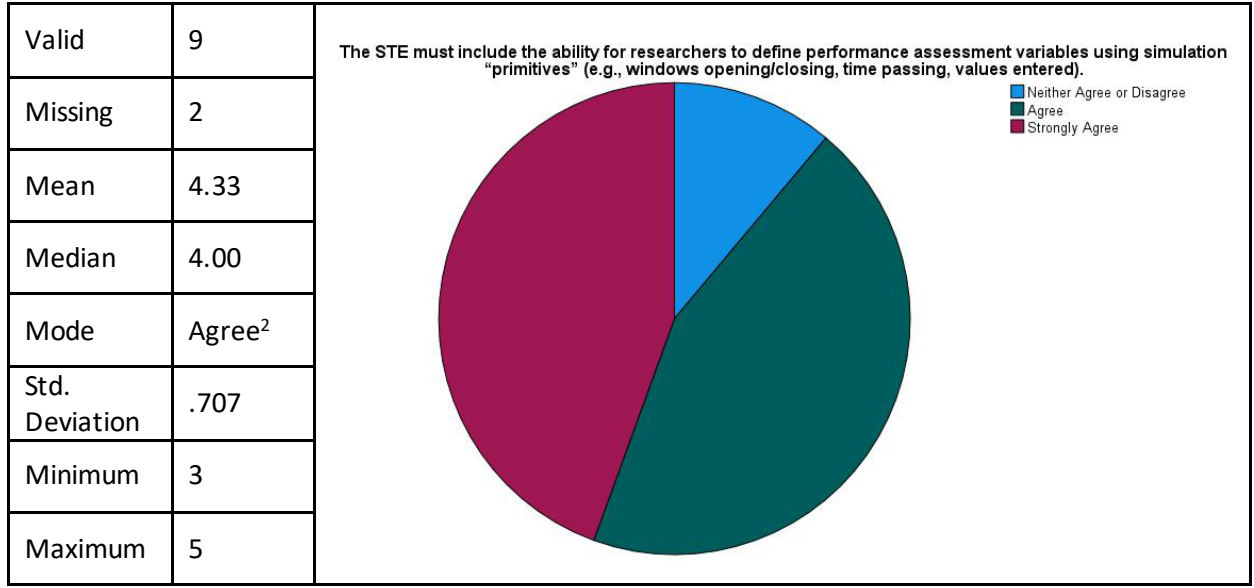

| | | Frequency | Percent | Valid Percent | Cumulative Percent |
|---|---|---|---|---|---|
| Valid | Strongly Disagree | | | | |
| | Disagree | | | | |
| | Neither Agree nor Disagree | 1 | 9.1 | 11.1 | 11.1 |
| | Agree | 4 | 36.4 | 44.4 | 55.6 |
| | Strongly Agree | 4 | 36.4 | 44.4 | 100.0 |
| | Total | 9 | 81.8 | 100.0 | |

The STE must include support for "freeze probe" assessments of situation awareness.

| | |
|---|---|
| Valid | 9 |
| Missing | 2 |
| Mean | 3.56 |
| Median | 3.00 |
| Mode | Neither Agree or Disagree |
| Std. Deviation | .882 |
| Minimum | 3 |
| Maximum | 5 |

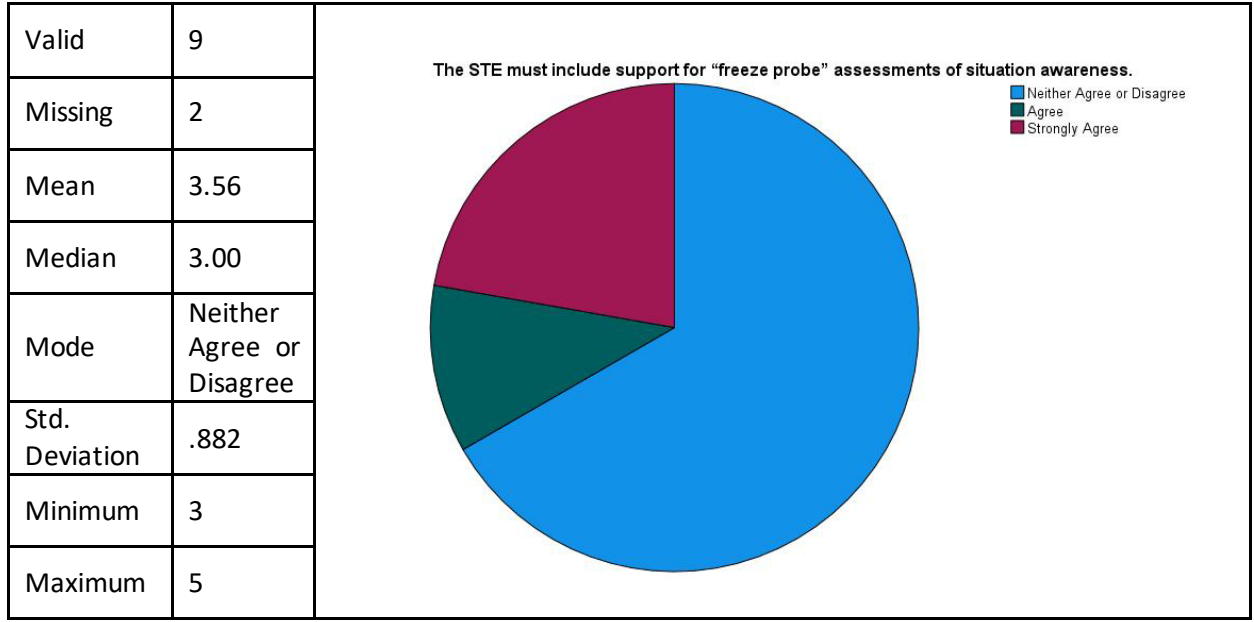

| | | Frequency | Percent | Valid Percent | Cumulative Percent |
|---|---|---|---|---|---|
| Valid | Strongly Disagree | | | | |
| | Disagree | | | | |
| | Neither Agree nor Disagree | 6 | 54.5 | 66.7 | 66.7 |
| | Agree | 1 | 9.1 | 11.1 | 77.8 |
| | Strongly Agree | 2 | 18.2 | 22.2 | 100.0 |
| | Total | 9 | 81.8 | 100.0 | |

The STE must include support for author-able surveys to assess issues such as trust; perceived team cohesion, potency, interdisciplinary, etc. and task difficulty.

| Valid | 9 |
|---|---|
| Missing | 2 |
| Mean | 4.00 |
| Median | 4.00 |
| Mode | Agree |
| Std. Deviation | .707 |
| Minimum | 3 |
| Maximum | 5 |

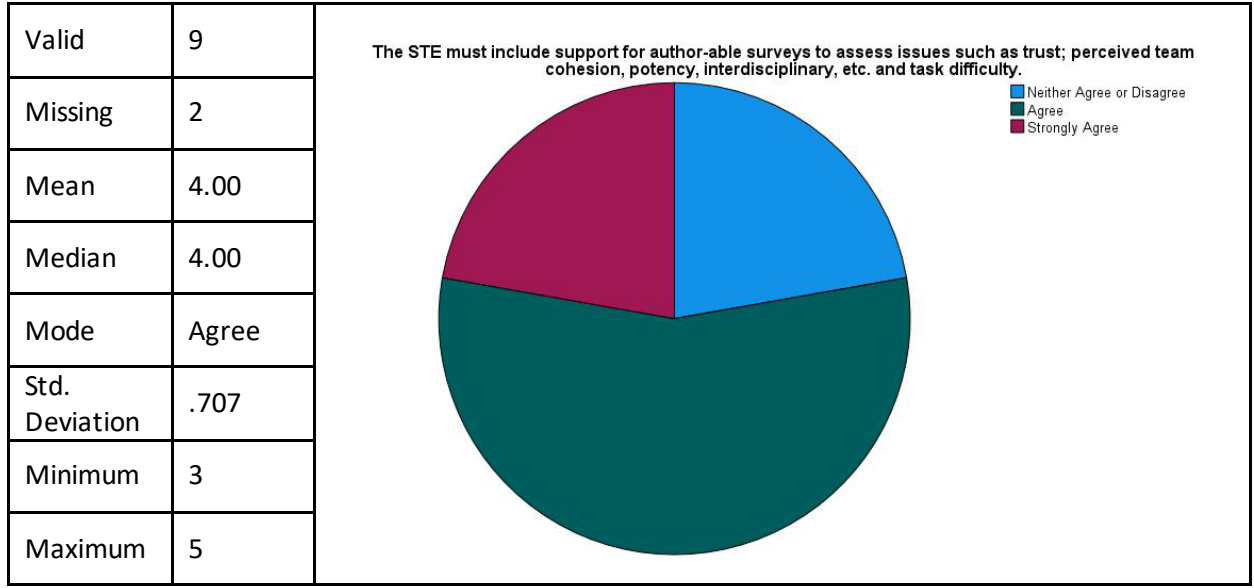

|  |  | Frequency | Percent | Valid Percent | Cumulative Percent |
|---|---|---|---|---|---|
| Valid | Strongly Disagree |  |  |  |  |
|  | Disagree |  |  |  |  |
|  | Neither Agree nor Disagree | 2 | 18.2 | 22.2 | 22.2 |
|  | Agree | 5 | 45.5 | 55.6 | 77.8 |
|  | Strongly Agree | 2 | 18.2 | 22.2 | 100.0 |
|  | Total | 9 | 81.8 | 100.0 |  |

The STE interfaces must be broadly "instrumented" to support action capture and performance assessment.

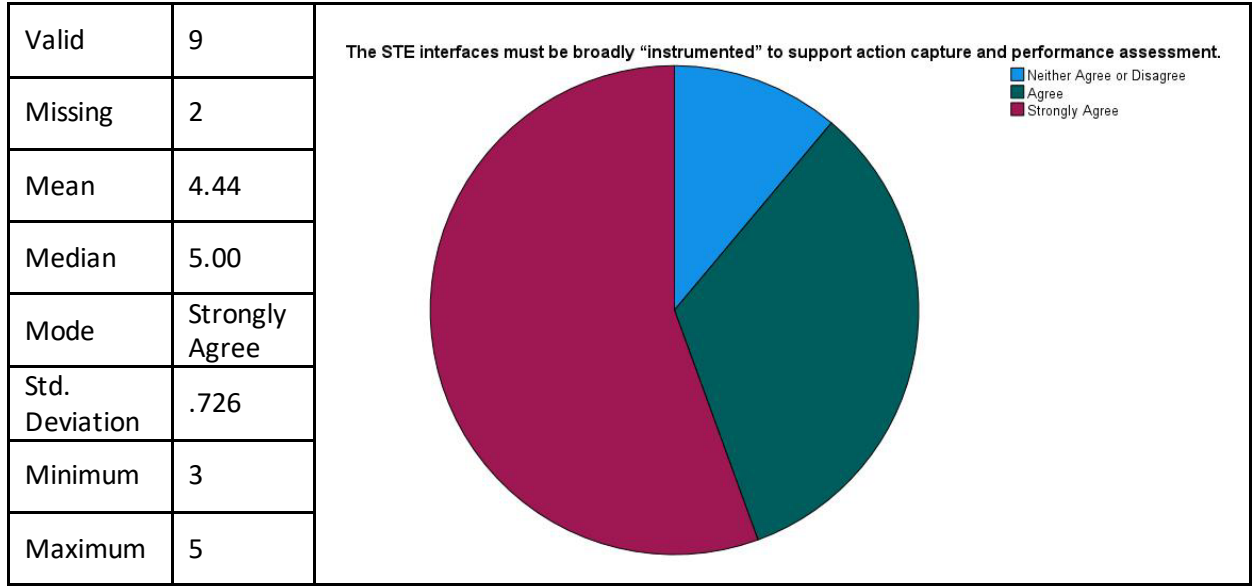

| | |
|---|---|
| Valid | 9 |
| Missing | 2 |
| Mean | 4.44 |
| Median | 5.00 |
| Mode | Strongly Agree |
| Std. Deviation | .726 |
| Minimum | 3 |
| Maximum | 5 |

| | | Frequency | Percent | Valid Percent | Cumulative Percent |
|---|---|---|---|---|---|
| Valid | Strongly Disagree | | | | |
| | Disagree | | | | |
| | Neither Agree nor Disagree | 1 | 9.1 | 11.1 | 11.1 |
| | Agree | 3 | 27.3 | 33.3 | 44.4 |
| | Strongly Agree | 5 | 45.5 | 55.6 | 100.0 |
| | Total | 9 | 81.8 | 100.0 | |

The STE must support the capture of communication activities among teammates.

| | |
|---|---|
| Valid | 9 |
| Missing | 2 |
| Mean | 4.78 |
| Median | 5.00 |
| Mode | Strongly Agree |
| Std. Deviation | .441 |
| Minimum | 4 |
| Maximum | 5 |

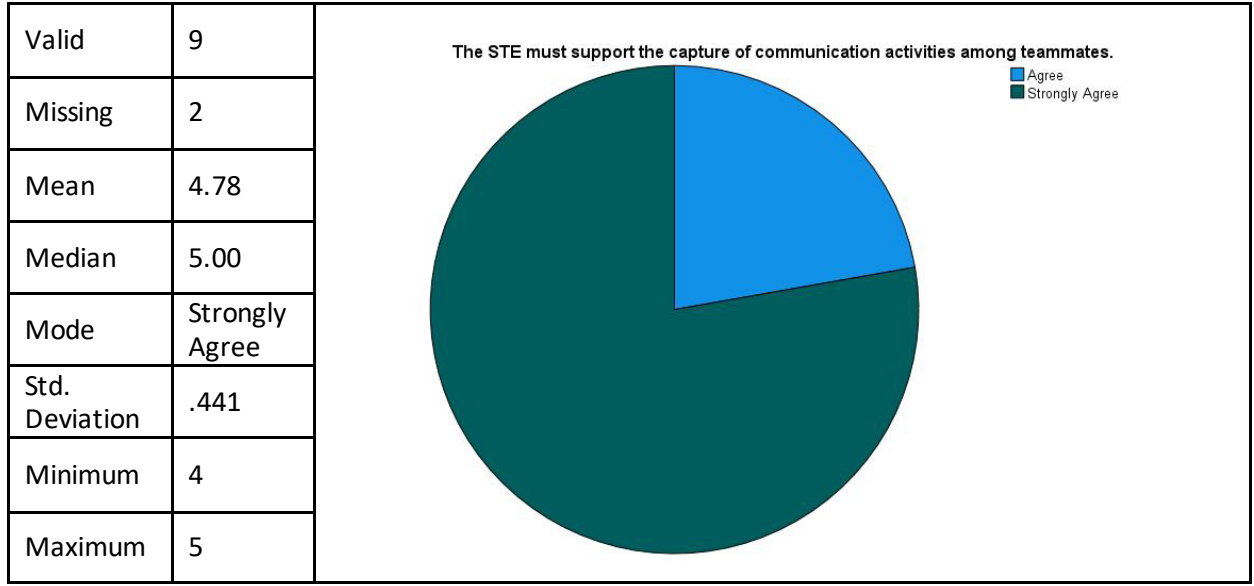

| | | Frequency | Percent | Valid Percent | Cumulative Percent |
|---|---|---|---|---|---|
| Valid | Strongly Disagree | | | | |
| | Disagree | | | | |
| | Neither Agree nor Disagree | | | | |
| | Agree | 2 | 18.2 | 22.2 | 22.2 |
| | Strongly Agree | 7 | 63.6 | 77.8 | 100.0 |
| | Total | 9 | 81.8 | 100.0 | |

This Page Intentionally Blank

*Appendix D*

**Results for Data Analysis and Visualization Likert Items**

The STE should provide descriptive stats (means, mode, median, range, standard deviation) for each variable.

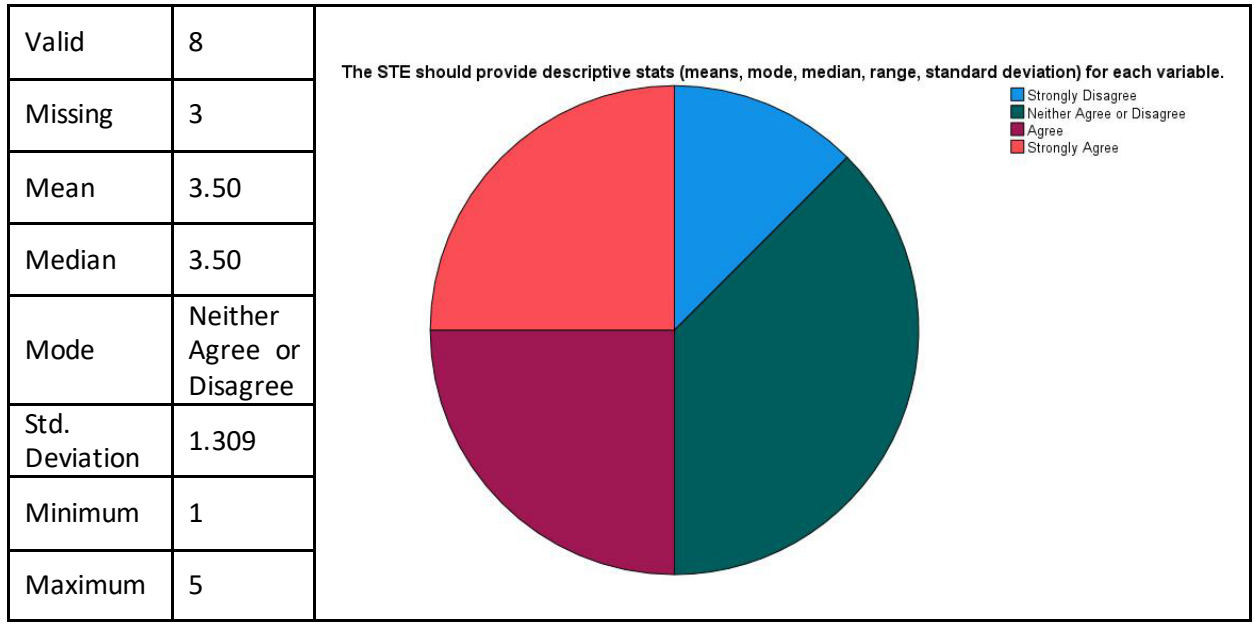

| Valid | 8 |
|---|---|
| Missing | 3 |
| Mean | 3.50 |
| Median | 3.50 |
| Mode | Neither Agree or Disagree |
| Std. Deviation | 1.309 |
| Minimum | 1 |
| Maximum | 5 |

|  |  | Frequency | Percent | Valid Percent | Cumulative Percent |
|---|---|---|---|---|---|
| Valid | Strongly Disagree | 1 | 9.1 | 12.5 | 12.5 |
|  | Disagree | 0 | 0 | 0 | 12.5 |
|  | Neither Agree nor Disagree | 3 | 27.3 | 37.5 | 50.0 |
|  | Agree | 2 | 18.2 | 25.0 | 75.0 |
|  | Strongly Agree | 2 | 18.2 | 25.0 | 100.0 |
|  | Total | 8 | 72.7 | 100.0 |  |

The STE should provide scatterplots that illustrate the association between pairs of variables.

| Valid | 8 |
|---|---|
| Missing | 3 |
| Mean | 3.38 |
| Median | 3.50 |
| Mode | Neither Agree or Disagree[2] |
| Std. Deviation | 1.188 |
| Minimum | 1 |
| Maximum | 5 |

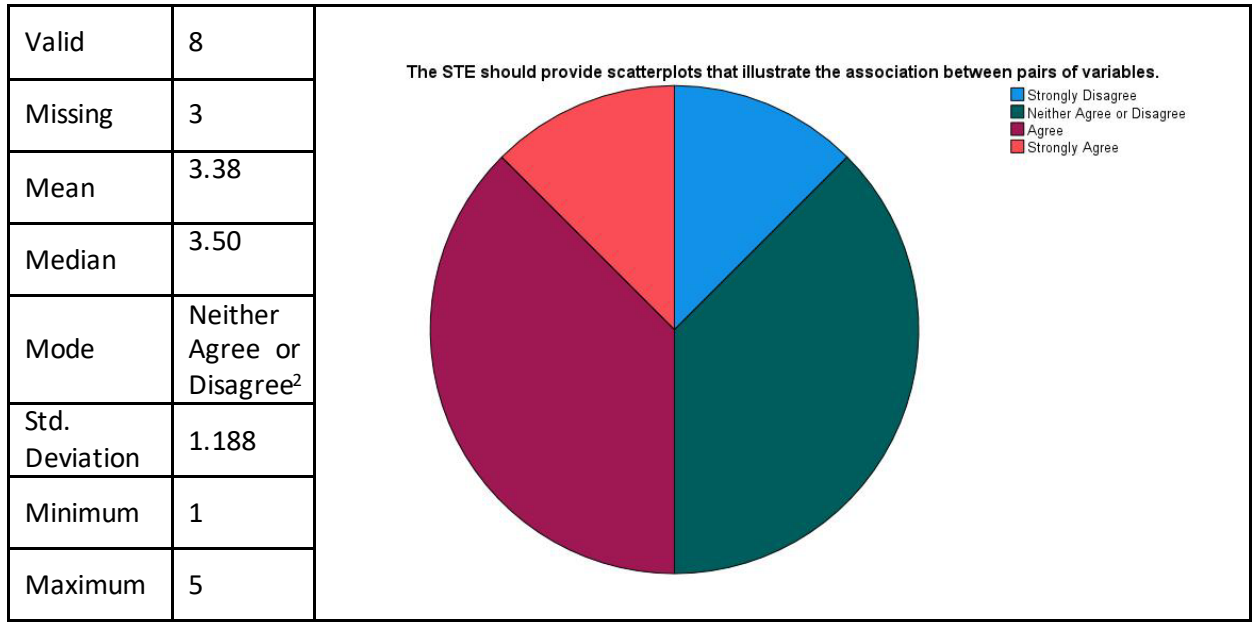

|  |  | Frequency | Percent | Valid Percent | Cumulative Percent |
|---|---|---|---|---|---|
| Valid | Strongly Disagree | 1 | 9.1 | 12.5 | 12.5 |
|  | Disagree | 0 | 0 | 0 | 12.5 |
|  | Neither Agree nor Disagree | 3 | 27.3 | 37.5 | 50.0 |
|  | Agree | 3 | 27.3 | 37.5 | 87.5 |
|  | Strongly Agree | 1 | 9.1 | 12.5 | 100.0 |
|  | Total | 8 | 72.7 | 100.0 |  |

The STE should provide bar charts of the means of variables.

| Valid | 8 |
|---|---|
| Missing | 3 |
| Mean | 3.38 |
| Median | 3.50 |
| Mode | Neither Agree or Disagree[2] |
| Std. Deviation | 1.188 |
| Minimum | 1 |
| Maximum | 5 |

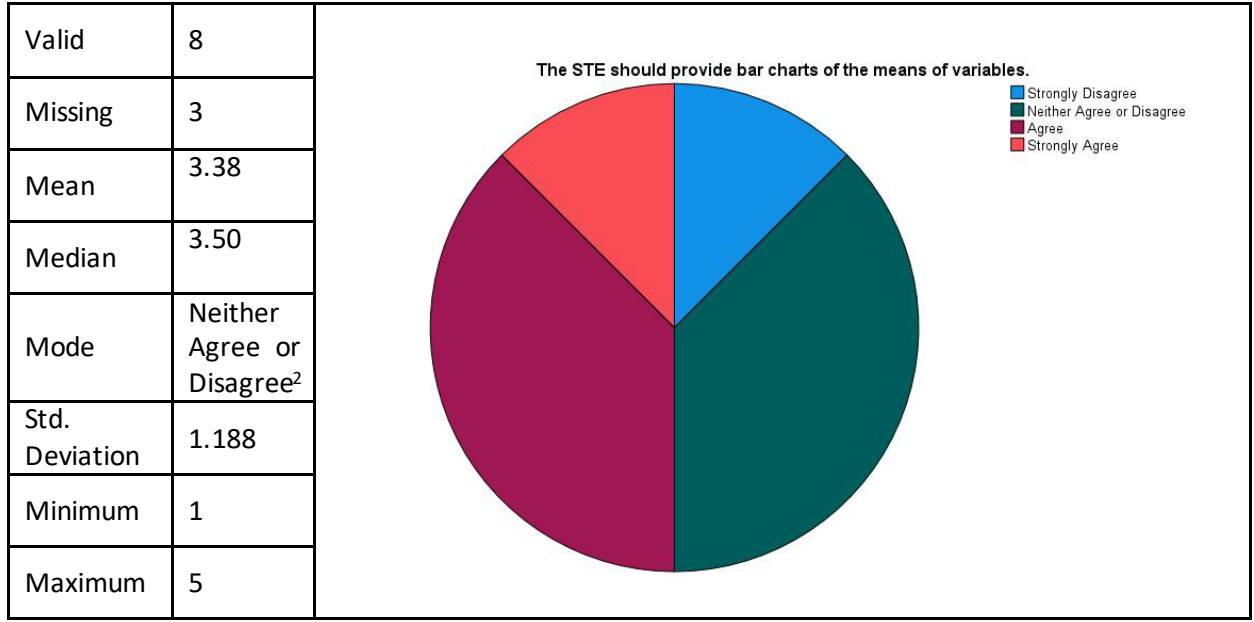

|  |  | Frequency | Percent | Valid Percent | Cumulative Percent |
|---|---|---|---|---|---|
| Valid | Strongly Disagree | 1 | 9.1 | 12.5 | 12.5 |
|  | Disagree | 0 | 0 | 0 | 12.5 |
|  | Neither Agree nor Disagree | 3 | 27.3 | 37.5 | 50.0 |
|  | Agree | 3 | 27.3 | 37.5 | 87.5 |
|  | Strongly Agree | 1 | 9.1 | 12.5 | 100.0 |
|  | Total | 8 | 72.7 | 100.0 |  |

The STE should provide histograms (bar charts of frequency/counts) for variables.

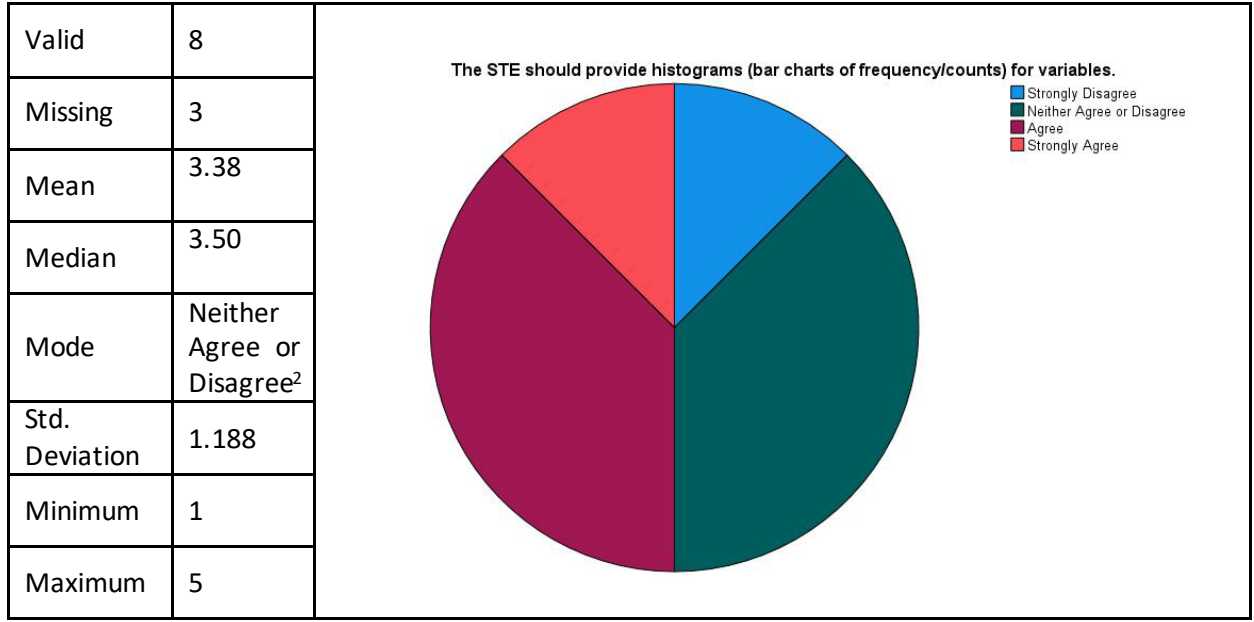

| Valid | 8 |
| --- | --- |
| Missing | 3 |
| Mean | 3.38 |
| Median | 3.50 |
| Mode | Neither Agree or Disagree[2] |
| Std. Deviation | 1.188 |
| Minimum | 1 |
| Maximum | 5 |

|  |  | Frequency | Percent | Valid Percent | Cumulative Percent |
| --- | --- | --- | --- | --- | --- |
| Valid | Strongly Disagree | 1 | 9.1 | 12.5 | 12.5 |
|  | Disagree | 0 | 0 | 0 | 12.5 |
|  | Neither Agree nor Disagree | 3 | 27.3 | 37.5 | 50.0 |
|  | Agree | 3 | 27.3 | 37.5 | 87.5 |
|  | Strongly Agree | 1 | 9.1 | 12.5 | 100.0 |
|  | Total | 8 | 72.7 | 100.0 |  |

The STE should enable t-test analysis.

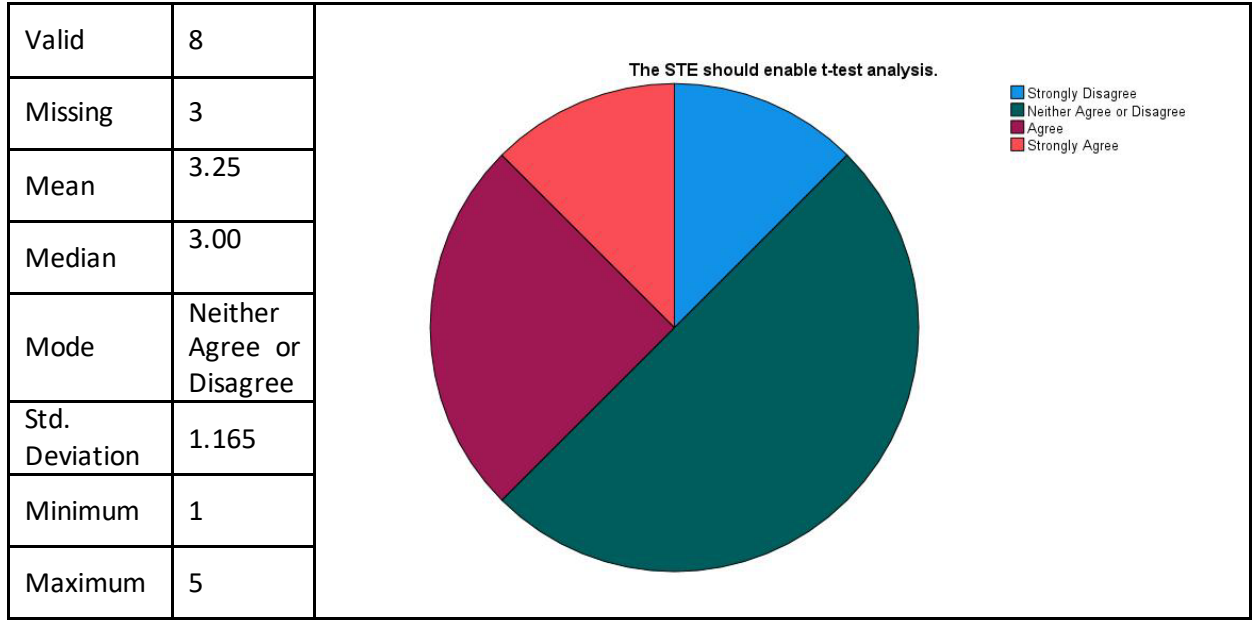

| Valid | 8 |
|---|---|
| Missing | 3 |
| Mean | 3.25 |
| Median | 3.00 |
| Mode | Neither Agree or Disagree |
| Std. Deviation | 1.165 |
| Minimum | 1 |
| Maximum | 5 |

|  |  | Frequency | Percent | Valid Percent | Cumulative Percent |
|---|---|---|---|---|---|
| Valid | Strongly Disagree | 1 | 9.1 | 12.5 | 12.5 |
|  | Disagree | 0 | 0 | 0 | 12.5 |
|  | Neither Agree nor Disagree | 4 | 36.4 | 50.0 | 62.5 |
|  | Agree | 2 | 18.2 | 25.0 | 87.5 |
|  | Strongly Agree | 1 | 9.1 | 12.5 | 100.0 |
|  | Total | 8 | 72.7 | 100.0 |  |

The STE should enable one-way ANOVA analysis.

| Valid | 8 |
|---|---|
| Missing | 3 |
| Mean | 3.13 |
| Median | 3.00 |
| Mode | Neither Agree or Disagree |
| Std. Deviation | 1.126 |
| Minimum | 1 |
| Maximum | 5 |

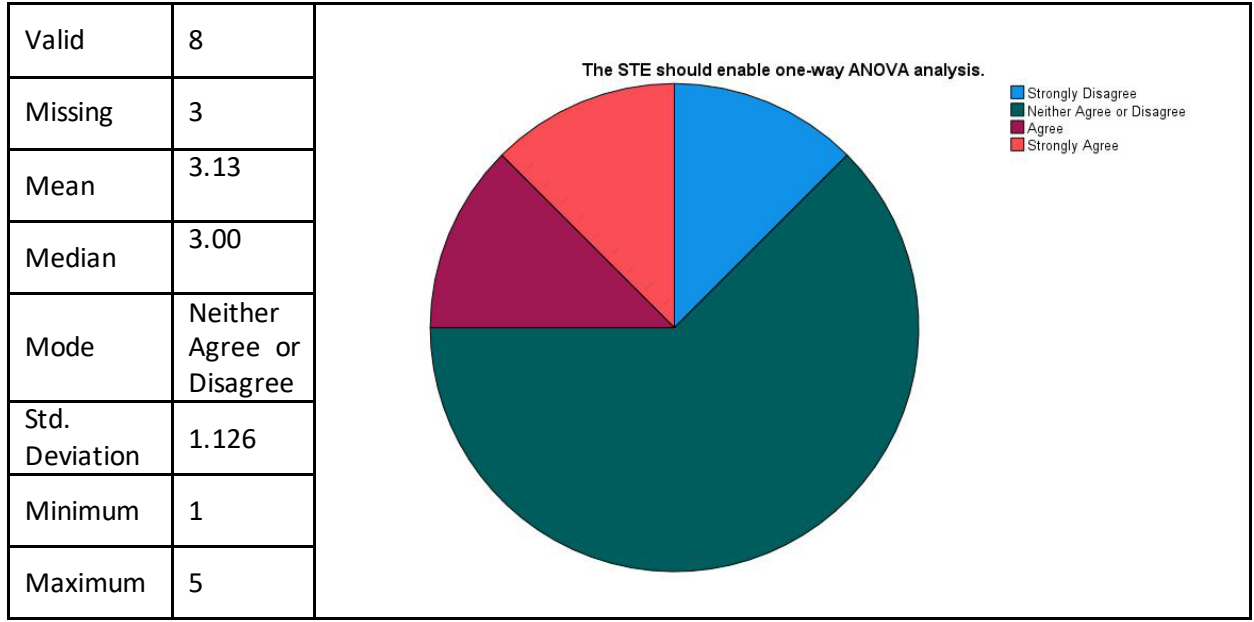

|  |  | Frequency | Percent | Valid Percent | Cumulative Percent |
|---|---|---|---|---|---|
| Valid | Strongly Disagree | 1 | 9.1 | 12.5 | 12.5 |
|  | Disagree | 0 | 0 | 0 | 12.5 |
|  | Neither Agree nor Disagree | 5 | 45.5 | 62.5 | 75.0 |
|  | Agree | 1 | 9.1 | 12.5 | 87.5 |
|  | Strongly Agree | 1 | 9.1 | 12.5 | 100.0 |
|  | Total | 8 | 72.7 | 100.0 |  |

The STE should enable regression analysis.

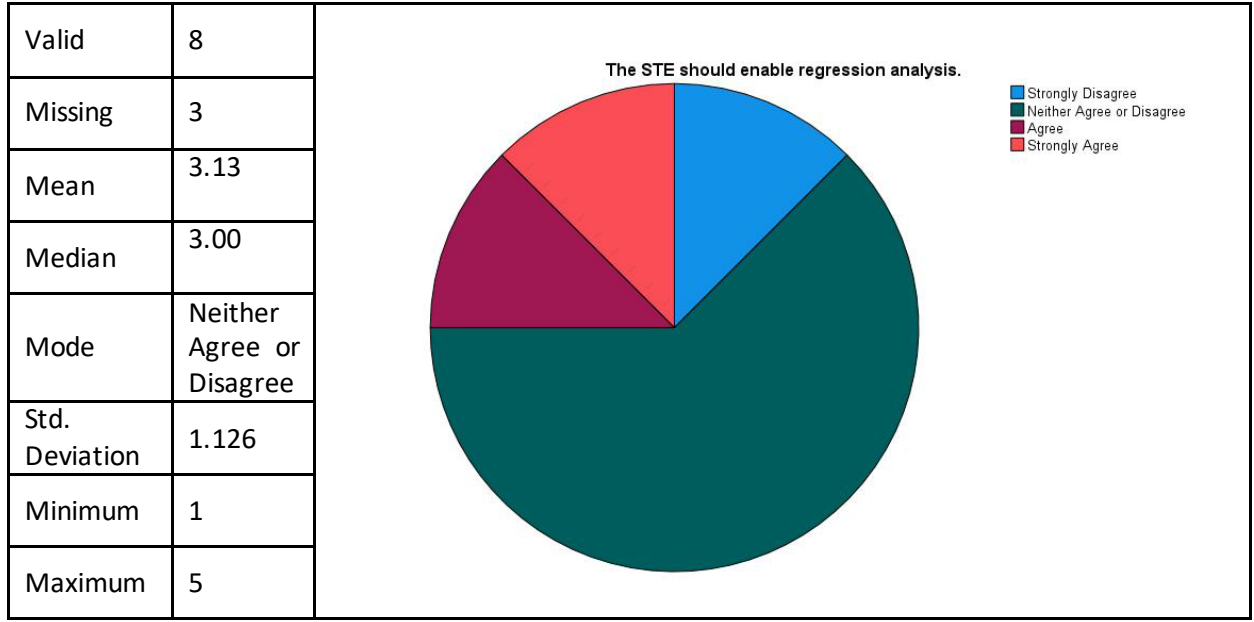

| | |
|---|---|
| Valid | 8 |
| Missing | 3 |
| Mean | 3.13 |
| Median | 3.00 |
| Mode | Neither Agree or Disagree |
| Std. Deviation | 1.126 |
| Minimum | 1 |
| Maximum | 5 |

| | | Frequency | Percent | Valid Percent | Cumulative Percent |
|---|---|---|---|---|---|
| Valid | Strongly Disagree | 1 | 9.1 | 12.5 | 12.5 |
| | Disagree | 0 | 0 | 0 | 12.5 |
| | Neither Agree nor Disagree | 5 | 45.5 | 62.5 | 75.0 |
| | Agree | 1 | 9.1 | 12.5 | 87.5 |
| | Strongly Agree | 1 | 9.1 | 12.5 | 100.0 |
| | Total | 8 | 72.7 | 100.0 | |

The STE should allow researchers to export identified variables to comma separated value files.

| | |
|---|---|
| Valid | 8 |
| Missing | 3 |
| Mean | 4.50 |
| Median | 5.00 |
| Mode | Strongly Agree |
| Std. Deviation | .756 |
| Minimum | 3 |
| Maximum | 5 |

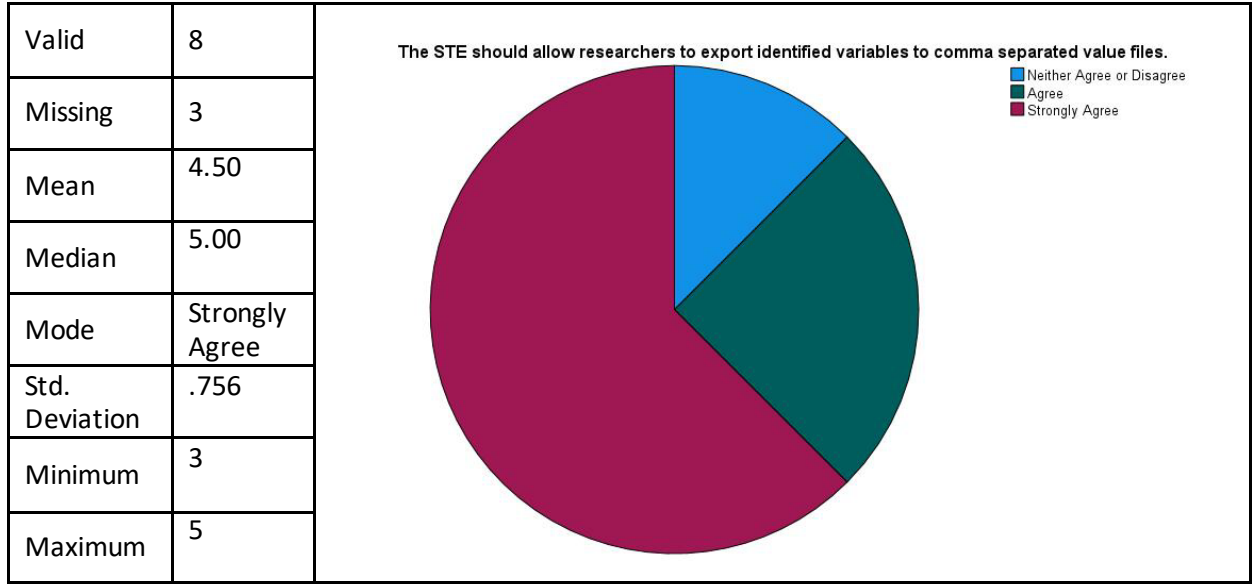

| | | Frequency | Percent | Valid Percent | Cumulative Percent |
|---|---|---|---|---|---|
| Valid | Strongly Disagree | | | | |
| | Disagree | | | | |
| | Neither Agree nor Disagree | 1 | 9.1 | 12.5 | 12.5 |
| | Agree | 2 | 18.2 | 25.0 | 37.5 |
| | Strongly Agree | 5 | 45.5 | 62.5 | 100.0 |
| | Total | 8 | 72.7 | 100.0 | |

The STE should allow researchers to export identified variables to Excel files.

| | |
|---|---|
| Valid | 8 |
| Missing | 3 |
| Mean | 4.00 |
| Median | 4.00 |
| Mode | Neither Agree or Disagree[2] |
| Std. Deviation | .926 |
| Minimum | 3 |
| Maximum | 5 |

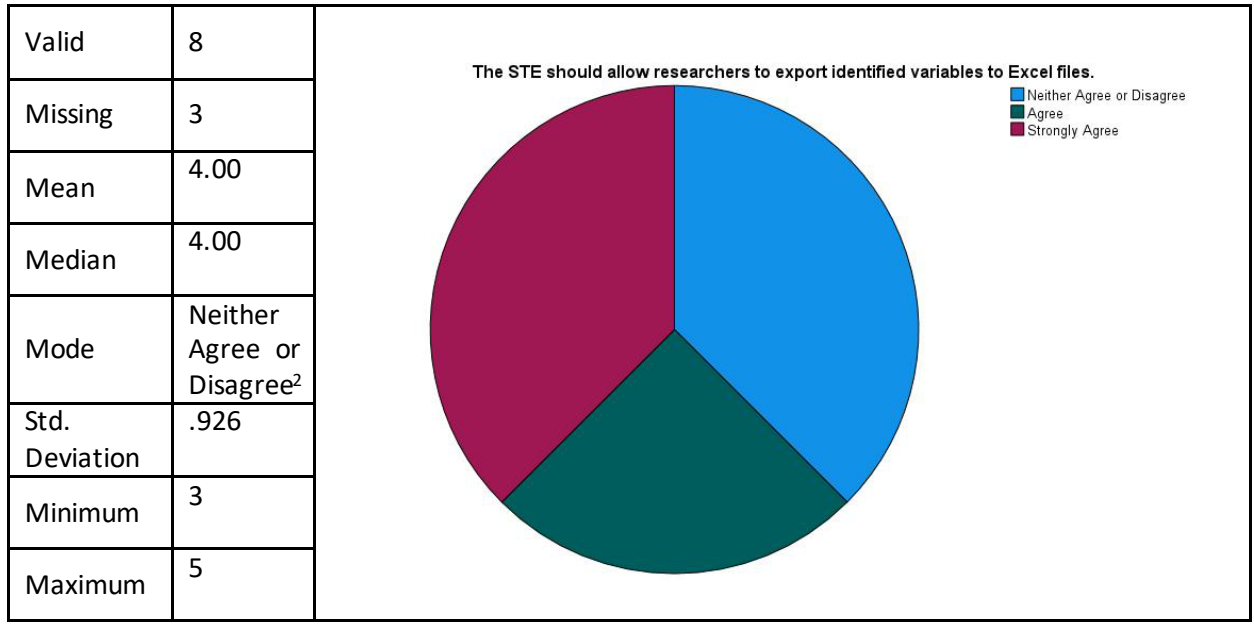

| | | Frequency | Percent | Valid Percent | Cumulative Percent |
|---|---|---|---|---|---|
| Valid | Strongly Disagree | | | | |
| | Disagree | | | | |
| | Neither Agree nor Disagree | 3 | 27.3 | 37.5 | 37.5 |
| | Agree | 2 | 18.2 | 25.0 | 62.5 |
| | Strongly Agree | 3 | 27.3 | 37.5 | 100.0 |
| | Total | 8 | 72.7 | 100.0 | |

The STE should allow researchers to export identified variables to database files.

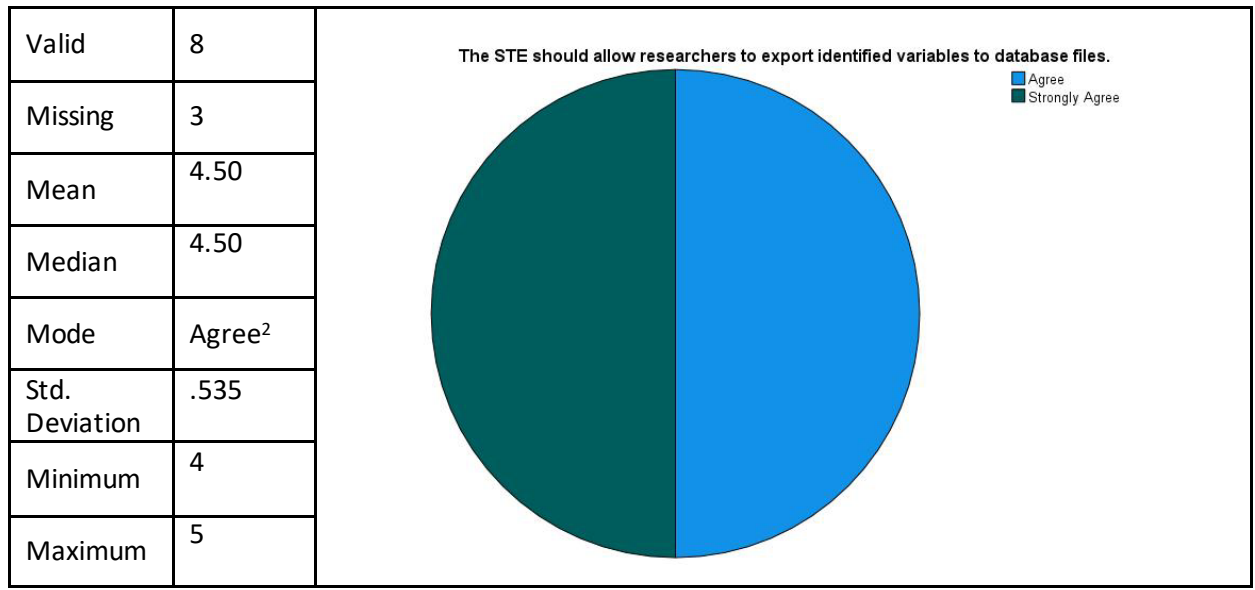

| Valid | 8 |
|---|---|
| Missing | 3 |
| Mean | 4.50 |
| Median | 4.50 |
| Mode | Agree[2] |
| Std. Deviation | .535 |
| Minimum | 4 |
| Maximum | 5 |

|  |  | Frequency | Percent | Valid Percent | Cumulative Percent |
|---|---|---|---|---|---|
| Valid | Strongly Disagree |  |  |  |  |
|  | Disagree |  |  |  |  |
|  | Neither Agree nor Disagree |  |  |  |  |
|  | Agree | 4 | 36.4 | 50.0 | 50.0 |
|  | Strongly Agree | 4 | 36.4 | 50.0 | 100.0 |
|  | Total | 8 | 72.7 | 100.0 |  |

This Page Intentionally Blank

*Appendix E*

*Results for Domain Feature Likert Items*

The task must be isomorphic with military operations.

| Valid | 11 |
|---|---|
| Missing | 0 |
| Mean | 3.27 |
| Median | 3.00 |
| Mode | Neither Agree or Disagree[2] |
| Std. Deviation | 1.272 |
| Minimum | 1 |
| Maximum | 5 |

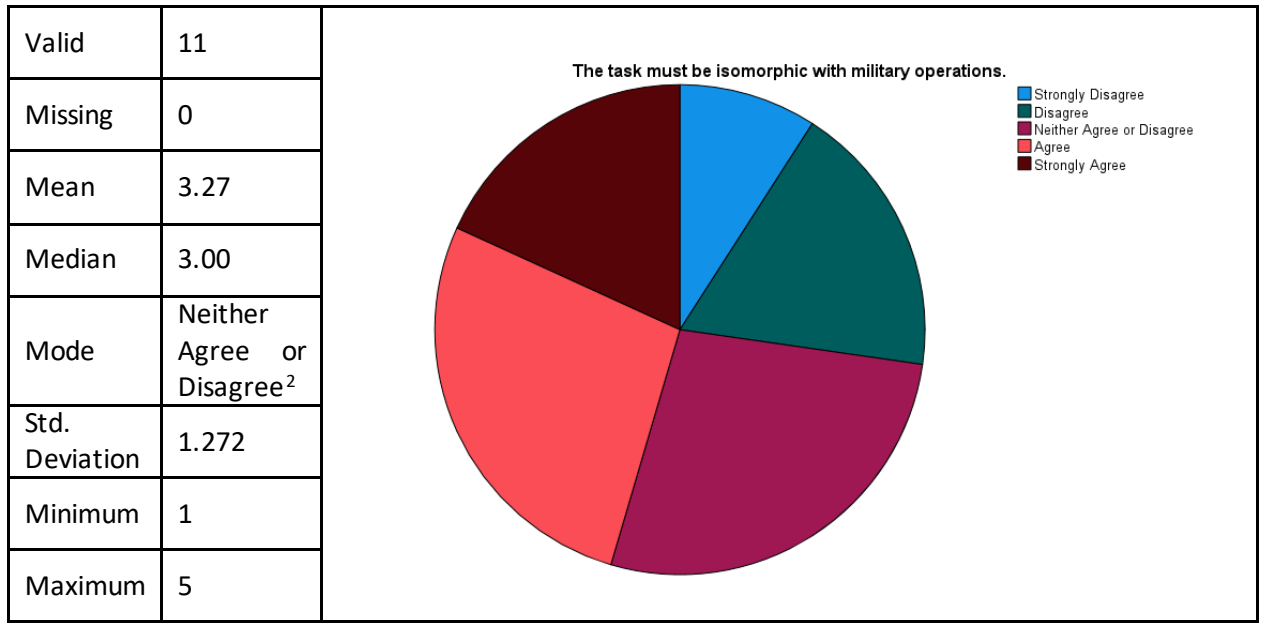

| | | Frequency | Percent | Valid Percent | Cumulative Percent |
|---|---|---|---|---|---|
| Valid | Strongly Disagree | 1 | 9.1 | 9.1 | 9.1 |
| | Disagree | 2 | 18.2 | 18.2 | 27.3 |
| | Neither Agree nor Disagree | 3 | 27.3 | 27.3 | 54.5 |
| | Agree | 3 | 27.3 | 27.3 | 81.8 |
| | Strongly Agree | 2 | 18.2 | 18.2 | 100.0 |
| | Total | 11 | 100.0 | 100.0 | |

---

[2] Multiple modes exist. The smallest value is shown.

The task environment must require the same type of data gathering, decision-making, and action-taking that characterizes any military operation that employs the Observe-Orient-Decide-Act (OODA) loop.

| Valid | 11 |
|---|---|
| Missing | 0 |
| Mean | 3.91 |
| Median | 4.00 |
| Mode | Neither Agree or Disagree[2] |
| Std. Deviation | .831 |
| Minimum | 3 |
| Maximum | 5 |

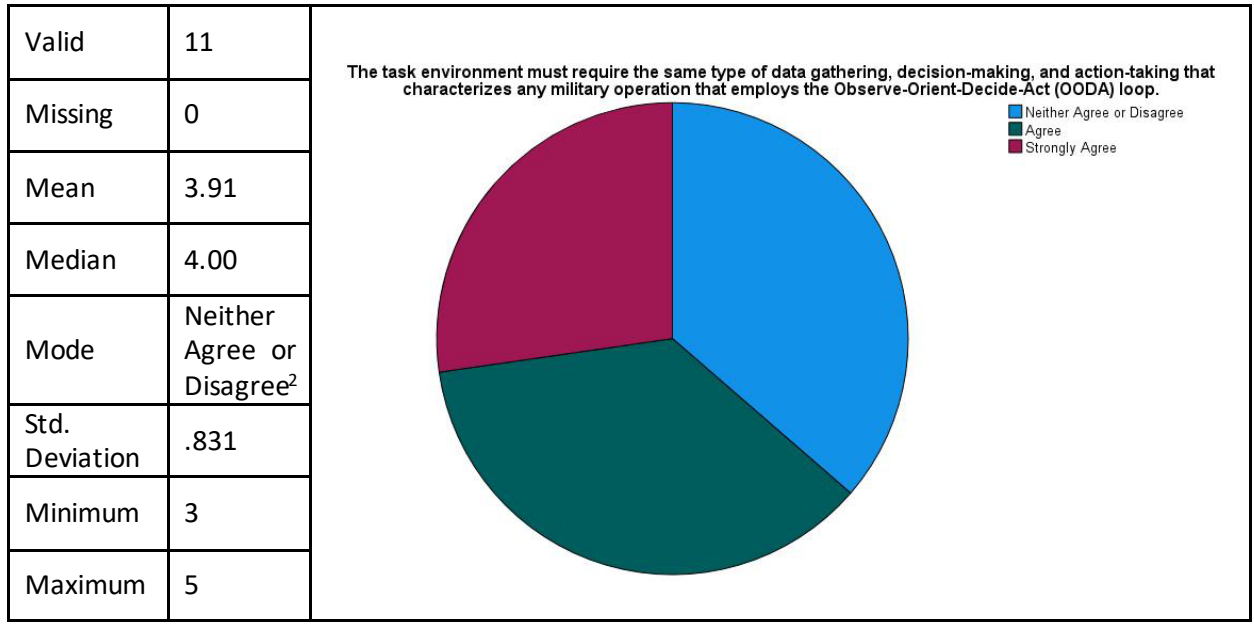

|  |  | Frequency | Percent | Valid Percent | Cumulative Percent |
|---|---|---|---|---|---|
| Valid | Strongly Disagree |  |  |  |  |
|  | Disagree |  |  |  |  |
|  | Neither Agree nor Disagree | 3.91 | 3.91 | 3.91 | 3.91 |
|  | Agree | 4.00 | 4.00 | 4.00 | 4.00 |
|  | Strongly Agree | 3 | 3 | 3 | 3 |
|  | Total | .831 | .831 | .831 | .831 |

The task environment must require the type of coordination and decision-making anticipated within a JADC2 context.

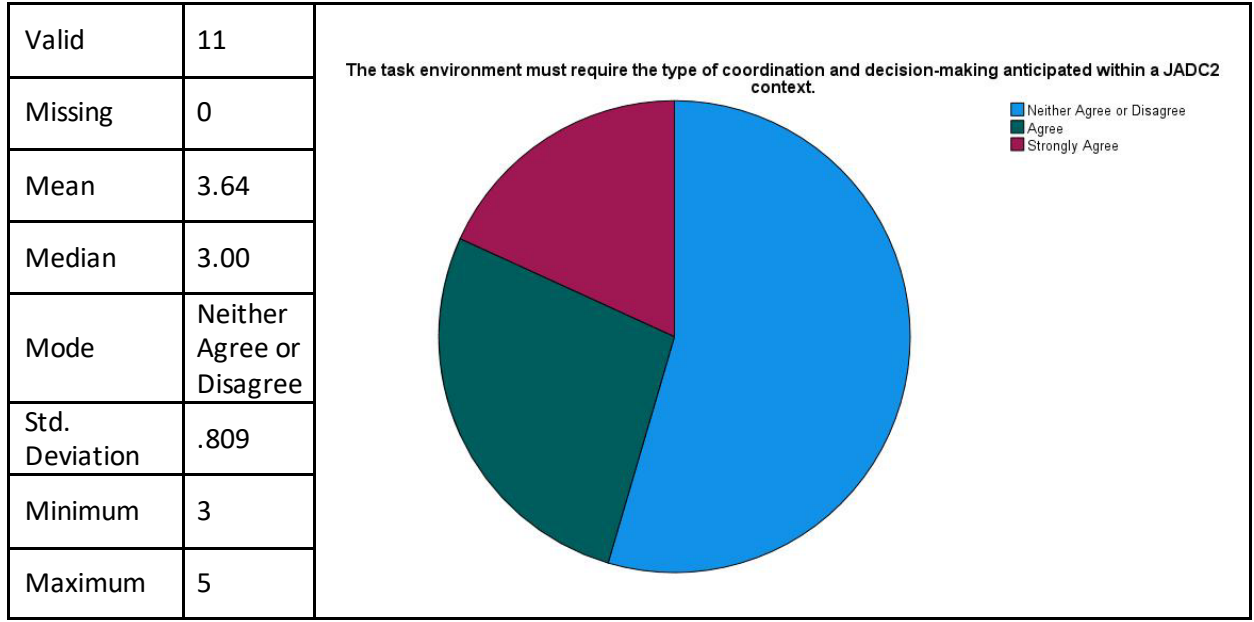

| Valid | 11 |
|---|---|
| Missing | 0 |
| Mean | 3.64 |
| Median | 3.00 |
| Mode | Neither Agree or Disagree |
| Std. Deviation | .809 |
| Minimum | 3 |
| Maximum | 5 |

|  |  | Frequency | Percent | Valid Percent | Cumulative Percent |
|---|---|---|---|---|---|
| Valid | Strongly Disagree |  |  |  |  |
|  | Disagree |  |  |  |  |
|  | Neither Agree nor Disagree | 6 | 54.5 | 54.5 | 54.5 |
|  | Agree | 3 | 27.3 | 27.3 | 81.8 |
|  | Strongly Agree | 2 | 18.2 | 18.2 | 100.0 |
|  | Total | 11 | 100.0 | 100.0 |  |

The task environment must be unclassified and avoid CUI.

| Valid | 11 |
|---|---|
| Missing | 0 |
| Mean | 3.18 |
| Median | 3.00 |
| Mode | Disagree[2] |
| Std. Deviation | 1.328 |
| Minimum | 1 |
| Maximum | 5 |

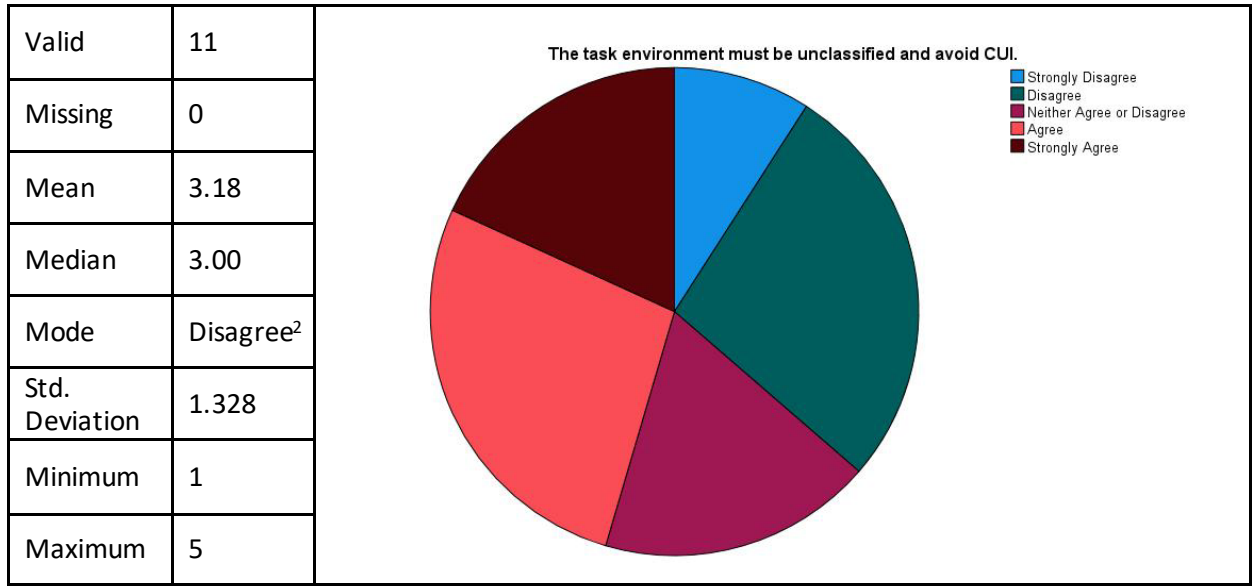

| | | Frequency | Percent | Valid Percent | Cumulative Percent |
|---|---|---|---|---|---|
| Valid | Strongly Disagree | 1 | 9.1 | 9.1 | 9.1 |
| | Disagree | 3 | 27.3 | 27.3 | 36.4 |
| | Neither Agree nor Disagree | 2 | 18.2 | 18.2 | 54.5 |
| | Agree | 3 | 27.3 | 27.3 | 81.8 |
| | Strongly Agree | 2 | 18.2 | 18.2 | 100.0 |
| | Total | 11 | 100.0 | 100.0 | |

The task environment must be easily learned by novice participants.

| | |
|---|---|
| Valid | 11 |
| Missing | 0 |
| Mean | 3.45 |
| Median | 4.00 |
| Mode | Agree |
| Std. Deviation | .934 |
| Minimum | 2 |
| Maximum | 5 |

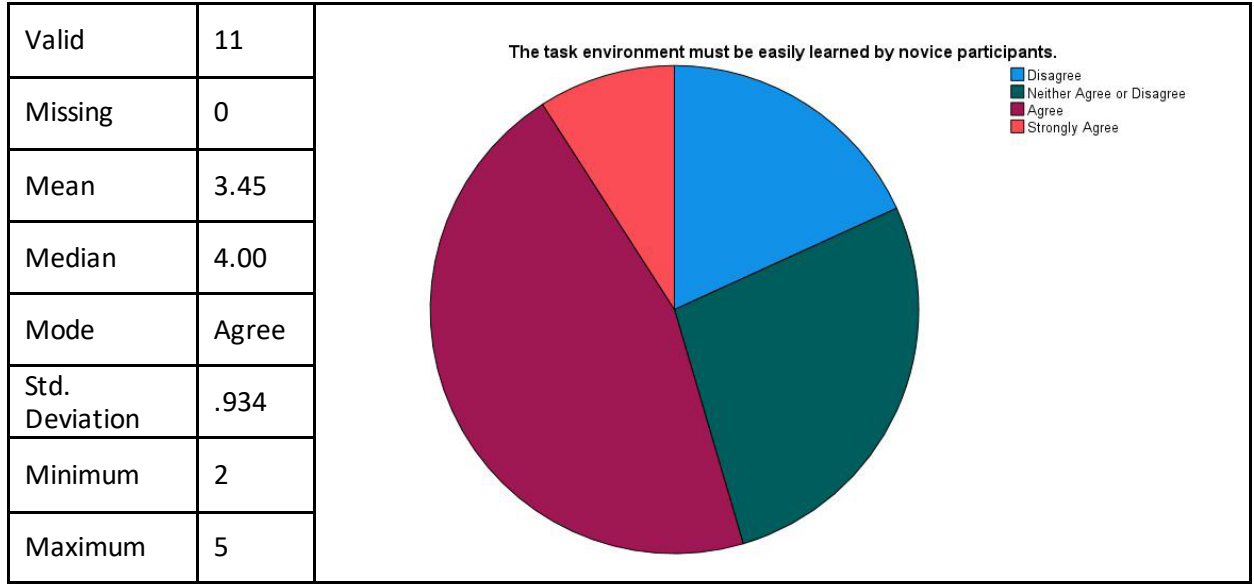

| | | Frequency | Percent | Valid Percent | Cumulative Percent |
|---|---|---|---|---|---|
| Valid | Strongly Disagree | | | | |
| | Disagree | 2 | 18.2 | 18.2 | 18.2 |
| | Neither Agree nor Disagree | 3 | 27.3 | 27.3 | 45.5 |
| | Agree | 5 | 45.5 | 45.5 | 90.9 |
| | Strongly Agree | 1 | 9.1 | 9.1 | 100.0 |
| | Total | 11 | 100.0 | 100.0 | |

The task environment must provide a venue for practicing and assessing coordination behaviors across the timeline (i.e., from "preparation" activities to "adjustment" activities).

| Valid | 11 |
|---|---|
| Missing | 0 |
| Mean | 4.18 |
| Median | 4.00 |
| Mode | Agree |
| Std. Deviation | .751 |
| Minimum | 3 |
| Maximum | 5 |

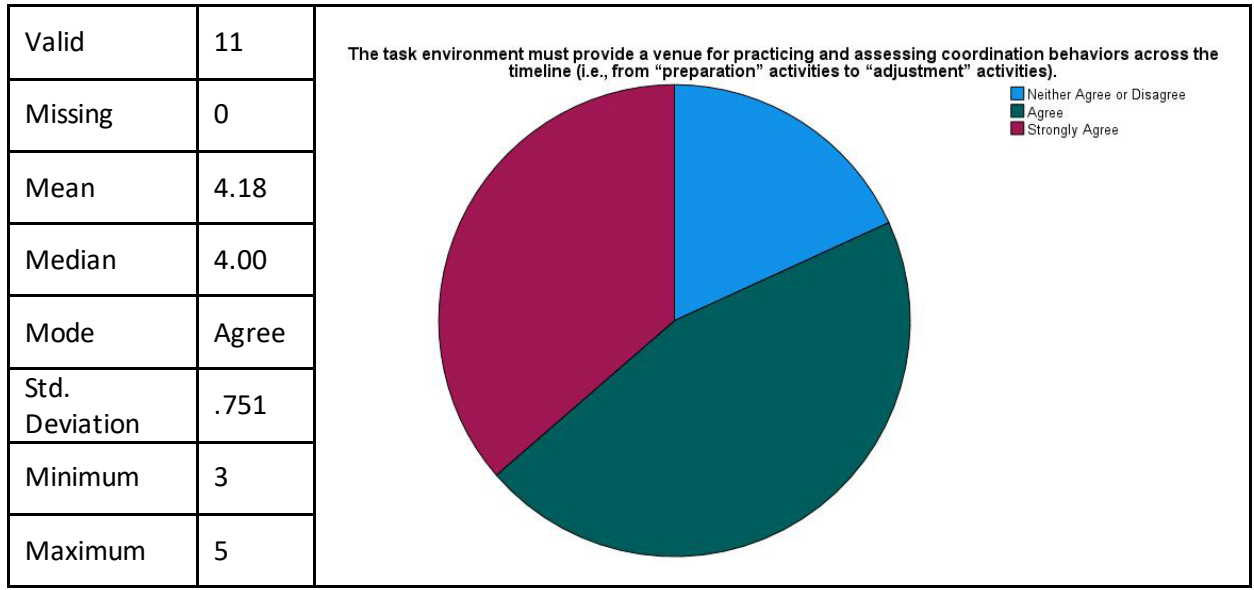

|   |   | Frequency | Percent | Valid Percent | Cumulative Percent |
|---|---|---|---|---|---|
| Valid | Strongly Disagree |   |   |   |   |
|   | Disagree |   |   |   |   |
|   | Neither Agree nor Disagree | 2 | 18.2 | 18.2 | 18.2 |
|   | Agree | 5 | 45.5 | 45.5 | 63.6 |
|   | Strongly Agree | 4 | 36.4 | 36.4 | 100.0 |
|   | Total | 11 | 100.0 | 100.0 |   |

The task environment must include opportunities for physical/logistical interdependency (e.g., you do this, then I do that).

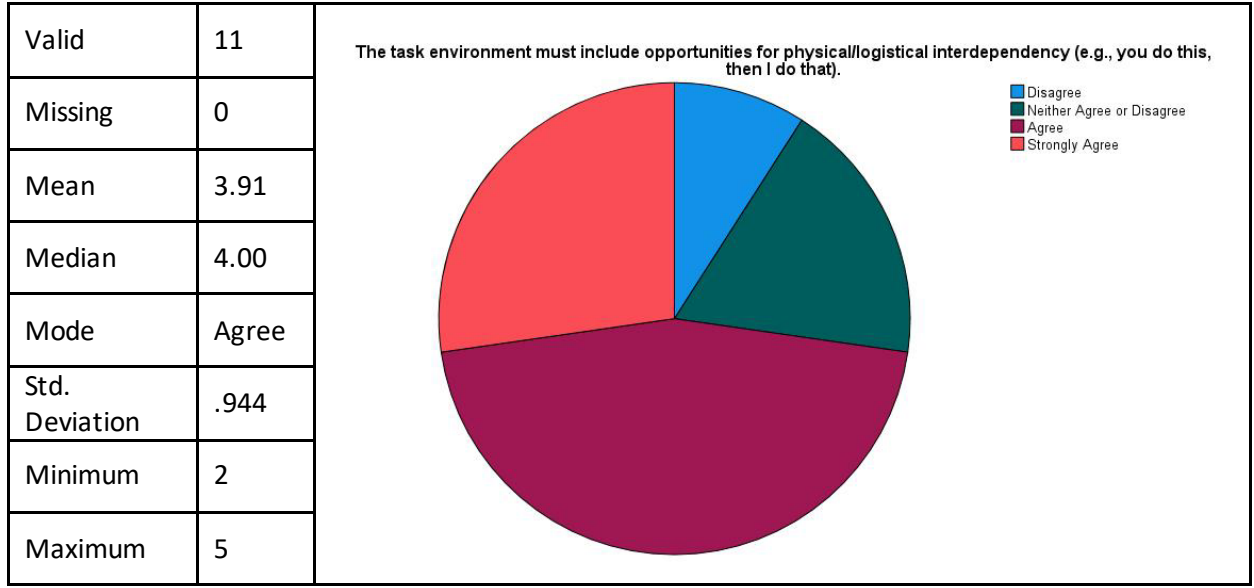

| Valid | 11 |
|---|---|
| Missing | 0 |
| Mean | 3.91 |
| Median | 4.00 |
| Mode | Agree |
| Std. Deviation | .944 |
| Minimum | 2 |
| Maximum | 5 |

| | | Frequency | Percent | Valid Percent | Cumulative Percent |
|---|---|---|---|---|---|
| Valid | Strongly Disagree | | | | |
| | Disagree | 1 | 9.1 | 9.1 | 9.1 |
| | Neither Agree nor Disagree | 2 | 18.2 | 18.2 | 27.3 |
| | Agree | 5 | 45.5 | 45.5 | 72.7 |
| | Strongly Agree | 3 | 27.3 | 27.3 | 100.0 |
| | Total | 11 | 100.0 | 100.0 | |

The task environment must include opportunities for cognitive coordination.

| | |
|---|---|
| Valid | 11 |
| Missing | 0 |
| Mean | 3.91 |
| Median | 4.00 |
| Mode | Neither Agree nor Disagree[2] |
| Std. Deviation | .831 |
| Minimum | 3 |
| Maximum | 5 |

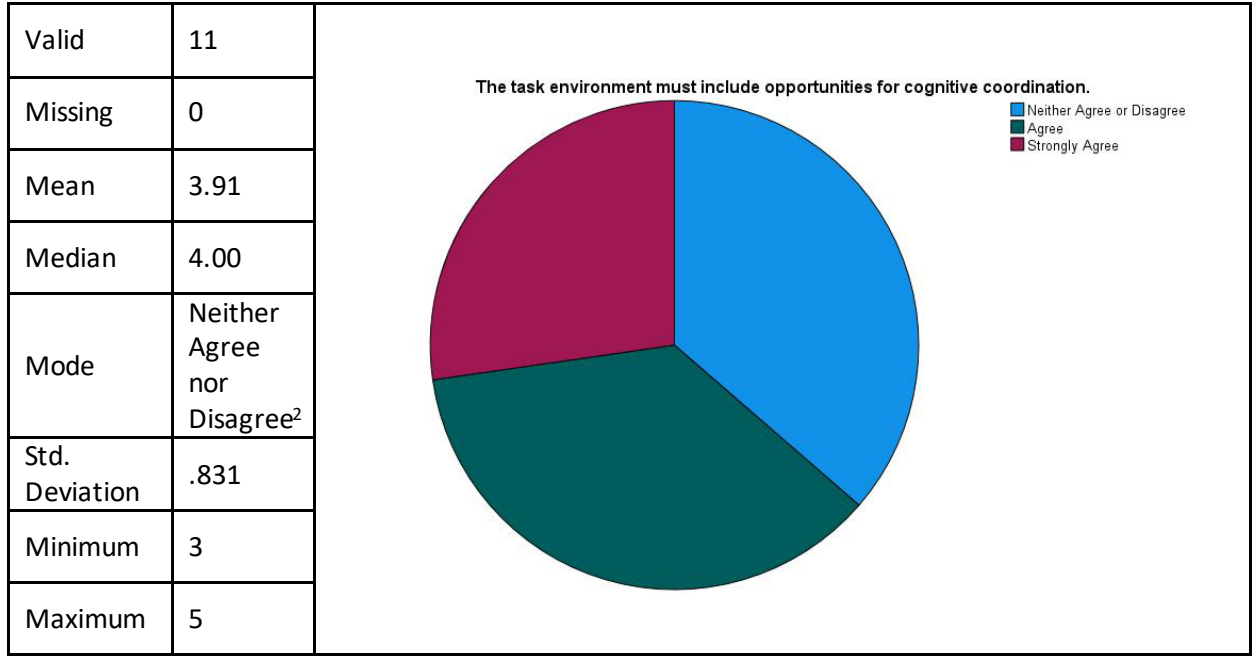

| | | Frequency | Percent | Valid Percent | Cumulative Percent |
|---|---|---|---|---|---|
| Valid | Strongly Disagree | | | | |
| | Disagree | | | | |
| | Neither Agree nor Disagree | 4 | 36.4 | 36.4 | 36.4 |
| | Agree | 4 | 36.4 | 36.4 | 72.7 |
| | Strongly Agree | 3 | 27.3 | 27.3 | 100.0 |
| | Total | 11 | 100.0 | 100.0 | |

This Page Intentionally Blank

*Appendix F*

*Results for Communication Likert Items*

The STE must enable spoken communication among teammates.

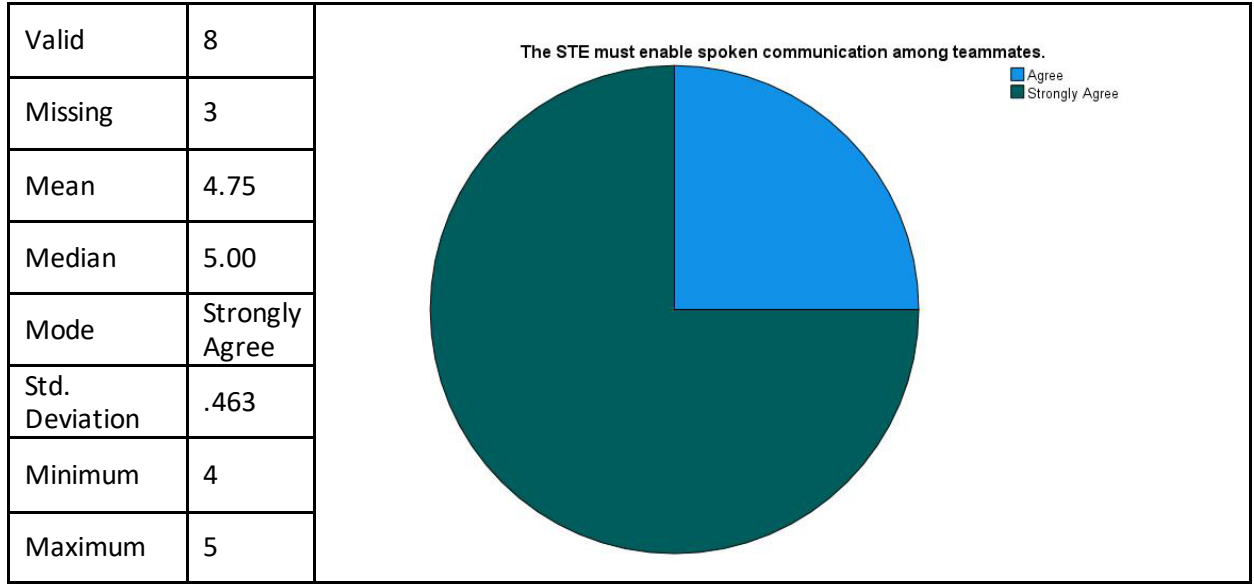

| | |
|---|---|
| Valid | 8 |
| Missing | 3 |
| Mean | 4.75 |
| Median | 5.00 |
| Mode | Strongly Agree |
| Std. Deviation | .463 |
| Minimum | 4 |
| Maximum | 5 |

| | | Frequency | Percent | Valid Percent | Cumulative Percent |
|---|---|---|---|---|---|
| Valid | Strongly Disagree | | | | |
| | Disagree | | | | |
| | Neither Agree nor Disagree | | | | |
| | Agree | 2 | 18.2 | 25.0 | 25.0 |
| | Strongly Agree | 6 | 54.5 | 75.0 | 100.0 |
| | Total | 8 | 72.7 | 100.0 | |

The STE must enable chat-based communication among teammates.

| Valid | 8 |
|---|---|
| Missing | 3 |
| Mean | 4.50 |
| Median | 4.50 |
| Mode | Agree[2] |
| Std. Deviation | .535 |
| Minimum | 4 |
| Maximum | 5 |

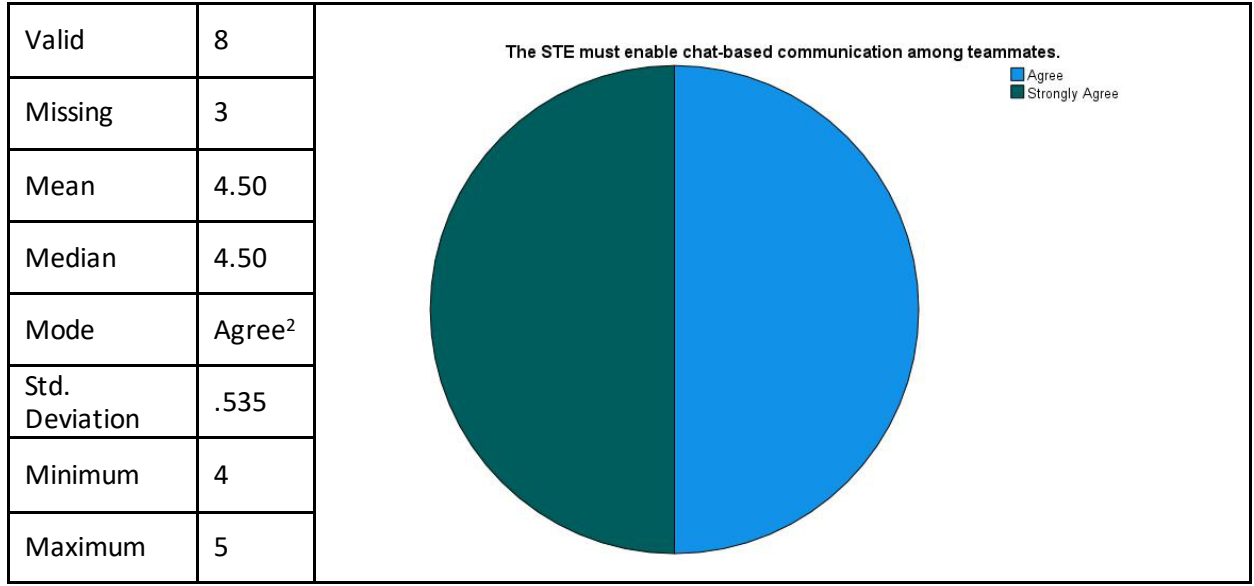

| | | Frequency | Percent | Valid Percent | Cumulative Percent |
|---|---|---|---|---|---|
| Valid | Strongly Disagree | | | | |
| | Disagree | | | | |
| | Neither Agree nor Disagree | | | | |
| | Agree | 4 | 36.4 | 50.0 | 50.0 |
| | Strongly Agree | 4 | 36.4 | 50.0 | 100.0 |
| | Total | 8 | 72.7 | 100.0 | |

The STE must enable email communications among teammates.

| | |
|---|---|
| Valid | 8 |
| Missing | 3 |
| Mean | 3.25 |
| Median | 3.00 |
| Mode | Neither Agree or Disagree |
| Std. Deviation | 1.165 |
| Minimum | 2 |
| Maximum | 5 |

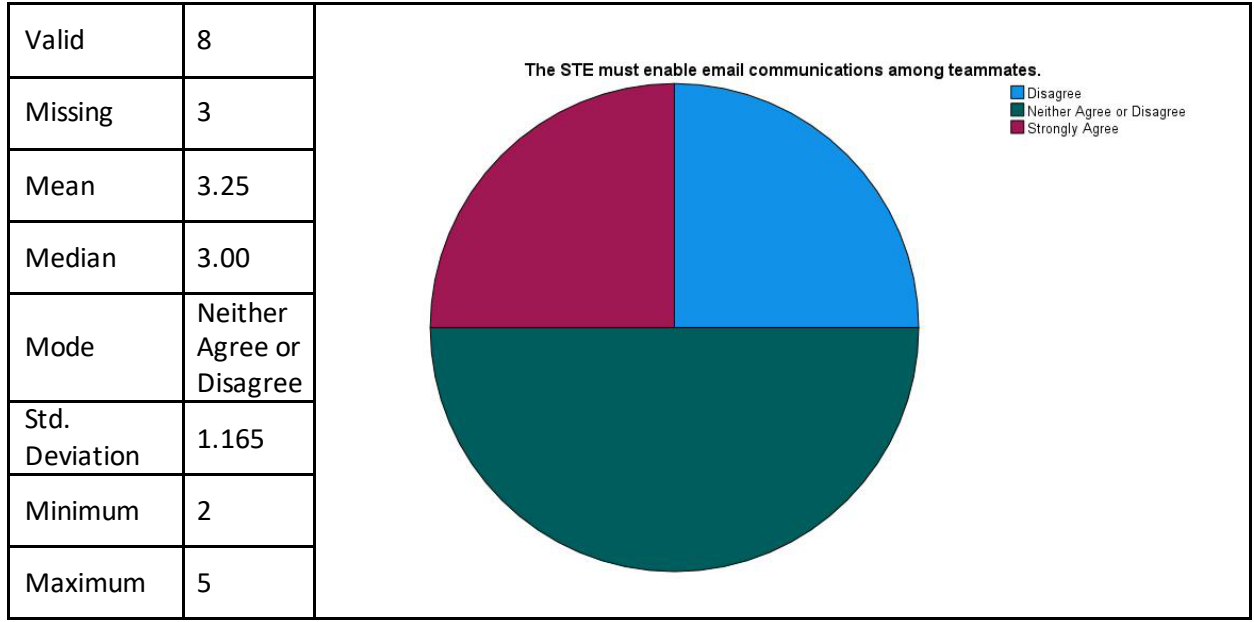

| | | Frequency | Percent | Valid Percent | Cumulative Percent |
|---|---|---|---|---|---|
| Valid | Strongly Disagree | | | | |
| | Disagree | 2 | 18.2 | 25.0 | 25.0 |
| | Neither Agree nor Disagree | 4 | 36.4 | 50.0 | 75.0 |
| | Agree | 0 | 0.00 | 0.00 | 75.0 |
| | Strongly Agree | 2 | 18.2 | 25.0 | 100.0 |
| | Total | 8 | 72.7 | 100.0 | |

The STE must enable video communications among teammates.

| Valid | 8 |
|---|---|
| Missing | 3 |
| Mean | 3.50 |
| Median | 3.00 |
| Mode | Neither Agree or Disagree |
| Std. Deviation | 1.069 |
| Minimum | 2 |
| Maximum | 5 |

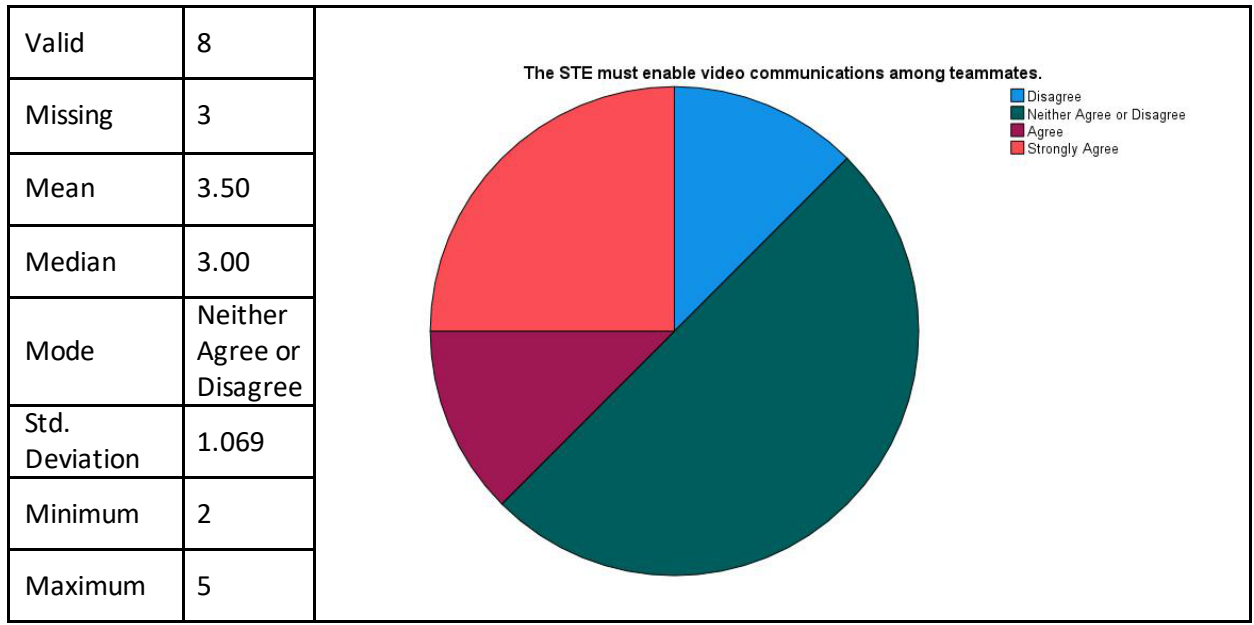

|  |  | Frequency | Percent | Valid Percent | Cumulative Percent |
|---|---|---|---|---|---|
| Valid | Strongly Disagree |  |  |  |  |
|  | Disagree | 1 | 9.1 | 12.5 | 12.5 |
|  | Neither Agree nor Disagree | 4 | 36.4 | 50.0 | 62.5 |
|  | Agree | 1 | 9.1 | 12.5 | 75.0 |
|  | Strongly Agree | 2 | 18.2 | 25.0 | 100.0 |
|  | Total | 8 | 72.7 | 100.0 |  |

The STE must enable paper-based communications among teammates.

| Valid | 8 |
|---|---|
| Missing | 3 |
| Mean | 2.88 |
| Median | 3.00 |
| Mode | Neither Agree or Disagree |
| Std. Deviation | 1.126 |
| Minimum | 1 |
| Maximum | 5 |

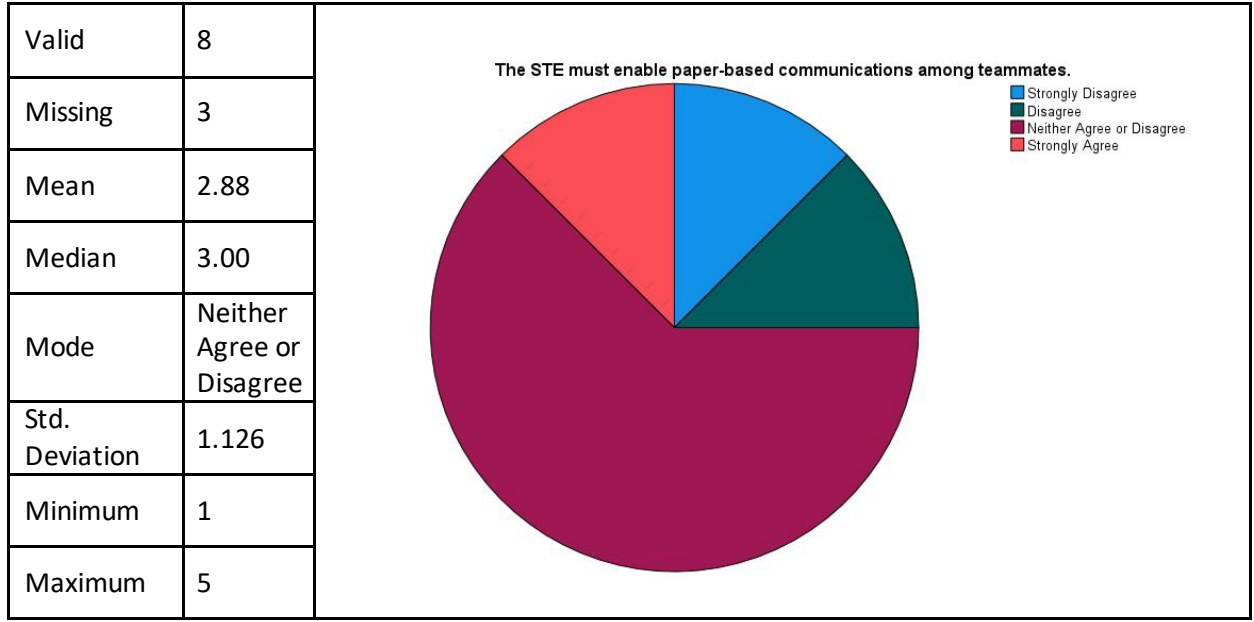

|  |  | Frequency | Percent | Valid Percent | Cumulative Percent |
|---|---|---|---|---|---|
| Valid | Strongly Disagree | 1 | 9.1 | 12.5 | 12.5 |
|  | Disagree | 1 | 9.1 | 12.5 | 25.0 |
|  | Neither Agree nor Disagree | 5 | 45.5 | 62.5 | 87.5 |
|  | Agree | 0 | 0.00 | 0.00 | 87.5 |
|  | Strongly Agree | 1 | 9.1 | 12.5 | 100.0 |
|  | Total | 8 | 72.7 | 100.0 |  |

The STE must include speech recognition, natural language processing, and text-to-speech to allow AI teammates to communicate with human teammates.

| | | |
|---|---|---|
| Valid | 8 | |
| Missing | 3 | |
| Mean | 3.88 | |
| Median | 4.00 | |
| Mode | Agree | |
| Std. Deviation | .641 | |
| Minimum | 3 | |
| Maximum | 5 | |

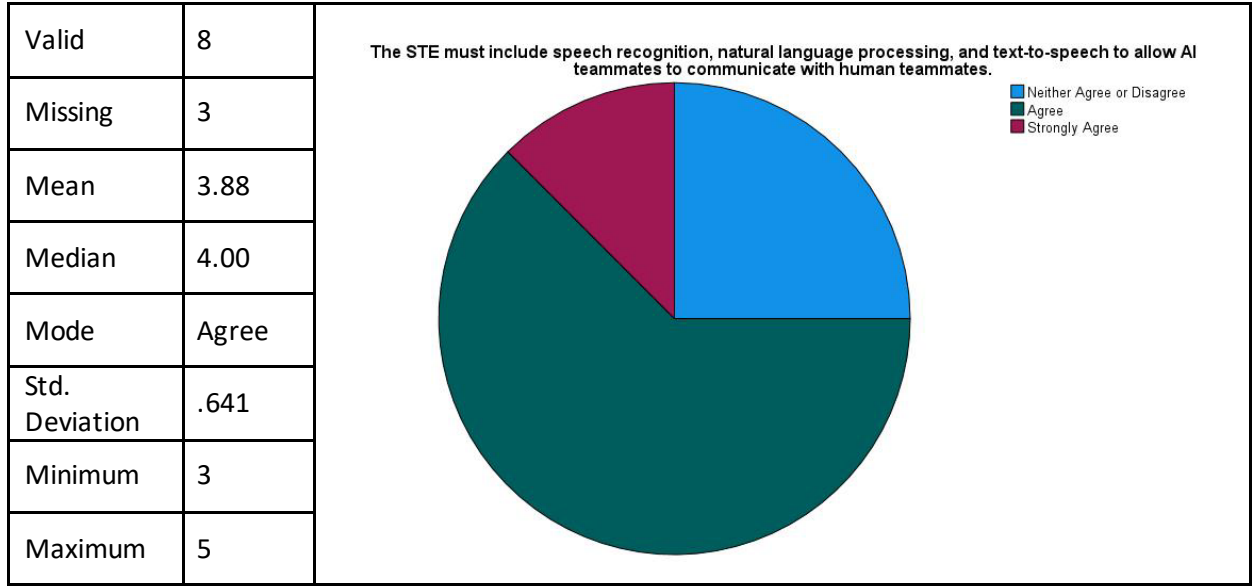

| | | Frequency | Percent | Valid Percent | Cumulative Percent |
|---|---|---|---|---|---|
| Valid | Strongly Disagree | | | | |
| | Disagree | | | | |
| | Neither Agree nor Disagree | 2 | 18.2 | 25.0 | 25.0 |
| | Agree | 5 | 45.5 | 62.5 | 87.5 |
| | Strongly Agree | 1 | 9.1 | 12.5 | 100.0 |
| | Total | 8 | 72.7 | 100.0 | |

To bypass the need for strong natural language processing capabilities, the STE must include a utility that allows researchers to serve as a "translator" between human and AI teammates.

| Valid | 8 |
|---|---|
| Missing | 3 |
| Mean | 3.50 |
| Median | 3.50 |
| Mode | Neither Agree or Disagree[2] |
| Std. Deviation | .926 |
| Minimum | 2 |
| Maximum | 5 |

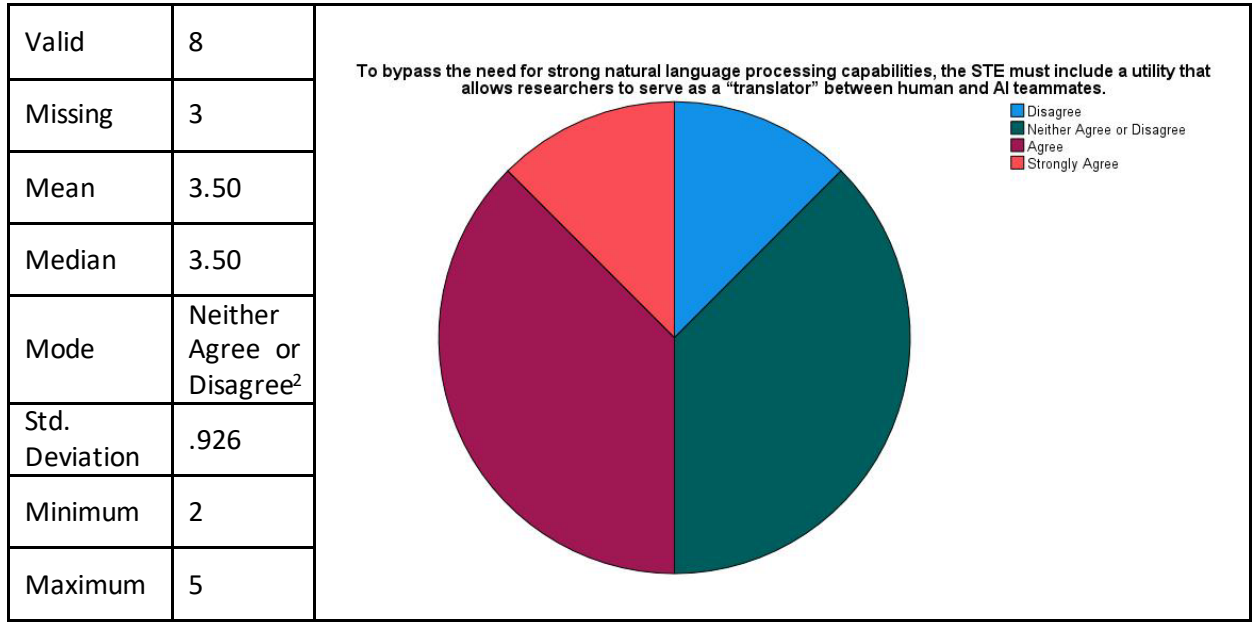

|  |  | Frequency | Percent | Valid Percent | Cumulative Percent |
|---|---|---|---|---|---|
| Valid | Strongly Disagree |  |  |  |  |
|  | Disagree | 1 | 9.1 | 12.5 | 12.5 |
|  | Neither Agree nor Disagree | 3 | 27.3 | 37.5 | 50.0 |
|  | Agree | 3 | 27.3 | 37.5 | 87.5 |
|  | Strongly Agree | 1 | 9.1 | 12.5 | 100.0 |
|  | Total | 8 | 72.7 | 100.0 |  |

The STE must include support for both "broadcast" and "station-to-station" communication among teammates.

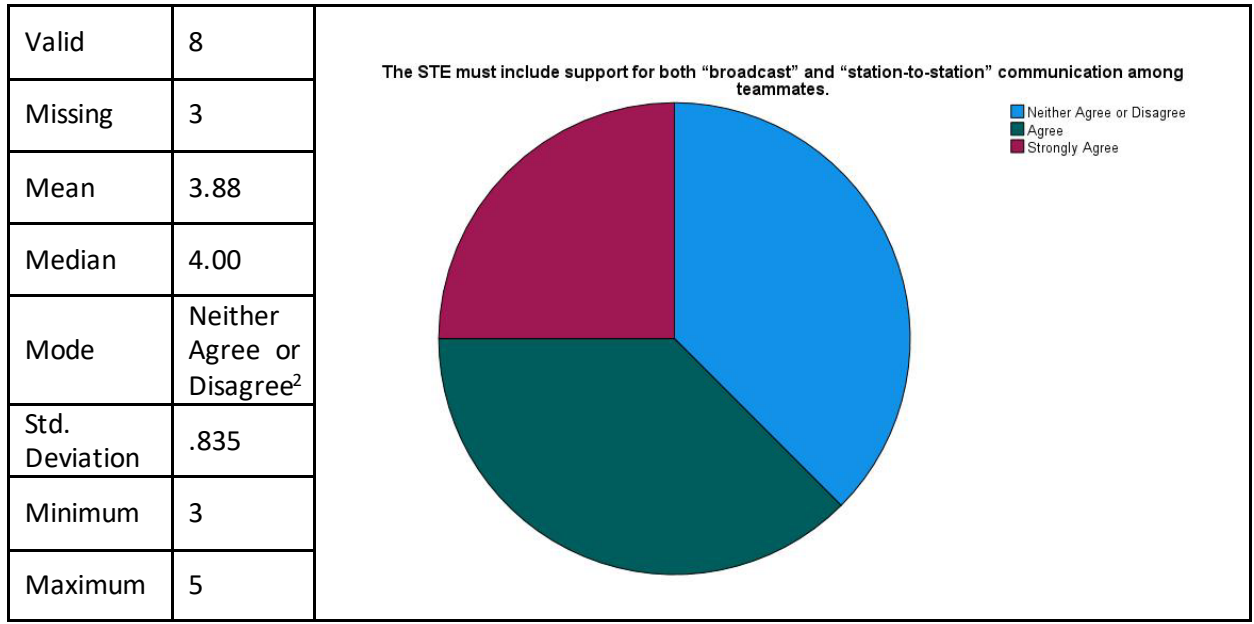

| Valid | 8 |
|---|---|
| Missing | 3 |
| Mean | 3.88 |
| Median | 4.00 |
| Mode | Neither Agree or Disagree[2] |
| Std. Deviation | .835 |
| Minimum | 3 |
| Maximum | 5 |

| | | Frequency | Percent | Valid Percent | Cumulative Percent |
|---|---|---|---|---|---|
| Valid | Strongly Disagree | | | | |
| | Disagree | | | | |
| | Neither Agree nor Disagree | 3 | 27.3 | 37.5 | 37.5 |
| | Agree | 3 | 27.3 | 37.5 | 75.0 |
| | Strongly Agree | 2 | 18.2 | 25.0 | 100.0 |
| | Total | 8 | 72.7 | 100.0 | |

The STE should include support for "Wizard of Oz" communication.

| | |
|---|---|
| Valid | 8 |
| Missing | 3 |
| Mean | 3.63 |
| Median | 3.50 |
| Mode | Neither Agree or Disagree |
| Std. Deviation | .744 |
| Minimum | 3 |
| Maximum | 5 |

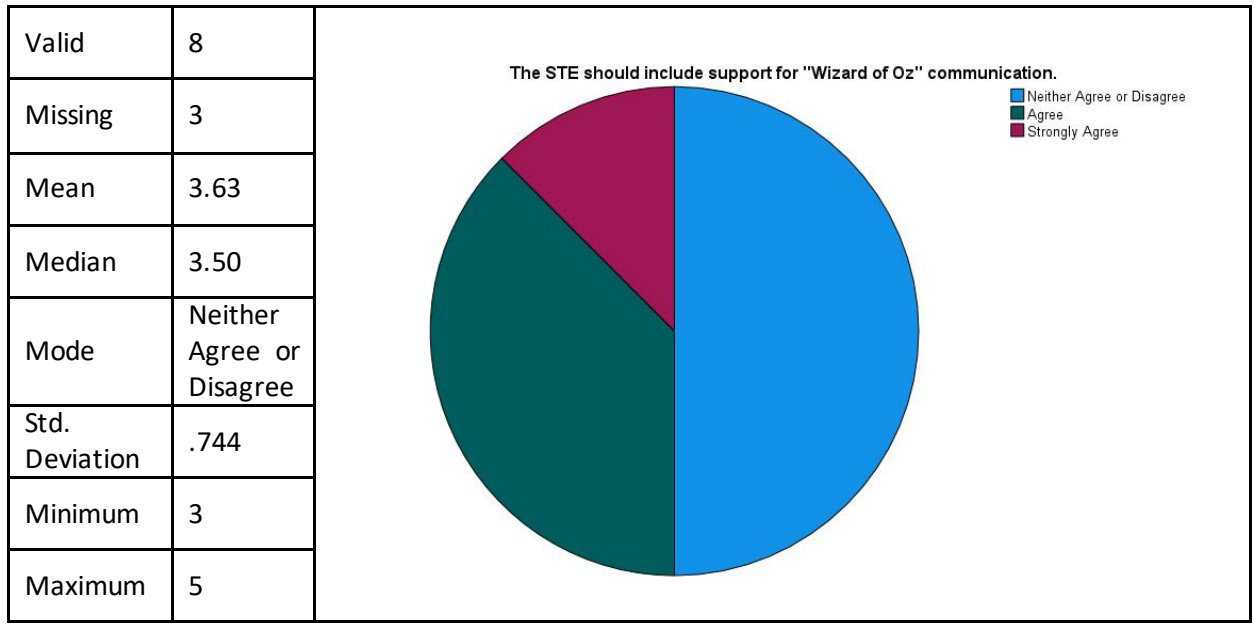

| | | Frequency | Percent | Valid Percent | Cumulative Percent |
|---|---|---|---|---|---|
| Valid | Strongly Disagree | | | | |
| | Disagree | | | | |
| | Neither Agree nor Disagree | 4 | 36.4 | 50.0 | 50.0 |
| | Agree | 3 | 27.3 | 37.5 | 87.5 |
| | Strongly Agree | 1 | 9.1 | 12.5 | 100.0 |
| | Total | 8 | 72.7 | 100.0 | |

*Appendix G*

*Results for Autonomous Agents Likert Items*

The STE must allow researchers to plug in externally developed autonomous teammates.

| | |
|---|---|
| Valid | 8 |
| Missing | 3 |
| Mean | 4.25 |
| Median | 4.00 |
| Mode | 4 |
| Std. Deviation | .707 |
| Minimum | 3 |
| Maximum | 5 |

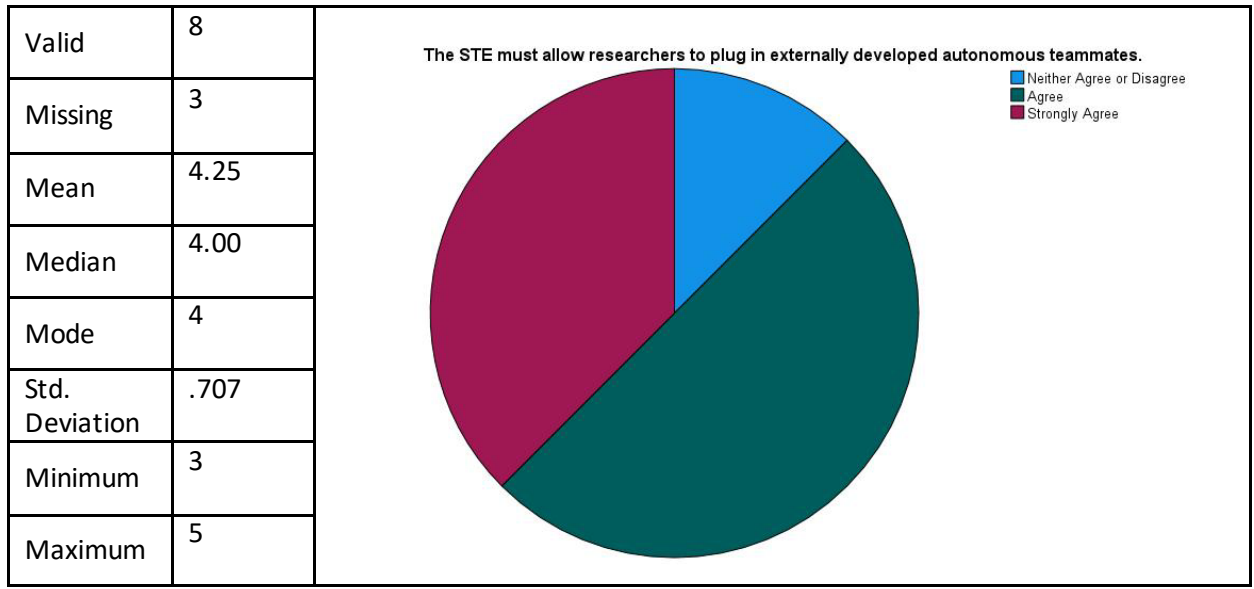

| | | Frequency | Percent | Valid Percent | Cumulative Percent |
|---|---|---|---|---|---|
| Valid | Strongly Disagree | | | | |
| | Disagree | | | | |
| | Neither Agree nor Disagree | 1 | 9.1 | 12.5 | 12.5 |
| | Agree | 4 | 36.4 | 50.0 | 62.5 |
| | Strongly Agree | 3 | 27.3 | 37.5 | 100.0 |
| | Total | 8 | 72.7 | 100.0 | |

Researchers using the STE must be able to staff different roles on the team with either human or autonomous teammates.

| | |
|---|---|
| Valid | 8 |
| Missing | 3 |
| Mean | 4.63 |
| Median | 5.00 |
| Mode | 5 |
| Std. Deviation | .518 |
| Minimum | 4 |
| Maximum | 5 |

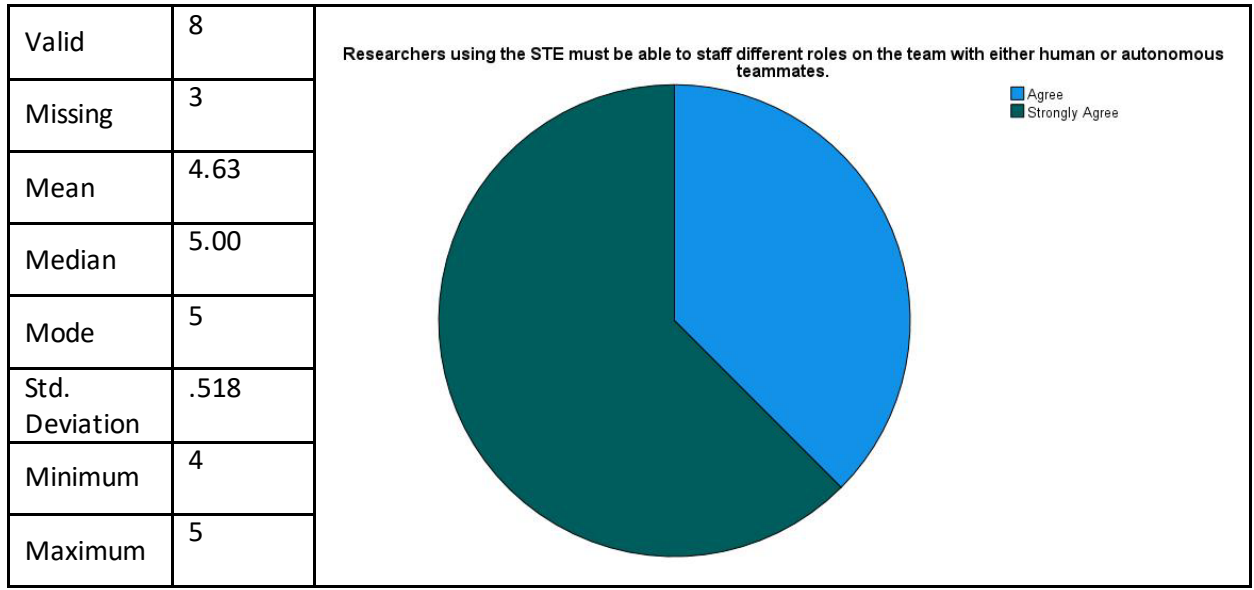

| | | Frequency | Percent | Valid Percent | Cumulative Percent |
|---|---|---|---|---|---|
| Valid | Strongly Disagree | | | | |
| | Disagree | | | | |
| | Neither Agree nor Disagree | | | | |
| | Agree | 3 | 27.3 | 37.5 | 37.5 |
| | Strongly Agree | 5 | 45.5 | 62.5 | 100.0 |
| | Total | 8 | 72.7 | 100.0 | |

Autonomous teammates must include both task work and team coordination functionality.

| Valid | 8 |
|---|---|
| Missing | 3 |
| Mean | 4.25 |
| Median | 4.50 |
| Mode | 5 |
| Std. Deviation | .886 |
| Minimum | 3 |
| Maximum | 5 |

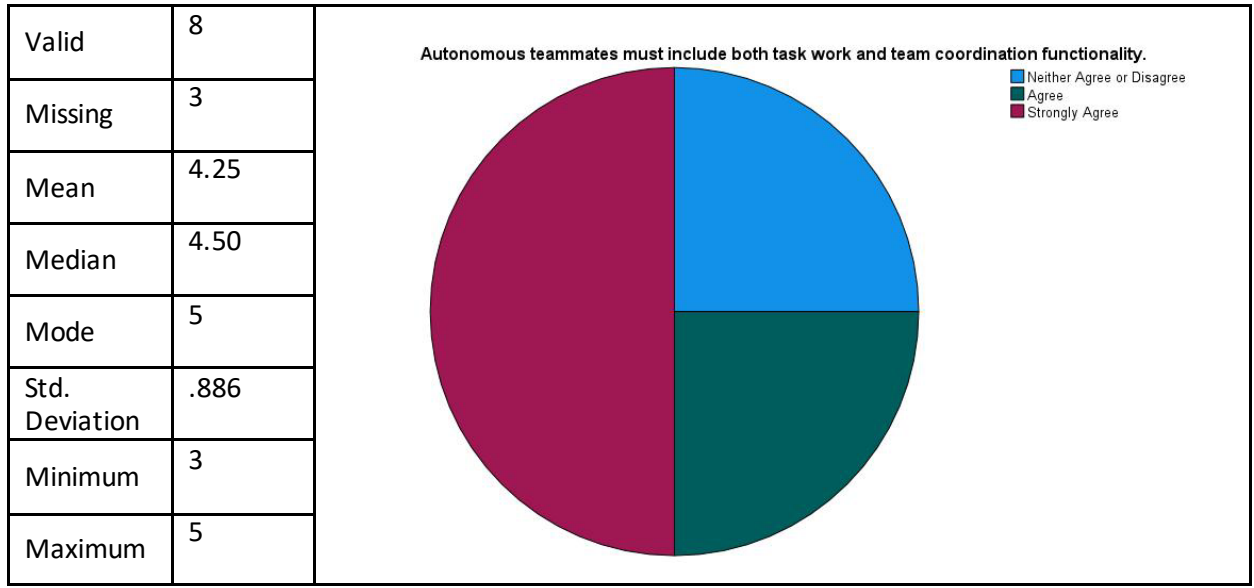

| | | Frequency | Percent | Valid Percent | Cumulative Percent |
|---|---|---|---|---|---|
| Valid | Strongly Disagree | | | | |
| | Disagree | | | | |
| | Neither Agree nor Disagree | 2 | 18.2 | 25.0 | 25.0 |
| | Agree | 2 | 18.2 | 25.0 | 50.0 |
| | Strongly Agree | 4 | 36.4 | 50.0 | 100.0 |
| | Total | 8 | 72.7 | 100.0 | |

Autonomous teammates within the STE must include domain knowledge similar to that possessed by humans who would staff that role.

| Valid | 8 |
|---|---|
| Missing | 3 |
| Mean | 3.63 |
| Median | 3.00 |
| Mode | 3 |
| Std. Deviation | .916 |
| Minimum | 3 |
| Maximum | 5 |

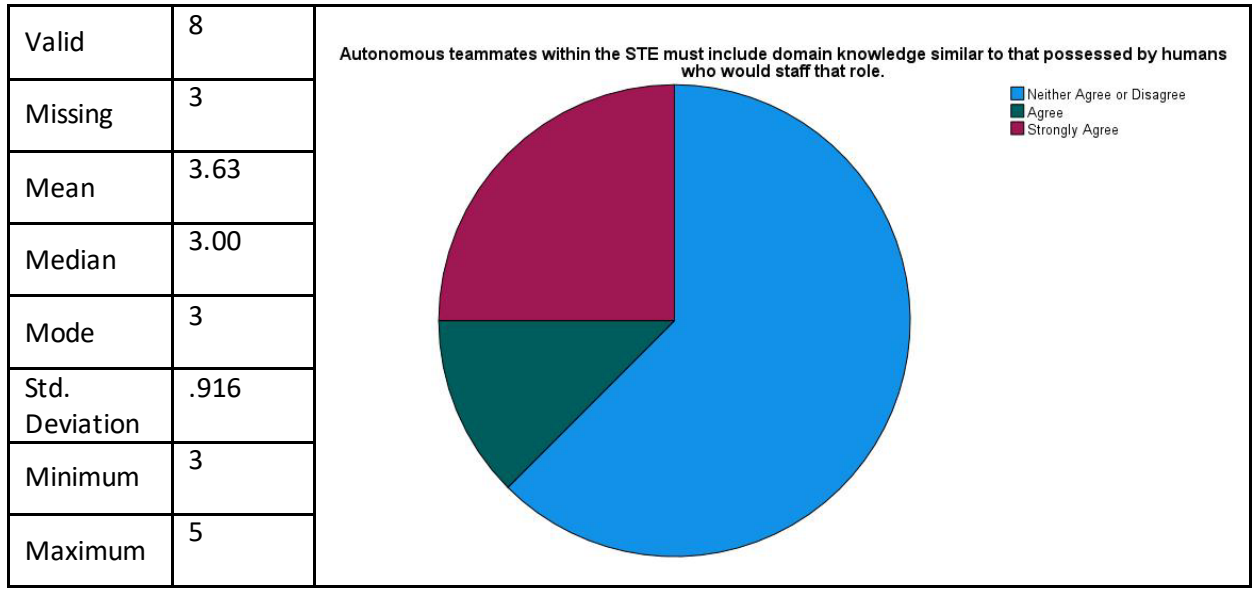

|  |  | Frequency | Percent | Valid Percent | Cumulative Percent |
|---|---|---|---|---|---|
| Valid | Strongly Disagree |  |  |  |  |
|  | Disagree |  |  |  |  |
|  | Neither Agree nor Disagree | 5 | 45.5 | 62.5 | 62.5 |
|  | Agree | 1 | 9.1 | 12.5 | 75.0 |
|  | Strongly Agree | 2 | 18.2 | 25.0 | 100.0 |
|  | Total | 8 | 72.7 | 100.0 |  |

Autonomous teammates within the STE must include knowledge of the capabilities and limitations of other autonomous teammates.

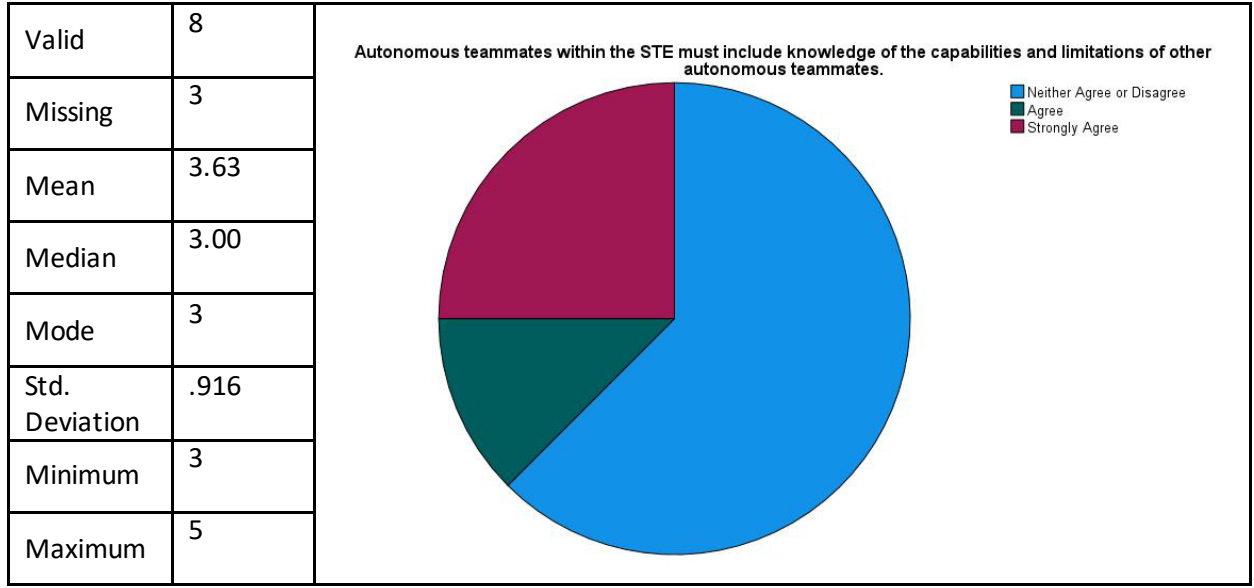

| Valid | 8 |
|---|---|
| Missing | 3 |
| Mean | 3.63 |
| Median | 3.00 |
| Mode | 3 |
| Std. Deviation | .916 |
| Minimum | 3 |
| Maximum | 5 |

|  |  | Frequency | Percent | Valid Percent | Cumulative Percent |
|---|---|---|---|---|---|
| Valid | Strongly Disagree |  |  |  |  |
|  | Disagree |  |  |  |  |
|  | Neither Agree nor Disagree | 5 | 45.5 | 62.5 | 62.5 |
|  | Agree | 1 | 9.1 | 12.5 | 75.0 |
|  | Strongly Agree | 2 | 18.2 | 25.0 | 100.0 |
|  | Total | 8 | 72.7 | 100.0 |  |

Autonomous teammates within the STE must be able to monitor the state of other autonomous teammates.

| Valid | 8 |
|---|---|
| Missing | 3 |
| Mean | 4.00 |
| Median | 4.00 |
| Mode | 3² |
| Std. Deviation | .926 |
| Minimum | 3 |
| Maximum | 5 |

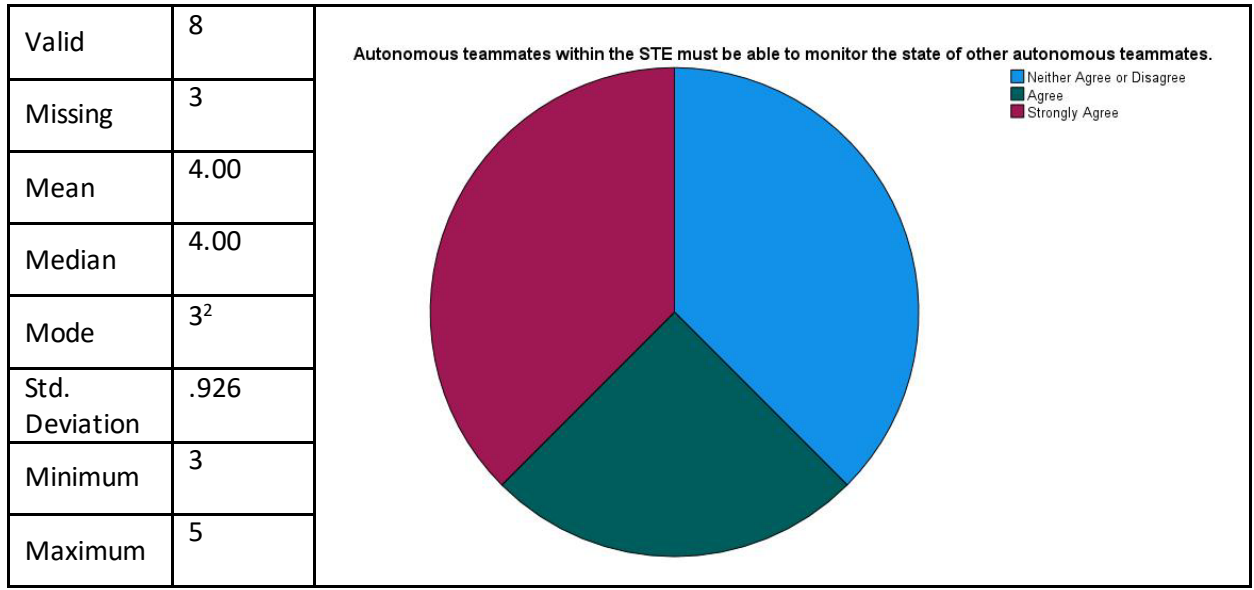

|  |  | Frequency | Percent | Valid Percent | Cumulative Percent |
|---|---|---|---|---|---|
| Valid | Strongly Disagree |  |  |  |  |
|  | Disagree |  |  |  |  |
|  | Neither Agree nor Disagree | 3 | 27.3 | 37.5 | 37.5 |
|  | Agree | 2 | 18.2 | 25.0 | 62.5 |
|  | Strongly Agree | 3 | 27.3 | 37.5 | 100.0 |
|  | Total | 8 | 72.7 | 100.0 |  |

Autonomous teammates within the STE must be able to monitor the state of the mission, task, and sub-tasks the team is performing.

| Valid | 8 |
|---|---|
| Missing | 3 |
| Mean | 4.38 |
| Median | 4.50 |
| Mode | 5 |
| Std. Deviation | .744 |
| Minimum | 3 |
| Maximum | 5 |

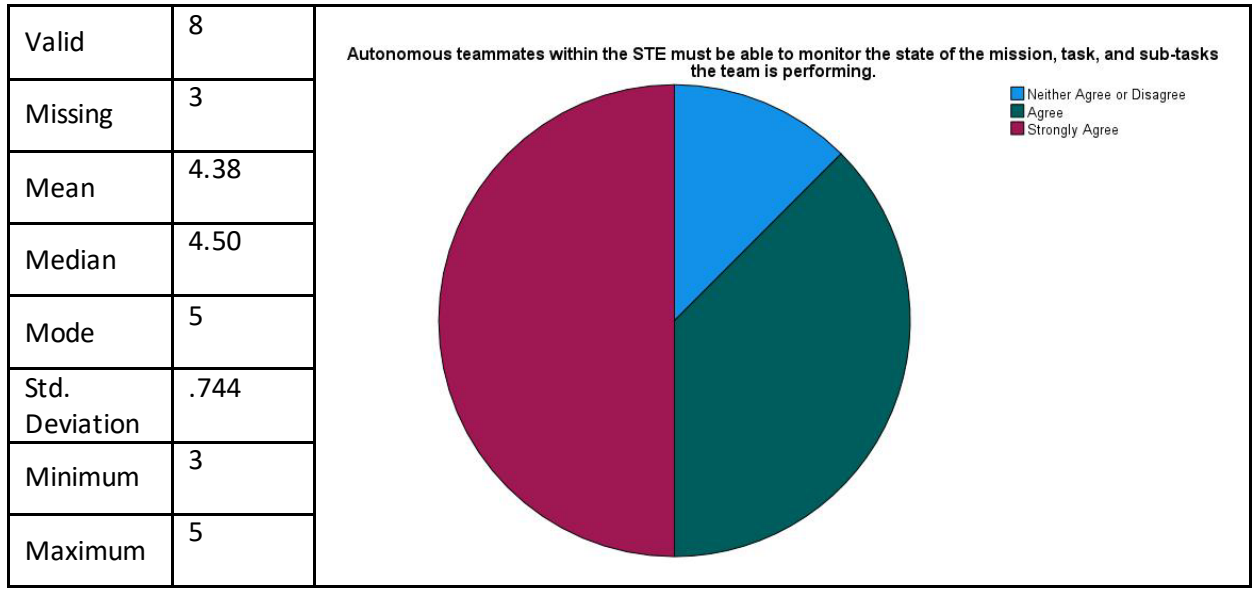

| | | Frequency | Percent | Valid Percent | Cumulative Percent |
|---|---|---|---|---|---|
| Valid | Strongly Disagree | | | | |
| | Disagree | | | | |
| | Neither Agree nor Disagree | 1 | 9.1 | 12.5 | 12.5 |
| | Agree | 3 | 27.3 | 37.5 | 50.0 |
| | Strongly Agree | 4 | 36.4 | 50.0 | 100.0 |
| | Total | 8 | 72.7 | 100.0 | |

Autonomous teammates within the STE must be able to perform all activities normally associated with the role to which they are assigned.

| | |
|---|---|
| Valid | 8 |
| Missing | 3 |
| Mean | 4.00 |
| Median | 4.00 |
| Mode | 4 |
| Std. Deviation | .756 |
| Minimum | 3 |
| Maximum | 5 |

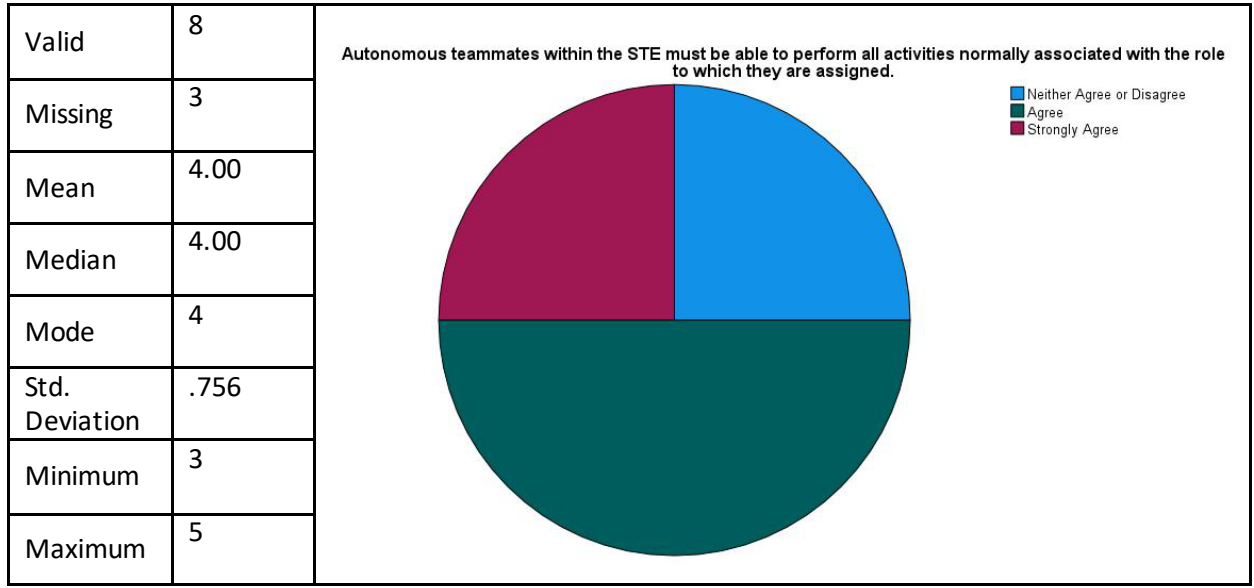

| | | Frequency | Percent | Valid Percent | Cumulative Percent |
|---|---|---|---|---|---|
| Valid | Strongly Disagree | | | | |
| | Disagree | | | | |
| | Neither Agree nor Disagree | 2 | 18.2 | 25.0 | 25.0 |
| | Agree | 4 | 36.4 | 50.0 | 75.0 |
| | Strongly Agree | 2 | 18.2 | 25.0 | 100.0 |
| | Total | 8 | 72.7 | 100.0 | |

Autonomous teammates within the STE must support "interpredictability" (i.e., the ability of the agent to predict the behavior of human teammates and of the humans to predict the ability of the autonomous teammates).

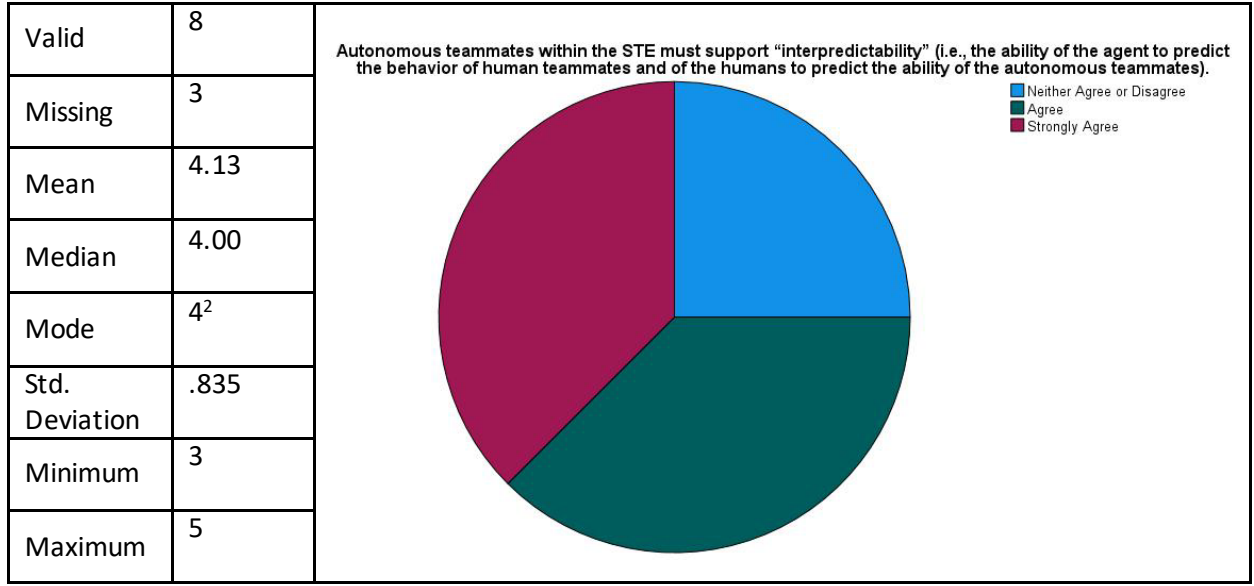

| | |
|---|---|
| Valid | 8 |
| Missing | 3 |
| Mean | 4.13 |
| Median | 4.00 |
| Mode | 4² |
| Std. Deviation | .835 |
| Minimum | 3 |
| Maximum | 5 |

| | | Frequency | Percent | Valid Percent | Cumulative Percent |
|---|---|---|---|---|---|
| Valid | Strongly Disagree | | | | |
| | Disagree | | | | |
| | Neither Agree nor Disagree | 2 | 18.2 | 25.0 | 25.0 |
| | Agree | 3 | 27.3 | 37.5 | 62.5 |
| | Strongly Agree | 3 | 27.3 | 37.5 | 100.0 |
| | Total | 8 | 72.7 | 100.0 | |

Autonomous teammates within the STE must be able to establish, maintain, and repair common ground with other teammates.

| | |
|---|---|
| Valid | 8 |
| Missing | 3 |
| Mean | 4.13 |
| Median | 4.50 |
| Mode | 5 |
| Std. Deviation | 1.126 |
| Minimum | 2 |
| Maximum | 5 |

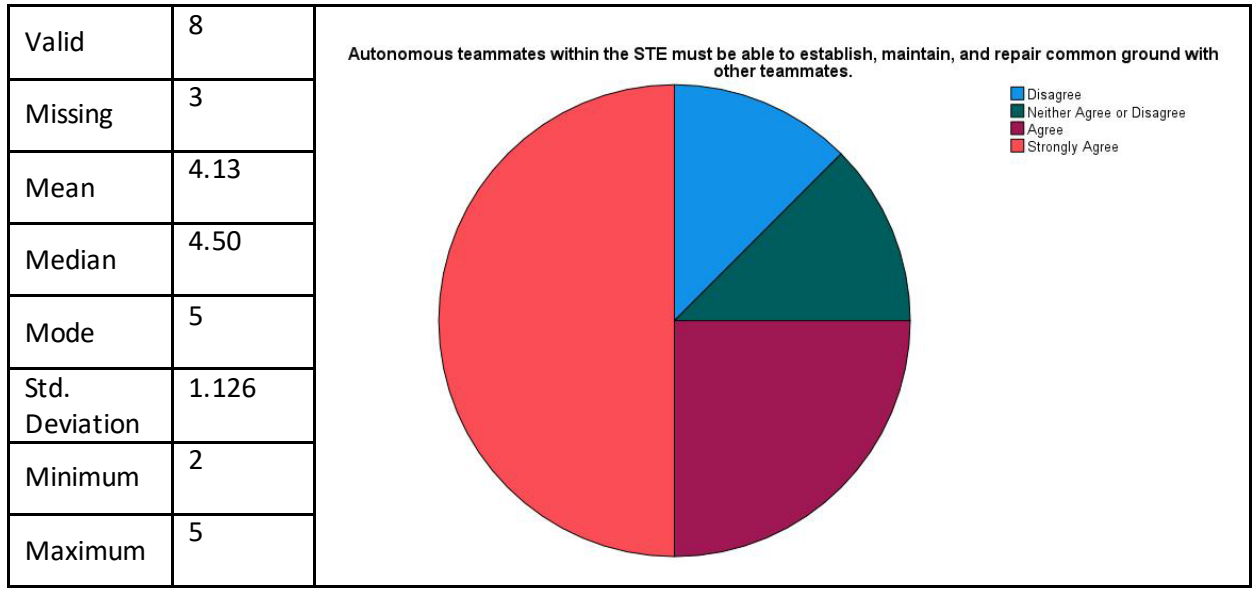

| | | Frequency | Percent | Valid Percent | Cumulative Percent |
|---|---|---|---|---|---|
| Valid | Strongly Disagree | | | | |
| | Disagree | 1 | 9.1 | 12.5 | 12.5 |
| | Neither Agree nor Disagree | 1 | 9.1 | 12.5 | 25.0 |
| | Agree | 2 | 18.2 | 25.0 | 50.0 |
| | Strongly Agree | 4 | 36.4 | 50.0 | 100.0 |
| | Total | 8 | 72.7 | 100.0 | |

Autonomous teammates within the STE must be able to accept guidance from teammates (human or non-human) as conditions and priorities change.

| | |
|---|---|
| Valid | 8 |
| Missing | 3 |
| Mean | 4.25 |
| Median | 4.00 |
| Mode | 4 |
| Std. Deviation | .707 |
| Minimum | 3 |
| Maximum | 5 |

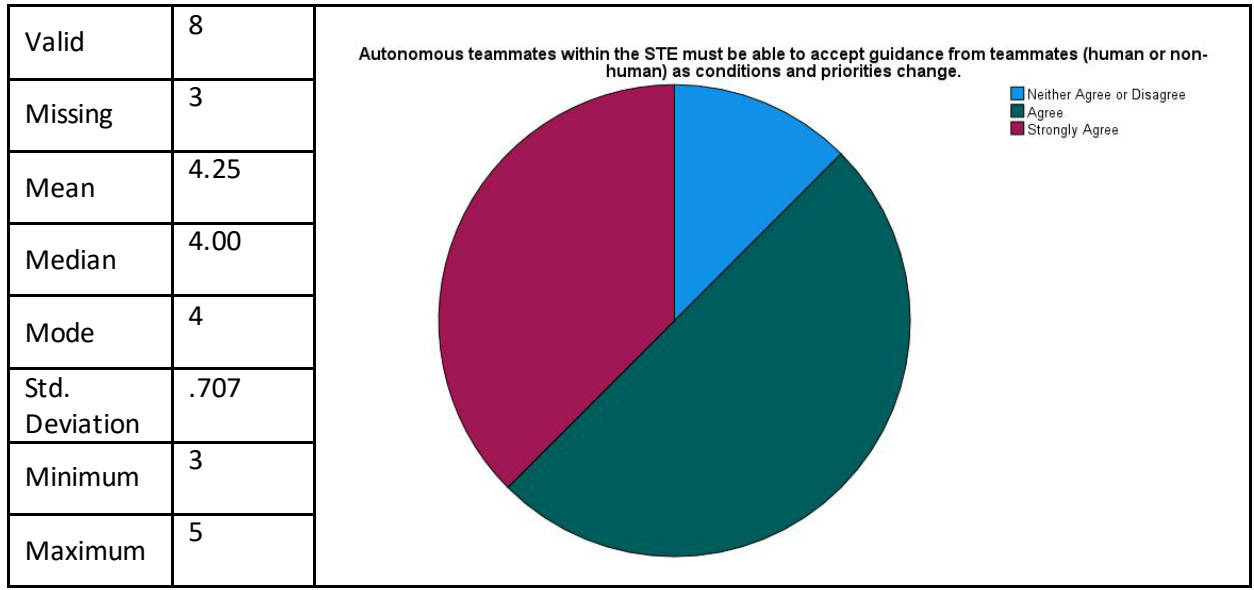

| | | Frequency | Percent | Valid Percent | Cumulative Percent |
|---|---|---|---|---|---|
| Valid | Strongly Disagree | | | | |
| | Disagree | | | | |
| | Neither Agree nor Disagree | 1 | 9.1 | 12.5 | 12.5 |
| | Agree | 4 | 36.4 | 50.0 | 62.5 |
| | Strongly Agree | 3 | 27.3 | 37.5 | 100.0 |
| | Total | 8 | 72.7 | 100.0 | |

Autonomous teammates within the STE must be able to monitor the evolving situation and respond proactively.

| Valid | 8 |
|---|---|
| Missing | 3 |
| Mean | 4.13 |
| Median | 4.00 |
| Mode | 4² |
| Std. Deviation | .835 |
| Minimum | 3 |
| Maximum | 5 |

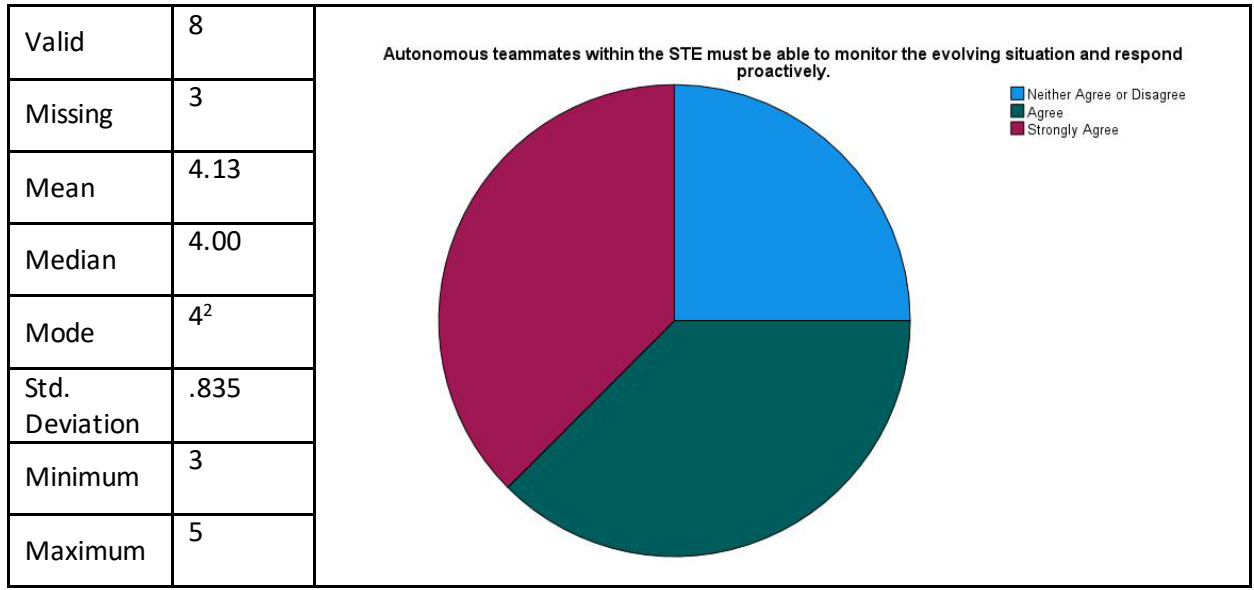

|  |  | Frequency | Percent | Valid Percent | Cumulative Percent |
|---|---|---|---|---|---|
| Valid | Strongly Disagree |  |  |  |  |
|  | Disagree |  |  |  |  |
|  | Neither Agree nor Disagree | 2 | 18.2 | 25.0 | 25.0 |
|  | Agree | 3 | 27.3 | 37.5 | 62.5 |
|  | Strongly Agree | 3 | 27.3 | 37.5 | 100.0 |
|  | Total | 8 | 72.7 | 100.0 |  |

Autonomous teammates within the STE must be able to adhere to a chain of command.

| | |
|---|---|
| Valid | 8 |
| Missing | 3 |
| Mean | 4.00 |
| Median | 4.00 |
| Mode | 4 |
| Std. Deviation | .756 |
| Minimum | 3 |
| Maximum | 5 |

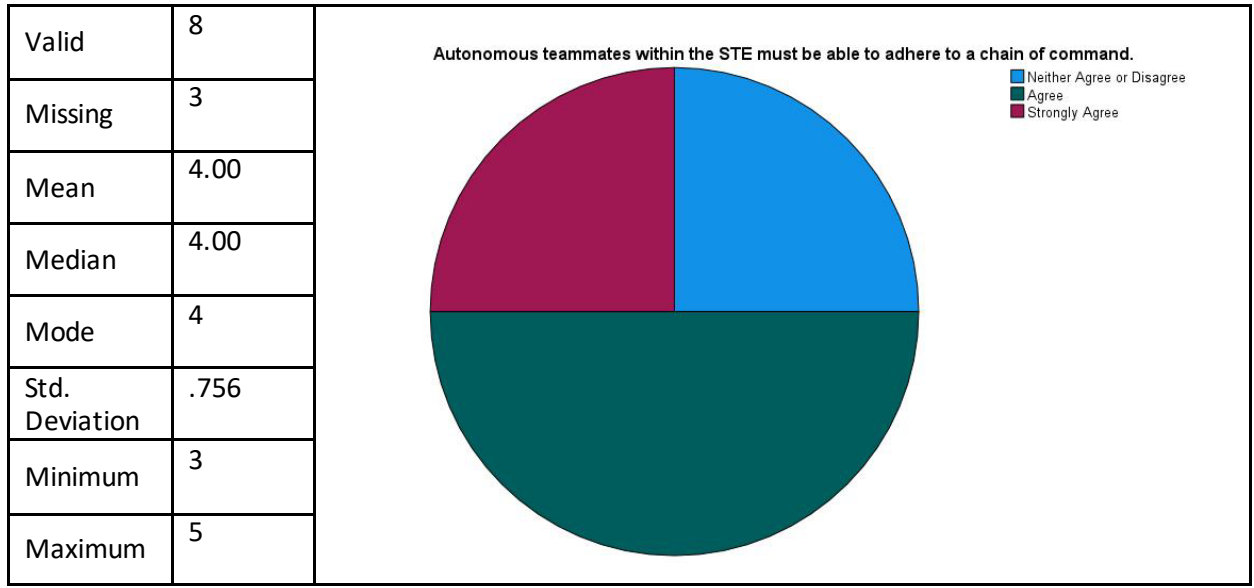

| | | Frequency | Percent | Valid Percent | Cumulative Percent |
|---|---|---|---|---|---|
| Valid | Strongly Disagree | | | | |
| | Disagree | | | | |
| | Neither Agree nor Disagree | 2 | 18.2 | 25.0 | 25.0 |
| | Agree | 4 | 36.4 | 50.0 | 75.0 |
| | Strongly Agree | 2 | 18.2 | 25.0 | 100.0 |
| | Total | 8 | 72.7 | 100.0 | |

Autonomous teammates within the STE must be able to notify teammates of on-going progress, state changes, etc.

| | |
|---|---|
| Valid | 8 |
| Missing | 3 |
| Mean | 4.38 |
| Median | 4.50 |
| Mode | 5 |
| Std. Deviation | .744 |
| Minimum | 3 |
| Maximum | 5 |

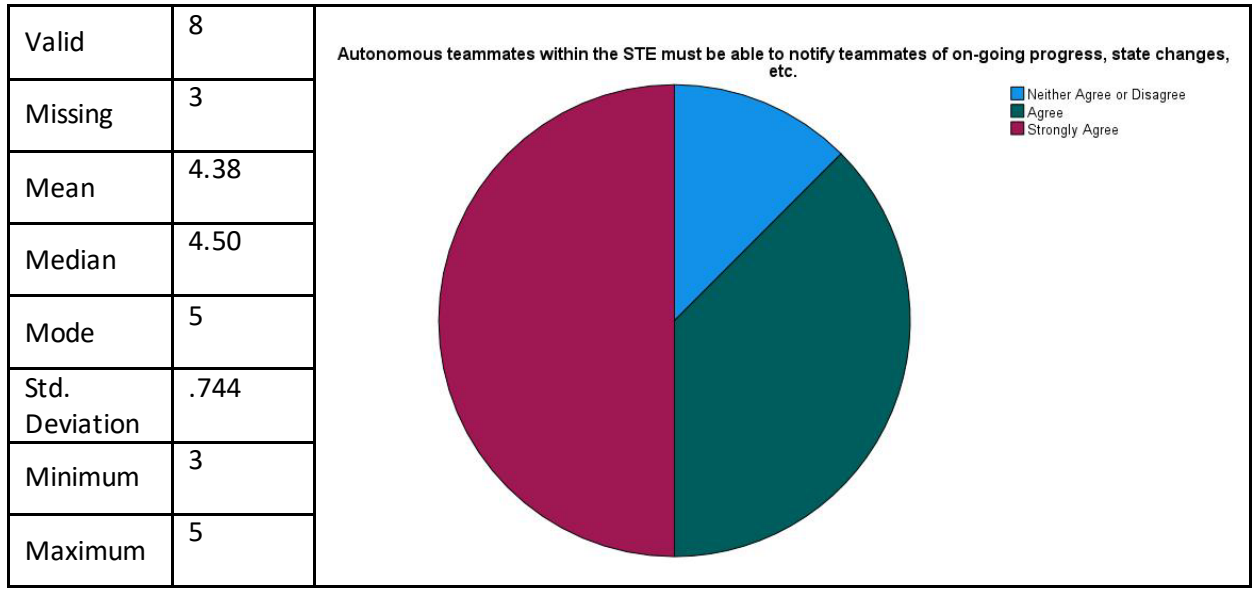

| | | Frequency | Percent | Valid Percent | Cumulative Percent |
|---|---|---|---|---|---|
| Valid | Strongly Disagree | | | | |
| | Disagree | | | | |
| | Neither Agree nor Disagree | 1 | 9.1 | 12.5 | 12.5 |
| | Agree | 3 | 27.3 | 37.5 | 50.0 |
| | Strongly Agree | 4 | 36.4 | 50.0 | 100.0 |
| | Total | 8 | 72.7 | 100.0 | |

Autonomous teammates within the STE must be able to acknowledge inputs from other teammates.

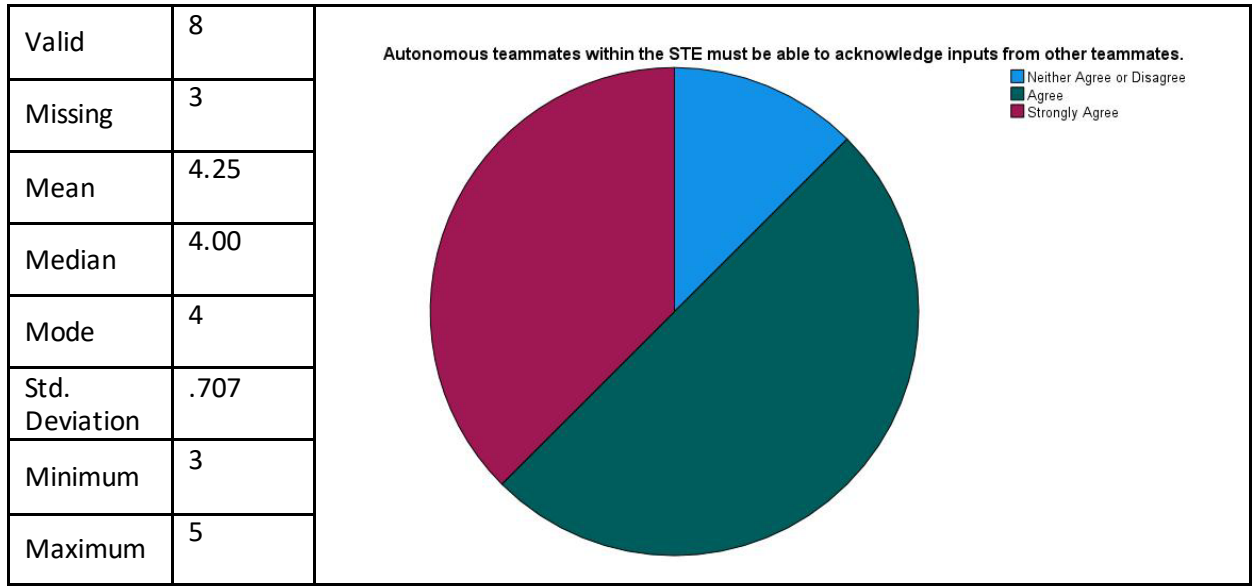

| Valid | 8 |
|---|---|
| Missing | 3 |
| Mean | 4.25 |
| Median | 4.00 |
| Mode | 4 |
| Std. Deviation | .707 |
| Minimum | 3 |
| Maximum | 5 |

|  |  | Frequency | Percent | Valid Percent | Cumulative Percent |
|---|---|---|---|---|---|
| Valid | Strongly Disagree |  |  |  |  |
|  | Disagree |  |  |  |  |
|  | Neither Agree nor Disagree | 1 | 9.1 | 12.5 | 12.5 |
|  | Agree | 4 | 36.4 | 50.0 | 62.5 |
|  | Strongly Agree | 3 | 27.3 | 37.5 | 100.0 |
|  | Total | 8 | 72.7 | 100.0 |  |

Autonomous teammates within the STE must be able to repair trust violations resulting from errors.

| | |
|---|---|
| Valid | 8 |
| Missing | 3 |
| Mean | 4.13 |
| Median | 4.00 |
| Mode | 4² |
| Std. Deviation | .835 |
| Minimum | 3 |
| Maximum | 5 |

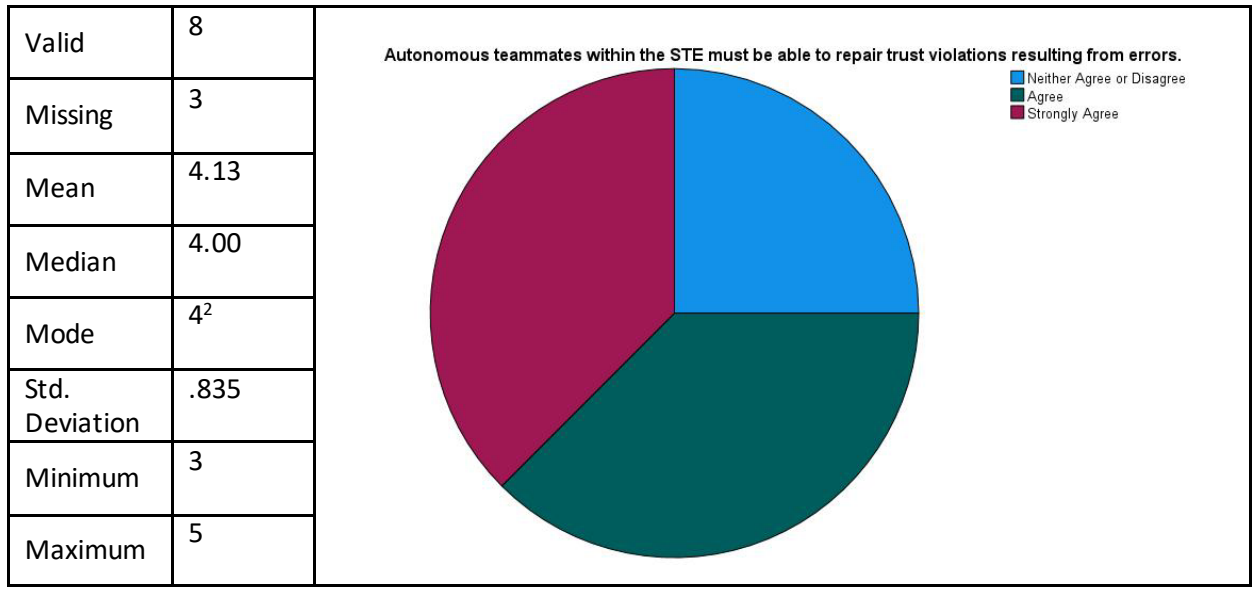

| | | Frequency | Percent | Valid Percent | Cumulative Percent |
|---|---|---|---|---|---|
| Valid | Strongly Disagree | | | | |
| | Disagree | | | | |
| | Neither Agree nor Disagree | 2 | 18.2 | 25.0 | 25.0 |
| | Agree | 3 | 27.3 | 37.5 | 62.5 |
| | Strongly Agree | 3 | 27.3 | 37.5 | 100.0 |
| | Total | 8 | 72.7 | 100.0 | |

Autonomous teammates within the STE must be able to temper trust by actively monitoring for conditions in which failure was likely and notifying teammates that they should not rely on the automation in such situations in the future.

| | |
|---|---|
| Valid | 8 |
| Missing | 3 |
| Mean | 4.25 |
| Median | 4.00 |
| Mode | 4 |
| Std. Deviation | .707 |
| Minimum | 3 |
| Maximum | 5 |

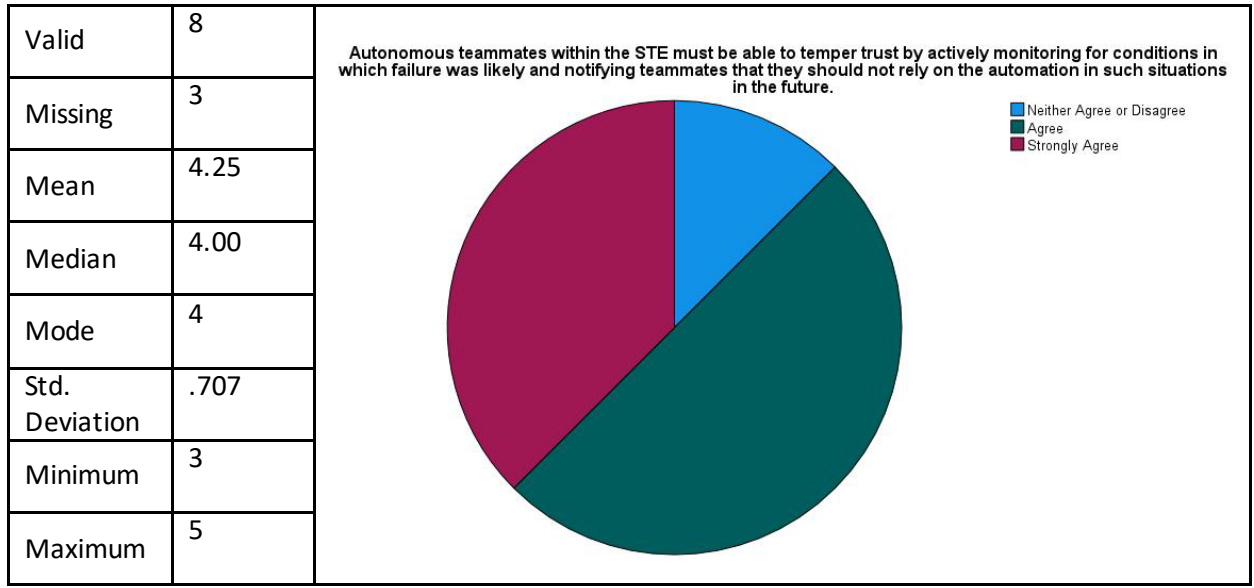

| | | Frequency | Percent | Valid Percent | Cumulative Percent |
|---|---|---|---|---|---|
| Valid | Strongly Disagree | | | | |
| | Disagree | | | | |
| | Neither Agree nor Disagree | 1 | 9.1 | 12.5 | 12.5 |
| | Agree | 4 | 36.4 | 50.0 | 62.5 |
| | Strongly Agree | 3 | 27.3 | 37.5 | 100.0 |
| | Total | 8 | 72.7 | 100.0 | |

Autonomous teammates within the STE must be able to accurately monitor and report their status.

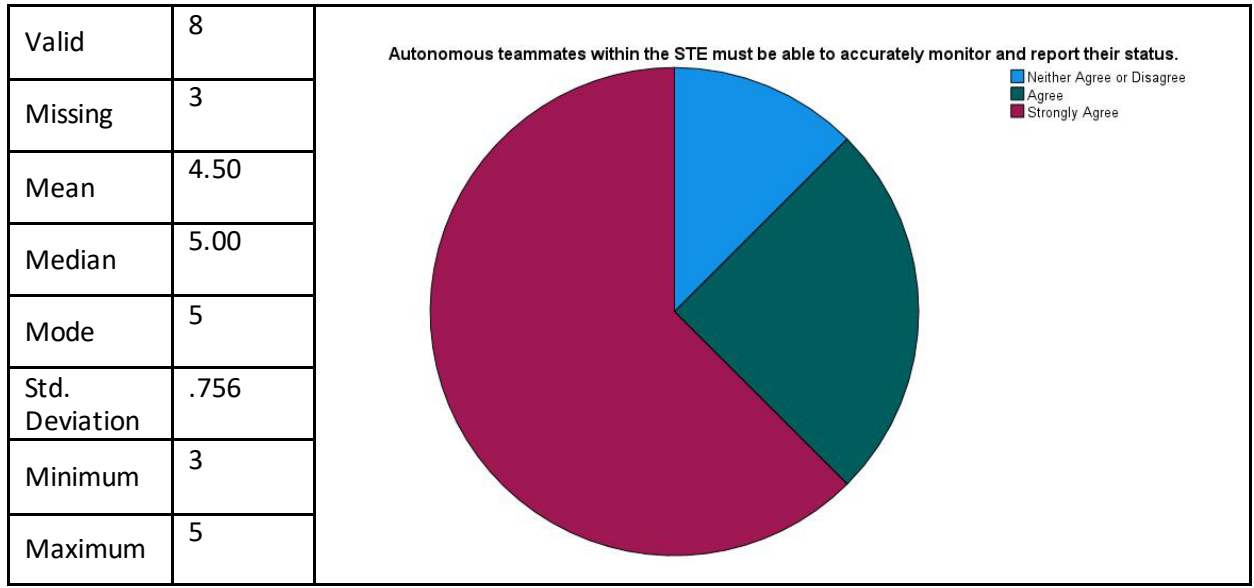

| | |
|---|---|
| Valid | 8 |
| Missing | 3 |
| Mean | 4.50 |
| Median | 5.00 |
| Mode | 5 |
| Std. Deviation | .756 |
| Minimum | 3 |
| Maximum | 5 |

| | | Frequency | Percent | Valid Percent | Cumulative Percent |
|---|---|---|---|---|---|
| Valid | Strongly Disagree | | | | |
| | Disagree | | | | |
| | Neither Agree nor Disagree | 1 | 9.1 | 12.5 | 12.5 |
| | Agree | 2 | 18.2 | 25.0 | 37.5 |
| | Strongly Agree | 5 | 45.5 | 62.5 | 100.0 |
| | Total | 8 | 72.7 | 100.0 | |

Autonomous teammates within the STE must be perceived as cognitively similar to their human teammates.

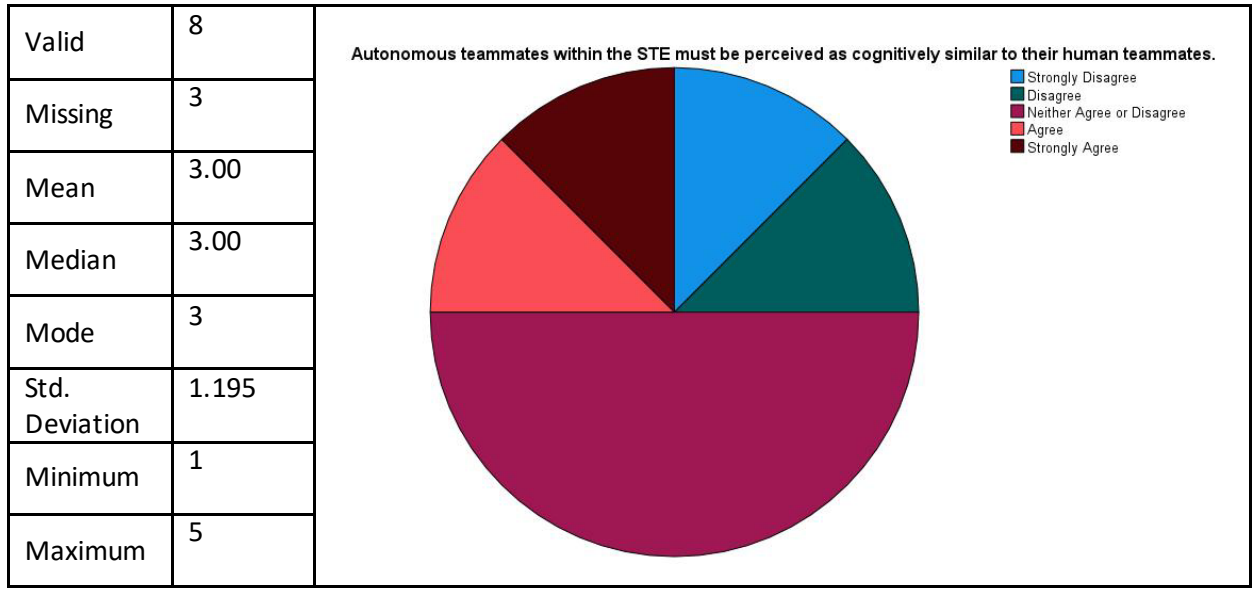

| Valid | 8 |
|---|---|
| Missing | 3 |
| Mean | 3.00 |
| Median | 3.00 |
| Mode | 3 |
| Std. Deviation | 1.195 |
| Minimum | 1 |
| Maximum | 5 |

| | | Frequency | Percent | Valid Percent | Cumulative Percent |
|---|---|---|---|---|---|
| Valid | Strongly Disagree | 1 | 9.1 | 12.5 | 12.5 |
| | Disagree | 1 | 9.1 | 12.5 | 25.0 |
| | Neither Agree nor Disagree | 4 | 36.4 | 50.0 | 75.0 |
| | Agree | 1 | 9.1 | 12.5 | 87.5 |
| | Strongly Agree | 1 | 9.1 | 12.5 | 100.0 |
| | Total | 8 | 72.7 | 100.0 | |

Autonomous teammates within the STE must be perceived as having complementary skills by their human teammates.

| | |
|---|---|
| Valid | 8 |
| Missing | 3 |
| Mean | 4.13 |
| Median | 4.00 |
| Mode | 4 |
| Std. Deviation | .641 |
| Minimum | 3 |
| Maximum | 5 |

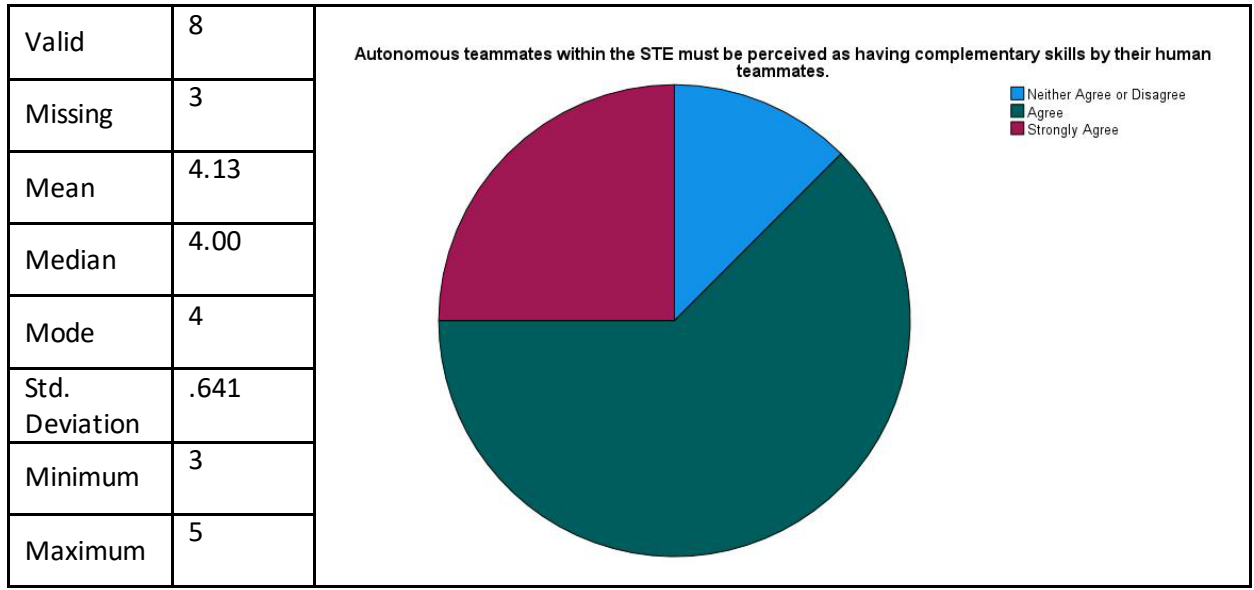

| | | Frequency | Percent | Valid Percent | Cumulative Percent |
|---|---|---|---|---|---|
| Valid | Strongly Disagree | | | | |
| | Disagree | | | | |
| | Neither Agree nor Disagree | 1 | 9.1 | 12.5 | 12.5 |
| | Agree | 5 | 45.5 | 62.5 | 75.0 |
| | Strongly Agree | 2 | 18.2 | 25.0 | 100.0 |
| | Total | 8 | 72.7 | 100.0 | |

This Page Intentionally Blank

*Appendix H*

**Results for Scenario Authoring Likert Items**

The authoring environment should make it easy to modulate the "relatedness" embodied in a task (i.e., the degree to which a given task can be performed by an individual).

| | |
|---|---|
| Valid | 7 |
| Missing | 4 |
| Mean | 3.86 |
| Median | 4.00 |
| Mode | Neither Agree or Disagree |
| Std. Deviation | .900 |
| Minimum | 3 |
| Maximum | 5 |

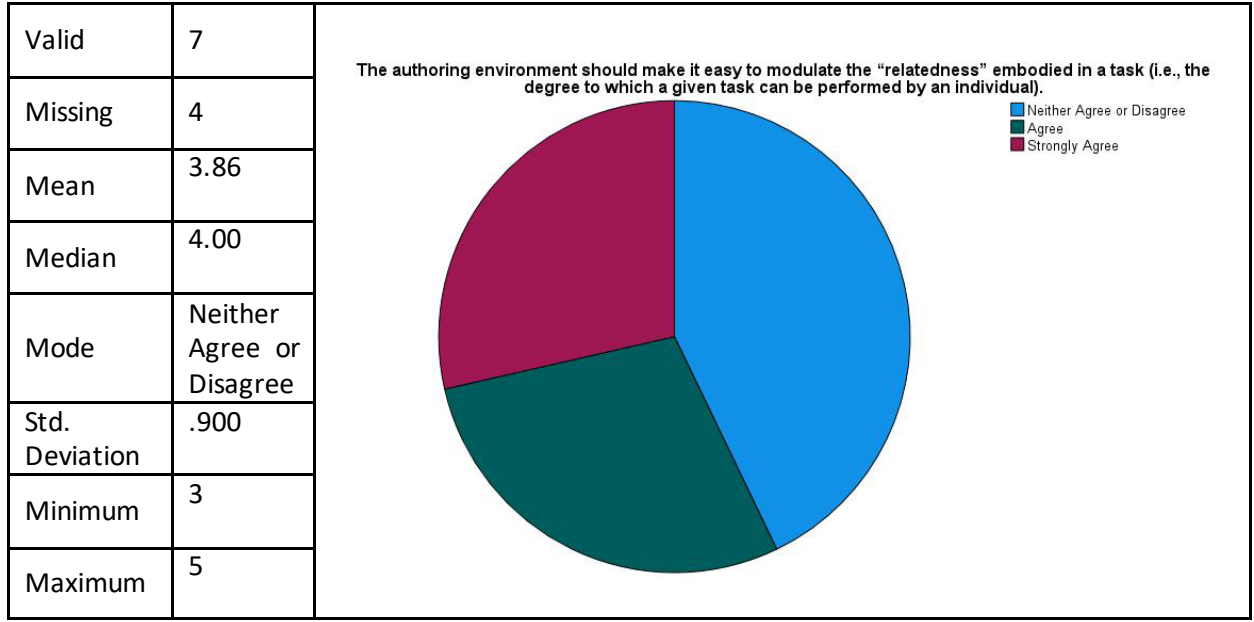

| | | Frequency | Percent | Valid Percent | Cumulative Percent |
|---|---|---|---|---|---|
| Valid | Strongly Disagree | | | | |
| | Disagree | | | | |
| | Neither Agree nor Disagree | 3 | 27.3 | 42.9 | 42.9 |
| | Agree | 2 | 18.2 | 28.6 | 71.4 |
| | Strongly Agree | 2 | 18.2 | 28.6 | 100.0 |
| | Total | 7 | 63.6 | 100.0 | |

The authoring environment should make it easy to modulate the "workflow" embodied in a task (i.e., the degree to which work products pass among members of a team).

| Valid | 7 |
|---|---|
| Missing | 4 |
| Mean | 4.00 |
| Median | 4.00 |
| Mode | Agree |
| Std. Deviation | .816 |
| Minimum | 3 |
| Maximum | 5 |

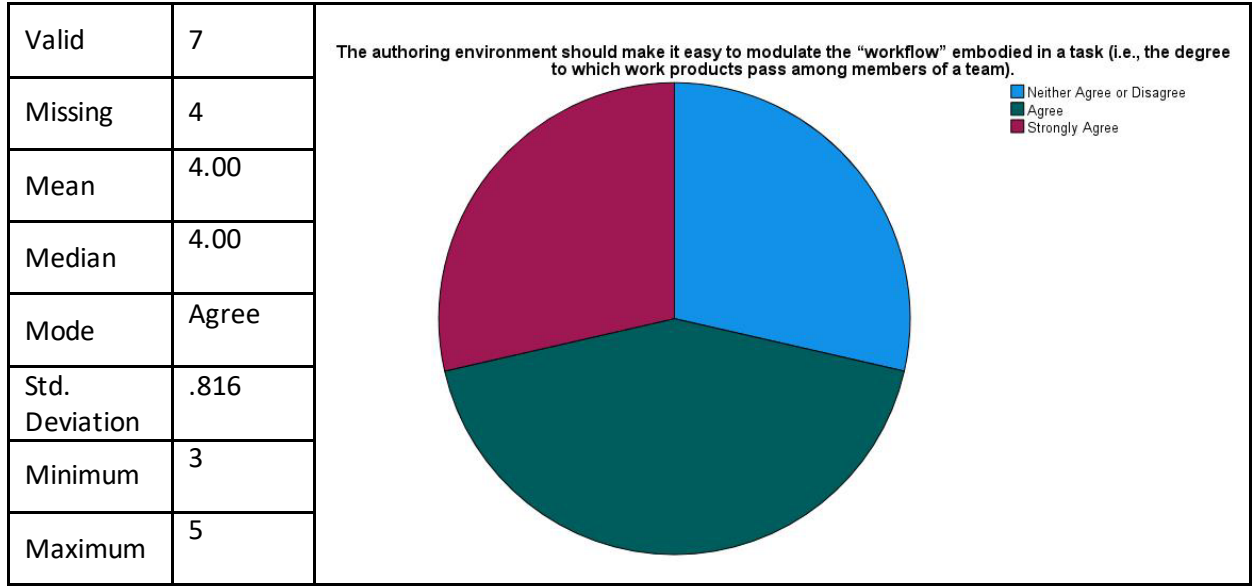

|  |  | Frequency | Percent | Valid Percent | Cumulative Percent |
|---|---|---|---|---|---|
| Valid | Strongly Disagree |  |  |  |  |
|  | Disagree |  |  |  |  |
|  | Neither Agree nor Disagree | 2 | 18.2 | 28.6 | 28.6 |
|  | Agree | 3 | 27.3 | 42.9 | 71.4 |
|  | Strongly Agree | 2 | 18.2 | 28.6 | 100.0 |
|  | Total | 7 | 63.6 | 100.0 |  |

The authoring environment should make it easy to modulate the "difficulty" of a task (i.e., the likelihood that a team will achieve a favorable outcome).

| | |
|---|---|
| Valid | 7 |
| Missing | 4 |
| Mean | 4.43 |
| Median | 4.00 |
| Mode | Agree |
| Std. Deviation | .535 |
| Minimum | 4 |
| Maximum | 5 |

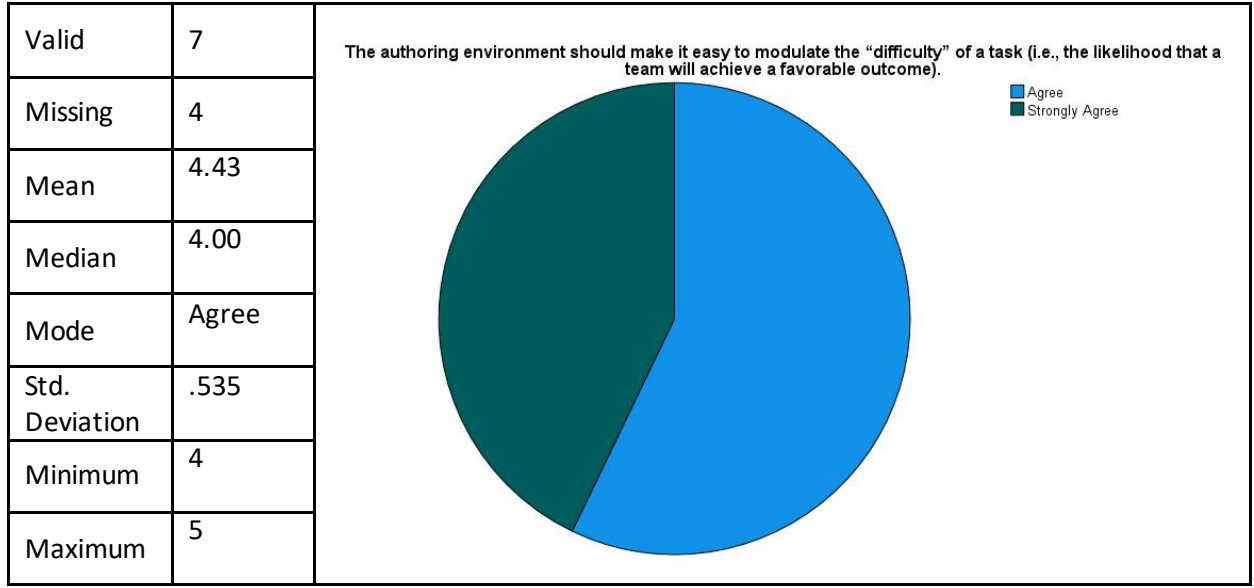

| | | Frequency | Percent | Valid Percent | Cumulative Percent |
|---|---|---|---|---|---|
| Valid | Strongly Disagree | | | | |
| | Disagree | | | | |
| | Neither Agree nor Disagree | | | | |
| | Agree | 4 | 36.4 | 57.1 | 57.1 |
| | Strongly Agree | 3 | 27.3 | 42.9 | 100.0 |
| | Total | 7 | 63.6 | 100.0 | |

The authoring environment should make it easy to modulate the "uncertainty" inherent in a task (i.e., the clarity or lack thereof regarding what is happening in the environment).

| | |
|---|---|
| Valid | 7 |
| Missing | 4 |
| Mean | 4.43 |
| Median | 4.00 |
| Mode | Agree |
| Std. Deviation | .535 |
| Minimum | 4 |
| Maximum | 5 |

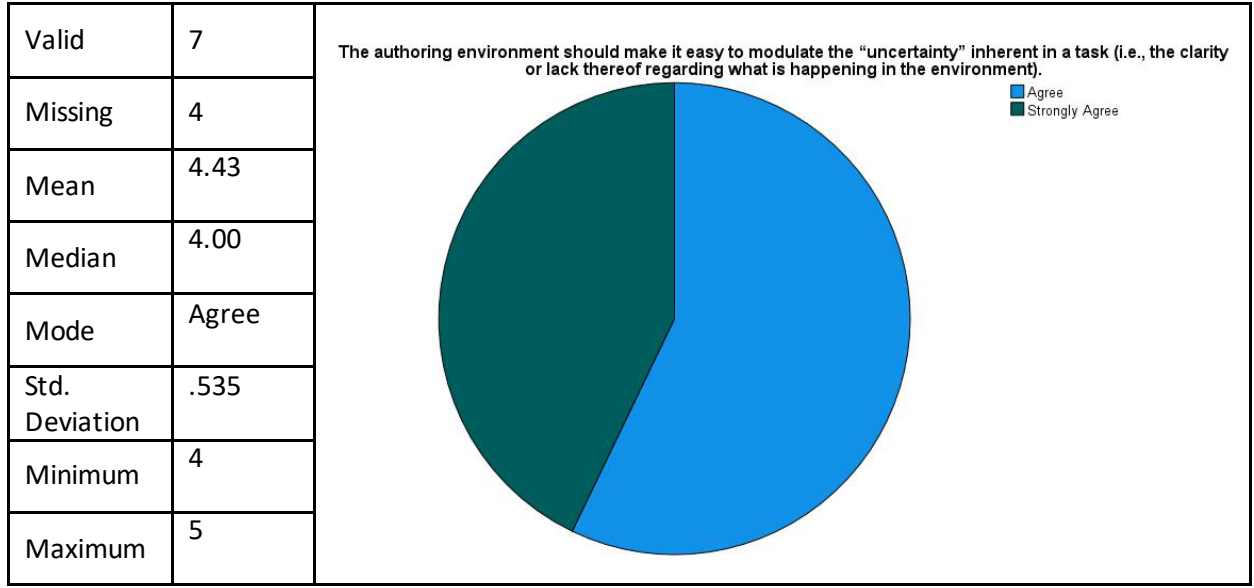

| | | Frequency | Percent | Valid Percent | Cumulative Percent |
|---|---|---|---|---|---|
| Valid | Strongly Disagree | | | | |
| | Disagree | | | | |
| | Neither Agree nor Disagree | | | | |
| | Agree | 4 | 36.4 | 57.1 | 57.1 |
| | Strongly Agree | 3 | 27.3 | 42.9 | 100.0 |
| | Total | 7 | 63.6 | 100.0 | |

The authoring environment should make it easy to modulate the "time pressure" inherent in a task (i.e., the time available for a task vs. the time the task requires).

| Valid | 7 |
|---|---|
| Missing | 4 |
| Mean | 4.00 |
| Median | 4.00 |
| Mode | Agree |
| Std. Deviation | .577 |
| Minimum | 3 |
| Maximum | 5 |

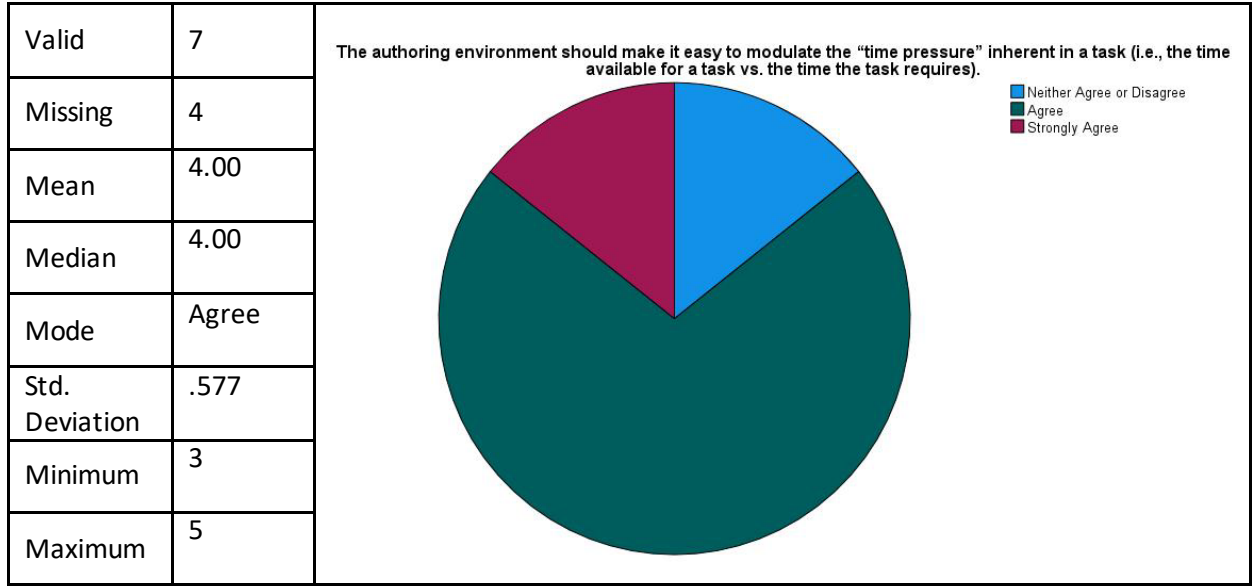

|  |  | Frequency | Percent | Valid Percent | Cumulative Percent |
|---|---|---|---|---|---|
| Valid | Strongly Disagree |  |  |  |  |
|  | Disagree |  |  |  |  |
|  | Neither Agree nor Disagree | 1 | 9.1 | 14.3 | 14.3 |
|  | Agree | 5 | 45.5 | 71.4 | 85.7 |
|  | Strongly Agree | 1 | 9.1 | 14.3 | 100.0 |
|  | Total | 7 | 63.6 | 100.0 |  |

The authoring environment should make it easy to define the number of roles that comprise a team.

| | | |
|---|---|---|
| Valid | 7 | |
| Missing | 4 | |
| Mean | 4.00 | |
| Median | 4.00 | |
| Mode | Agree | |
| Std. Deviation | .577 | |
| Minimum | 3 | |
| Maximum | 5 | |

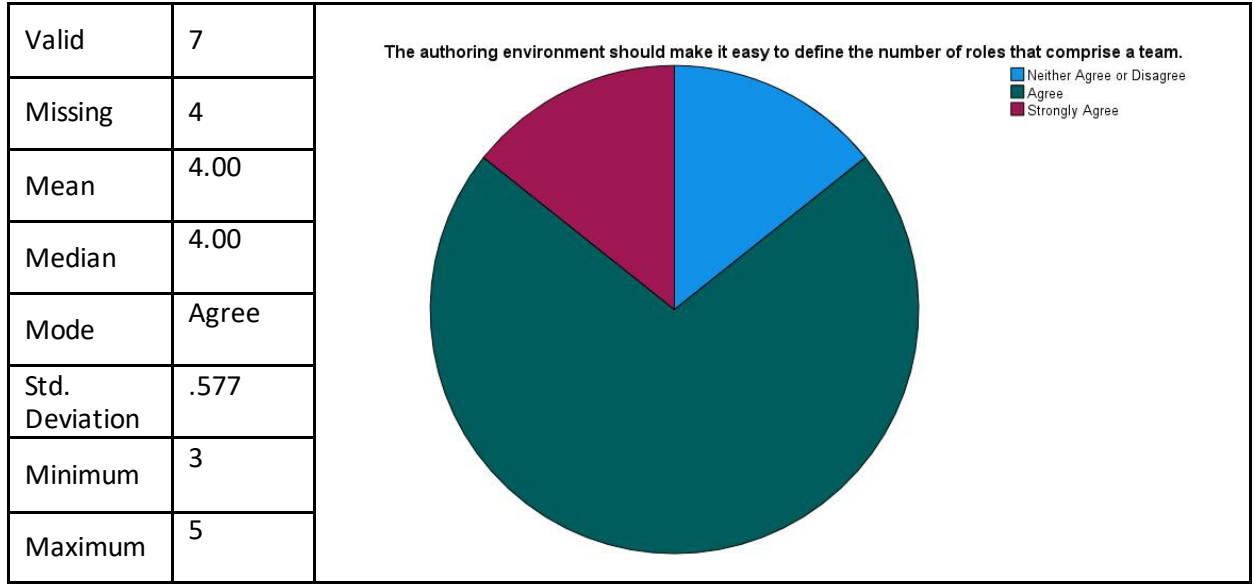

| | | Frequency | Percent | Valid Percent | Cumulative Percent |
|---|---|---|---|---|---|
| Valid | Strongly Disagree | | | | |
| | Disagree | | | | |
| | Neither Agree nor Disagree | 1 | 9.1 | 14.3 | 14.3 |
| | Agree | 5 | 45.5 | 71.4 | 85.7 |
| | Strongly Agree | 1 | 9.1 | 14.3 | 100.0 |
| | Total | 7 | 63.6 | 100.0 | |

The authoring environment should make it easy to specify which roles should be staffed by humans and which should be staffed by agents.

| | |
|---|---|
| Valid | 7 |
| Missing | 4 |
| Mean | 4.00 |
| Median | 4.00 |
| Mode | Agree |
| Std. Deviation | 1.000 |
| Minimum | 2 |
| Maximum | 5 |

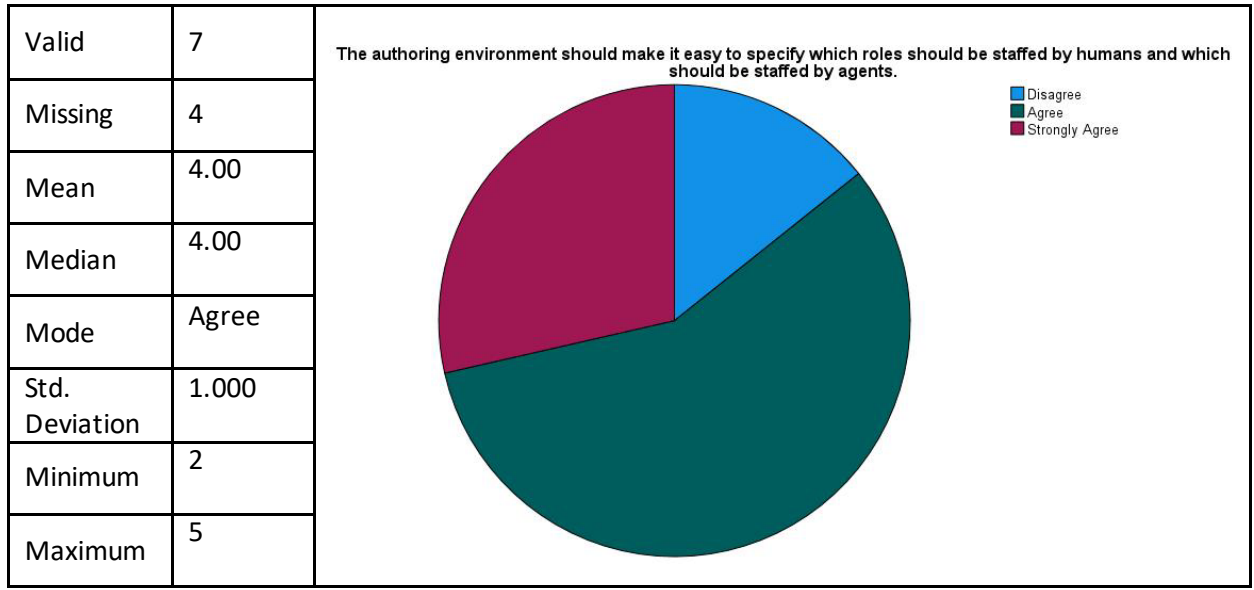

| | | Frequency | Percent | Valid Percent | Cumulative Percent |
|---|---|---|---|---|---|
| Valid | Strongly Disagree | | | | |
| | Disagree | 1 | 9.1 | 14.3 | 14.3 |
| | Neither Agree nor Disagree | 0 | 0.0 | 0.0 | 14.3 |
| | Agree | 4 | 36.4 | 57.1 | 71.4 |
| | Strongly Agree | 2 | 18.2 | 28.6 | 100.0 |
| | Total | 7 | 63.6 | 100.0 | |

*Appendix I*

**Results for System Architecture Likert Items**

The STE must allow geographically distributed labs to replicate a given task environment.

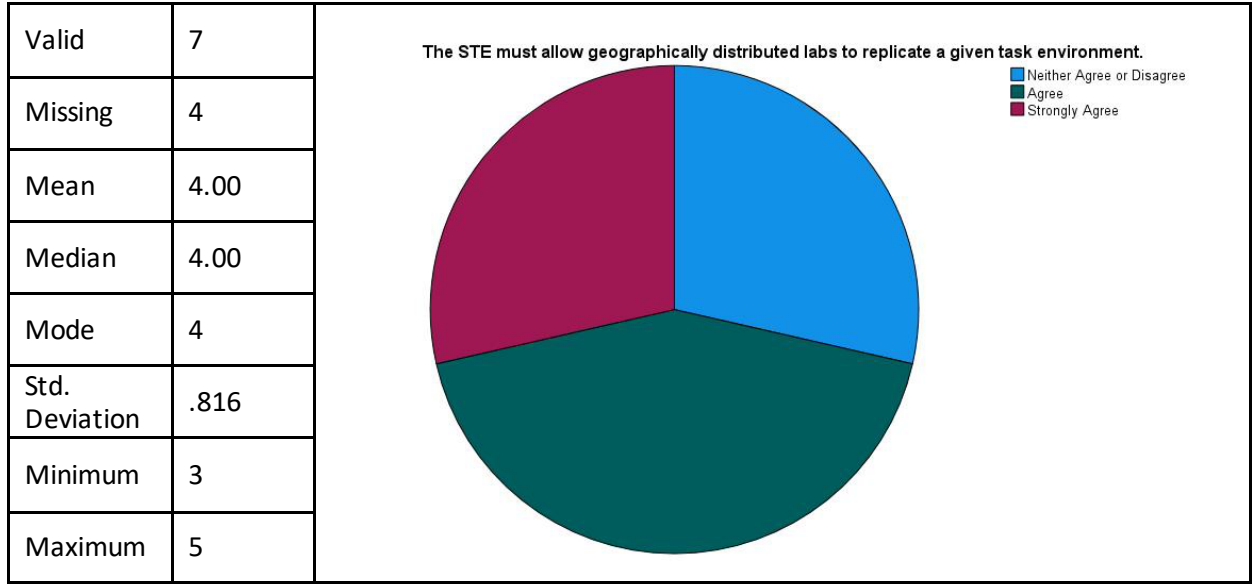

| Valid | 7 |
|---|---|
| Missing | 4 |
| Mean | 4.00 |
| Median | 4.00 |
| Mode | 4 |
| Std. Deviation | .816 |
| Minimum | 3 |
| Maximum | 5 |

|  |  | Frequency | Percent | Valid Percent | Cumulative Percent |
|---|---|---|---|---|---|
| Valid | Strongly Disagree |  |  |  |  |
|  | Disagree |  |  |  |  |
|  | Neither Agree nor Disagree | 2 | 18.2 | 28.6 | 28.6 |
|  | Agree | 3 | 27.3 | 42.9 | 71.4 |
|  | Strongly Agree | 2 | 18.2 | 28.6 | 100.0 |
|  | Total | 7 | 63.6 | 100.0 |  |

The STE must support research on distributed teams (i.e., interdependent but physically separated teams).

| Valid | 8 |
|---|---|
| Missing | 3 |
| Mean | 4.38 |
| Median | 4.00 |
| Mode | Agree |
| Std. Deviation | .518 |
| Minimum | 4 |
| Maximum | 5 |

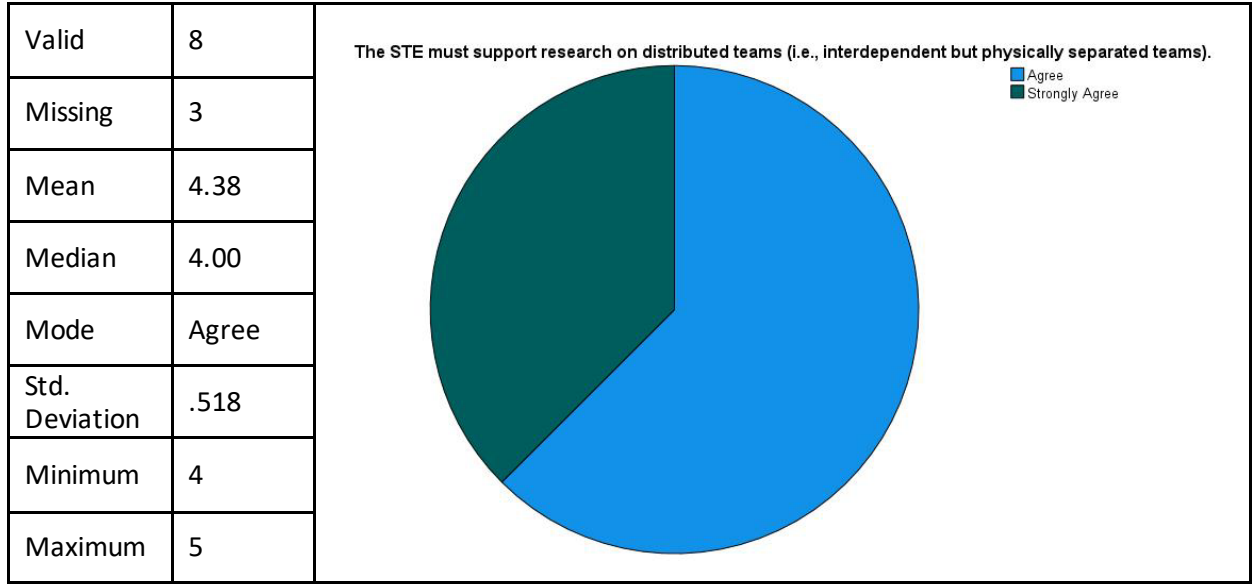

| | | Frequency | Percent | Valid Percent | Cumulative Percent |
|---|---|---|---|---|---|
| Valid | Strongly Disagree | | | | |
| | Disagree | | | | |
| | Neither Agree nor Disagree | | | | |
| | Agree | 5 | 45.5 | 62.5 | 62.5 |
| | Strongly Agree | 3 | 27.3 | 37.5 | 100.0 |
| | Total | 8 | 72.7 | 100.0 | |

My research team includes individuals with the software engineering skills required to make modifications to an open-source STE.

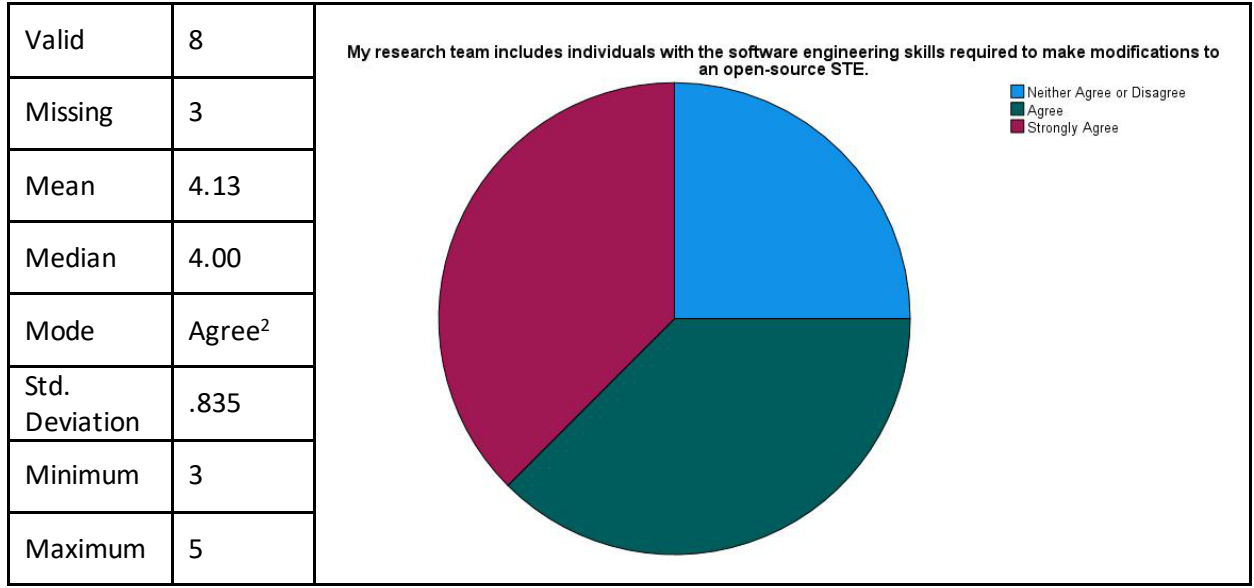

| | |
|---|---|
| Valid | 8 |
| Missing | 3 |
| Mean | 4.13 |
| Median | 4.00 |
| Mode | Agree[2] |
| Std. Deviation | .835 |
| Minimum | 3 |
| Maximum | 5 |

| | | Frequency | Percent | Valid Percent | Cumulative Percent |
|---|---|---|---|---|---|
| Valid | Strongly Disagree | | | | |
| | Disagree | | | | |
| | Neither Agree nor Disagree | 2 | 18.2 | 25.0 | 25.0 |
| | Agree | 3 | 27.3 | 37.5 | 62.5 |
| | Strongly Agree | 3 | 27.3 | 37.5 | 100.0 |
| | Total | 8 | 72.7 | 100.0 | |